%% file: survey_cr_v2.tex
\documentclass[journal]{IEEEtran}

\usepackage[english]{babel}
\usepackage[T1]{fontenc}
\usepackage{cite,url,color} 
\usepackage{graphics,amsfonts}
\usepackage{epstopdf}
\usepackage[pdftex]{graphicx}
\usepackage[cmex10]{amsmath}
\usepackage[utf8]{inputenc}
\usepackage{enumitem}

\usepackage{tikz}
    \usetikzlibrary{shapes.arrows,automata,positioning,chains,shapes,arrows}
\usepackage{pgfplots}
\usepgfplotslibrary{fillbetween}
\usetikzlibrary{plotmarks}
\newlength\fheight
\newlength\fwidth
\pgfplotsset{compat=newest}
\pgfplotsset{plot coordinates/math parser=false}

\usepackage{colortbl}
\usepackage{tabularx}
\usepackage{array}
\usepackage{dblfloatfix}
\usepackage{subfig}

\usepackage{indentfirst}

\usepackage{times}

\usepackage{breqn}
\usepackage{booktabs}
\usepackage[nomain,xindy,acronym,nonumberlist]{glossaries}

\definecolor{mycolor1}{rgb}{0.00000,0.44700,0.74100}%
\definecolor{mycolor2}{rgb}{0.85000,0.32500,0.09800}%
\definecolor{mycolor3}{rgb}{0.92900,0.69400,0.12500}%
\definecolor{mycolor4}{rgb}{0.49400,0.18400,0.55600}%
\definecolor{mycolor5}{rgb}{0.46600,0.67400,0.18800}%
\definecolor{mycolor6}{rgb}{0.30100,0.74500,0.93300}%

\definecolor{violet}{rgb}{0.6,0,0.6}%
\definecolor{orange_D}{rgb}{1.0000,0.5,0}%
\definecolor{cyan}{rgb}{0,0.67,0.64}%
\definecolor{red}{rgb}{0.9,0,0}%
\definecolor{green}{rgb}{0,0.8,0}%

\def \fwidth{0.8\columnwidth}
\def \fheight {0.54\columnwidth}

\makeglossaries

\input{acronyms.tex}
\makeglossaries

\usepackage{placeins}

\newcommand{\rev}[1]{{#1}}

\begin{document}
\title{A Survey on Recent Advances\\ in Transport Layer Protocols}

\author{\IEEEauthorblockN{Michele Polese, \IEEEmembership{Student Member, IEEE}, Federico Chiariotti, \IEEEmembership{Member, IEEE}, Elia Bonetto, Filippo Rigotto, Andrea Zanella, \IEEEmembership{Senior Member, IEEE}, Michele Zorzi, \IEEEmembership{Fellow, IEEE}}
\thanks{The authors are with the Department of Information Engineering (DEI), University of Padova, Padova, 35131, Italy.
Email:\{polesemi, chiariot, bonettoe, rigottof, zanella, zorzi\}@dei.unipd.it}
\thanks{This work was partially supported by the program Supporting Talent in Research@University of Padua: STARS Grants, through the project "Cognition-Based Networks: Building the Next Generation of Wireless Communications Systems Using Learning and Distributed Intelligence," and by the US Army Research Office under Grant no. W911NF1910232: "Towards Intelligent Tactical Ad hoc Networks (TITAN)."}

}

\maketitle

\begin{abstract}
Over the years, the Internet has been enriched with new available communication technologies, for both fixed and mobile networks and devices, exhibiting an impressive growth in terms of performance, with steadily increasing available data rates. The Internet research community has kept trying to evolve the transport layer protocols to match the capabilities of modern networks, in order to fully reap the benefits of the new communication technologies. This paper surveys the main novelties related to transport protocols that have been recently proposed, identifying three main research trends: (i) the evolution of congestion control algorithms, to target optimal performance in challenging scenarios, possibly with the application of machine learning techniques; (ii) the proposal of brand new transport protocols, alternative to the \gls{tcp} and implemented in the user-space; and (iii) the introduction of multipath capabilities at the transport layer.
\end{abstract}

\begin{picture}(0,0)(0,-330)
\put(0,0){
\put(0,0){\small This paper has been accepted for publication in IEEE Communications Surveys \& Tutorials, DOI 10.1109/COMST.2019.2932905.}
\put(0,-10){\small This is a preprint version of the accepted paper. Copyright (c) 2019 IEEE. Personal use of this material is permitted.}
}
\end{picture}


\glsresetall

\section{Introduction}\label{sec:intro}

The communication technologies that provide access and backhaul connectivity to the Internet have dramatically changed since the 1980s, when the protocols that are part of today's TCP/IP stack were first introduced~\cite{rfc793,rfc791,rfc768}. Traffic generated on mobile devices is expected to exceed desktop and server traffic by 2021~\cite{ciscoVni}. New communication standards are being proposed and launched to market every few years. Driven by the increase in web and multimedia traffic demand, mobile and fixed networks are rapidly evolving. 3GPP NR~\cite{38300,8258595} will bring ultra-high data rates with low latency to future 5G devices, and, similarly, the IEEE 802.11 standard will target ultra-dense deployments~\cite{7422404} and multi-gigabit-per-second throughput~\cite{8088544}. Modern devices are capable of connecting to heterogeneous networks~\cite{7432153}, and fixed networks are using new optical technologies and \gls{sdn} for unprecedented rates and low latency~\cite{7947156}.

The increasing capabilities of the network make new kinds of applications possible; the exponential growth of multimedia or real-time traffic~\cite{ciscoVni} would have been impossible without the recent technological advances. As networks progress towards 5G, new kinds of applications, such as \gls{ar}\footnote{\rev{A comprehensive list of acronyms is provided at the end of the paper.}} and \gls{vr} or autonomous cooperative driving, are going to require more from the network and impose ever more stringent \gls{qos} constraints. This, along with the increasing heterogeneity of the network, makes the role of transport protocols more important, and, at the same time, more challenging. Indeed, the end-to-end performance and the \gls{qoe} of the users largely depend on the interaction among the applications, the transport layer and the underlying network~\cite{liu2008experiences}. \rev{In particular, the transport layer, which is responsible for the management of the end-to-end connection over any number of network hops, has to adapt and evolve in order to let users fully benefit from the aforementioned innovations.} However,
a number of factors prevent new solutions at the transport layer from being widely adopted, and, in recent years, the research community has been forced to cope with these limitations and identify innovative solutions in order to have significant effects on Internet performance.

In particular, the deployment of alternative transport protocols, such as the \gls{sctp}~\cite{rfc4960}, is slowed down by the widespread use of middleboxes\cite{rfc3303,rfc3234}, which often drop packets from protocols which are different from the \gls{tcp} and/or the \gls{udp}~\cite{8064352}. Moreover, the socket \gls{api} (offered by the Operating System kernel and supported by \gls{tcp}/\gls{udp}) is \rev{almost universally used}~\cite{7738442}, thus limiting the interfacing options between application and transport layers to what the \gls{api} supports. Finally, the most widespread Operating Systems implement the transport functionalities (i.e., \gls{tcp} and \gls{udp}) in the kernel, making the deployment of new solutions difficult. These elements define what is called transport layer ossification~\cite{7738442}, a phenomenon which has pushed developers and researchers to only use legacy \gls{tcp} (for reliable traffic and congestion control), even though it may not be the best performing protocol for the desired use case. \gls{tcp} has indeed some performance issues in specific scenarios, e.g., on wireless links with high variability~\cite{6082645,zhang2018will}, \gls{hol} blocking with web traffic~\cite{4150963}, and bufferbloat~\cite{Gettys:2011:BDB:2063166.2071893}.

This survey focuses on three directions in transport layer research (i.e., new transport protocols, congestion control innovations and multipath approaches) that have emerged in the last fifteen years to solve the aforementioned problems. New congestion protocols have been proposed to target low latency and full bandwidth utilization~\cite{BBR,8109356}. Novel transport protocols have been discussed by the \gls{ietf}, with technical novelties such as multipath capabilities, to exploit the multiple interfaces available in modern smartphones, computers and servers, and user space implementations, to overcome the ossification that prevents a widespread adoption of novel algorithms at the transport layer. Therefore, in this survey, we first review the main new transport protocols that have been proposed or standardized by the \gls{ietf} since 2006: we provide a brief overview on \gls{sctp} and \gls{dccp}~\cite{rfc4340}, and then delve into a more recent contribution, i.e., the \gls{quic}~\cite{draftquicktr} protocol. Then, we review the research on congestion control. We describe both new mechanisms using classic approaches, e.g., \gls{bbr} and \gls{lola} for \gls{tcp}, and some novel proposals for using machine learning techniques for congestion control.
Finally, the third trend we discuss in this survey is related to the adoption of multipath solutions at the transport layer, mainly with \gls{mptcp}~\cite{rfc6182,rfc6356}, but also with multipath extensions for \gls{sctp}~\cite{wallace2014concurrent}, \gls{quic}~\cite{deconinck-multipath-quic-00,8422951}, and with the \gls{leap}~\cite{LEAP}. 

Our goal in this survey is to offer the interested reader a comprehensive point of view on the up-to-date research on transport protocols, which is lacking in other recent surveys that separately focus on ossification~\cite{7738442}, multipath transmissions~\cite{7501871}, or congestion control schemes for \gls{mptcp}~\cite{7460213} \rev{or in data centers~\cite{7096919}}.

The remainder of the paper is organized as follows. In Sec.~\ref{sec:issues} we describe the main issues and limitations of \rev{transport protocols} in modern networks. 
Then, in Sec.~\ref{sec:transport}, we describe three recently proposed transport protocols, focusing in particular on \gls{quic}. In Sec.~\ref{sec:cc} we report the latest proposals in terms of congestion control algorithms for \gls{tcp} and other transport protocols, classifying them as loss-, delay- or capacity-based. We also discuss hybrid mechanisms, machine-learning-based algorithms and cross-layer approaches. 
In Sec.~\ref{sec:mp} we investigate the different approaches to multipath transport protocols recently proposed. Finally, Sec.~\ref{sec:conclusion} concludes the paper and summarizes the main future research directions. A comprehensive list of acronyms is also provided at the end of the paper.

\section{\rev{Transport Layer} Limitations in Modern Networks}
\label{sec:issues}

\rev{As mentioned in Sec.~\ref{sec:intro}, transport layer protocols have an end-to-end view of the connection: they do not consider each individual hop, but only a single logical link between the two endpoints.} For this reason, the most important link in the connection is the slowest one, that is the so-called bottleneck. The service provided by \gls{tcp}, the \emph{de facto} standard transport protocol of the modern Internet, can then be roughly modeled as a single pipe with the capacity of the bottleneck link and a certain \gls{rtt}, i.e., the time from the instant the sender transmits a packet to the instant it receives the corresponding acknowledgment. 

However, the particular features of the individual links and the behavior of lower layers do influence \gls{tcp}'s behavior,  as well as that of other transport protocols: several properties of the links composing the end-to-end connection (e.g., latency, packet loss, buffer state and size, and volatility of the capacity) can affect the transport layer performance~\cite{10gComparison}. In order to overcome this problem, more and more complex \rev{congestion control mechanisms} have been proposed, going far beyond the original simple \gls{aimd} principle. We will describe the main design philosophies in greater detail in Sec.~\ref{sec:cc}.
In fact, researchers are questioning \gls{tcp}'s extensive use as a one-size-fits-all solution, because all these factors can cause performance issues that are becoming more and more apparent. Some of these issues are fundamental problems of the transport layer abstraction, while others depend on the specific features of the protocol and can be avoided by designing the protocol correctly. This section provides a short review of some of the most important issues, while the rest of the paper is dedicated to the discussion of the several solutions that have been proposed to address these challenges.

\subsection{Bufferbloat}\label{ssec:bufferbloat}

\rev{As we will discuss in depth in Sec.~\ref{sec:cc}, congestion control mechanisms} exploit an abstract view of the underlying network to tune the amount of data to be sent. As shown in~\cite{Gettys:2011:BDB:2063166.2071893}, however, this abstraction in some instances fails to provide accurate information on the links connecting the two hosts and leads to degraded performance. In particular, when large buffers are deployed before a bottleneck in order to prevent packet losses, then loss-based \gls{tcp} probing mechanisms increase the queue occupancy, thus causing a spike in latency. Moreover, since the currently implemented versions of \gls{tcp} will keep increasing the sending rate until the first packet loss, they will often overshoot the capacity of the channel, increasing congestion and causing multiple retransmissions when the queue is eventually filled. \rev{Other protocols such as \gls{quic}, \gls{sctp}, and \gls{dccp} face the same issue, since it is a fundamental problem of congestion control with large buffers and not a protocol-specific problem.}

This phenomenon, known as \textit{bufferbloat}, degrades the \gls{qos} of applications, in particular when video or file transfer flows share the buffer with web browsing flows, and it has worsened in recent years mainly due to loss-preventing design strategies that place large buffers in front of low capacity access links (either wired or wireless)~\cite{Bufferbloat}.

The research in this area aims at solving this issue with local \gls{aqm} techniques or end-to-end flow and congestion control for transport protocols.

The problem, while not being hard to detect, is hard to solve without a significant overhead cost~\cite{BufferbloatAQM}. The congestion control protocols at the endpoints might also be different, making \gls{aqm} more complex; this information is often not even available to routers, which might not be able to predict the consequences of discarding a packet on congestion, making the algorithms extremely complex and sensitive to the parameter settings.

\gls{aqm} is not a recent idea: numerous techniques were proposed, such as \gls{red}, and we refer the reader to the extensive literature cited in \cite{BufferbloatAQM,8000042} for further information on this subject.
Despite these many proposals, they have encountered limited adoption, partly due to the above-mentioned issue of parameter tuning and the computational cost of the algorithms\cite{Bufferbloat}. More recently, new and easier ways to tune and deploy \gls{aqm} schemes, such as \gls{codel}~\cite{Nichols:2012:CQD:2209249.2209264}, have been adopted in several commercial products. \gls{codel} is an \gls{aqm} algorithm that limits the buffering latency by monitoring the queueing delay $D$ in an interval (typically of 100 ms) and dropping packets when the minimum value of $D$ is larger than 5 ms.
Nonetheless, as we will discuss in Sec.~\ref{sec:wireless}, the bufferbloat issue remains relevant in the wireless domain (e.g., at mmWave frequencies~\cite{zhang2018will}).


\subsection{The Incast issue}\label{ssec:incast}

Data centers are restricted areas containing servers and systems monitoring server's activity, web traffic and performance. The data exchange between servers generally relies on APIs based on the \gls{http}, making \gls{tcp} a widely used transport protocol in data centers. Some activities, such as virtual machine migrations, also generate a high volume of traffic between servers. Therefore, the links in a data center generally have high bandwidth and low latency and delay, while switches have small buffers~\cite{IncastCurbing}, contrary to what usually happens in access links, as mentioned in the previous section. 

\begin{figure}[!t]
\centering
\includegraphics[width=\columnwidth]{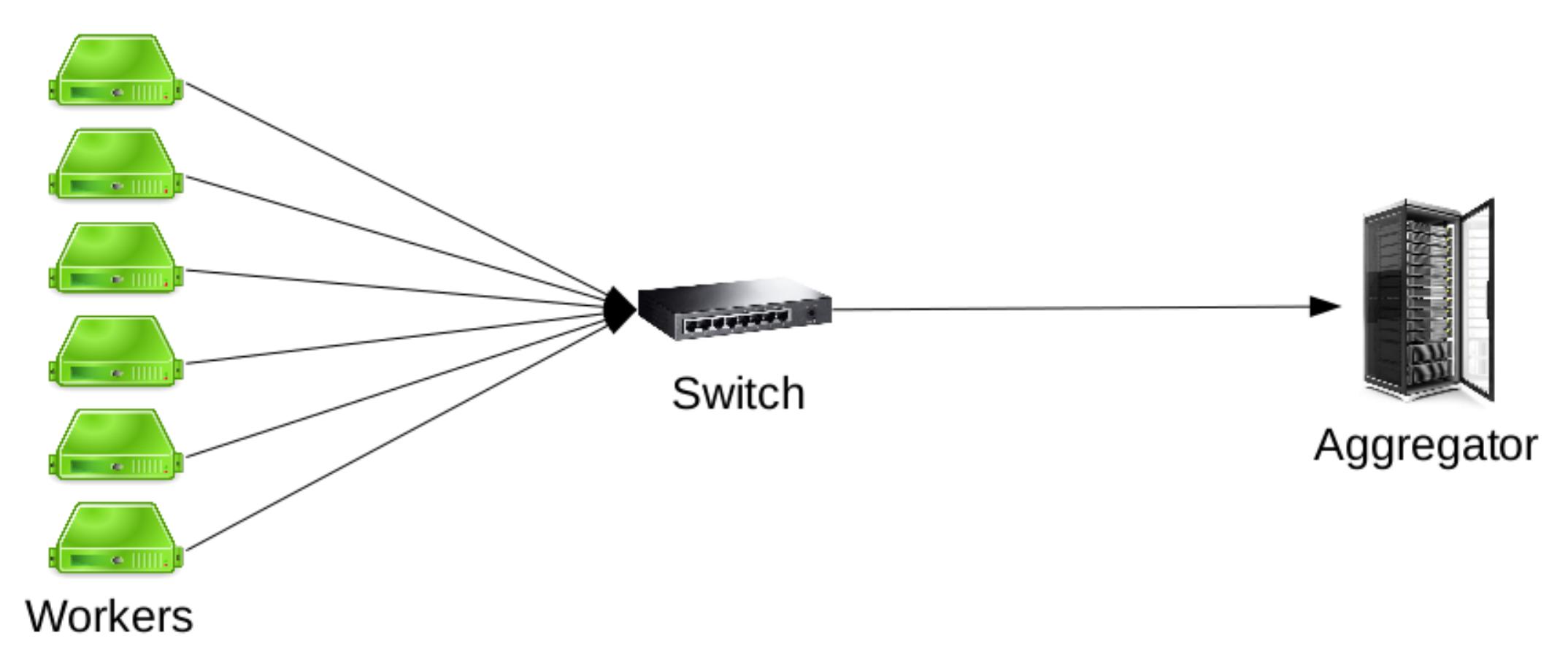}
\caption{A typical scenario in which the \gls{tcp} Incast problem arises.}
\label{fig:Incast}
\end{figure}

Cloud computing frameworks are also widely deployed in large data centers and generate very high traffic loads. For example, MapReduce (which uses a partition/aggregation design pattern)~\cite{MapReduce} or PageRank (used for web search)~\cite{PageRank} often involve many-to-one traffic patterns, where \rev{multiple workers send data simultaneously to a single aggregator node, as shown in Fig.~\ref{fig:Incast}. In this many-to-one scenario, if all the multiple incast flows go through a single switch, its buffer might be insufficient, leading to congestion. The \gls{tcp} loss recovery mechanism will then become less efficient, triggering multiple timeouts and causing throughput collapse and long delays~\cite{IncastUnderstanding}.}

Many attempts have been made to analyze and solve this problem, called the \textit{Incast issue}, that degrades network performance and user experience~\cite{ren2014survey}.
Detailed throughput estimation analysis can be found in \cite{IncastUnderstanding} and \cite{IncastModeling}. Solutions may be classified into four categories, as mentioned in~\cite{IncastUnderstanding}:
(i) system parameters adjustments, like disabling slow start to avoid massive and sudden buffer overflows that cause retransmission timeout;
(ii) enhanced in-network and client-side algorithm design, to reduce waste of bandwidth, minimize retransmission timeouts and the number of packet losses, and to improve the quick recovery of lost packets \cite{IncastCurbing};
(iii) replacement of loss-based congestion control algorithms with better implementations that adjust the congestion window size according to the delay measured from \glspl{rtt}, like Vegas;
and (iv) design of completely new algorithms for this particular environment, like \gls{dctcp} \cite{DCTCP}, that uses \gls{ecn} to provide window-based control methods, or IATCP \cite{IATCP}, a rate-based approach that counts the total number of packets injected to constantly meet the \gls{bdp} of the network \cite{IncastUnderstanding}.
Additional information regarding the transport protocol control in data centers can be found in \cite{SurveyDatacenter}.

\subsection{Latency and Head of Line Blocking}\label{ssec:hol}
\gls{hol} is a phenomenon that happens when two or more independent data flows share \rev{the same \gls{tcp} connection}, as for example with web traffic over \gls{tcp}: since this transport protocol interprets the data it receives as a single continuous stream of bytes, and it requires in-order delivery, one missing packet delays all subsequent packets for any other flows, causing significant delays. \gls{hol} is a well-known problem of any protocol that requires in-order delivery, and the obvious solution of opening one connection for every data flow suffers from significant overhead in connection setup and error recovery. Moreover, with this option, the congestion control is less stable, since each connection performs it independently~\cite{4150963}. 

\input{img/tp-table.tex}

\begin{figure}[!t]
\centering
\includegraphics[width=\columnwidth]{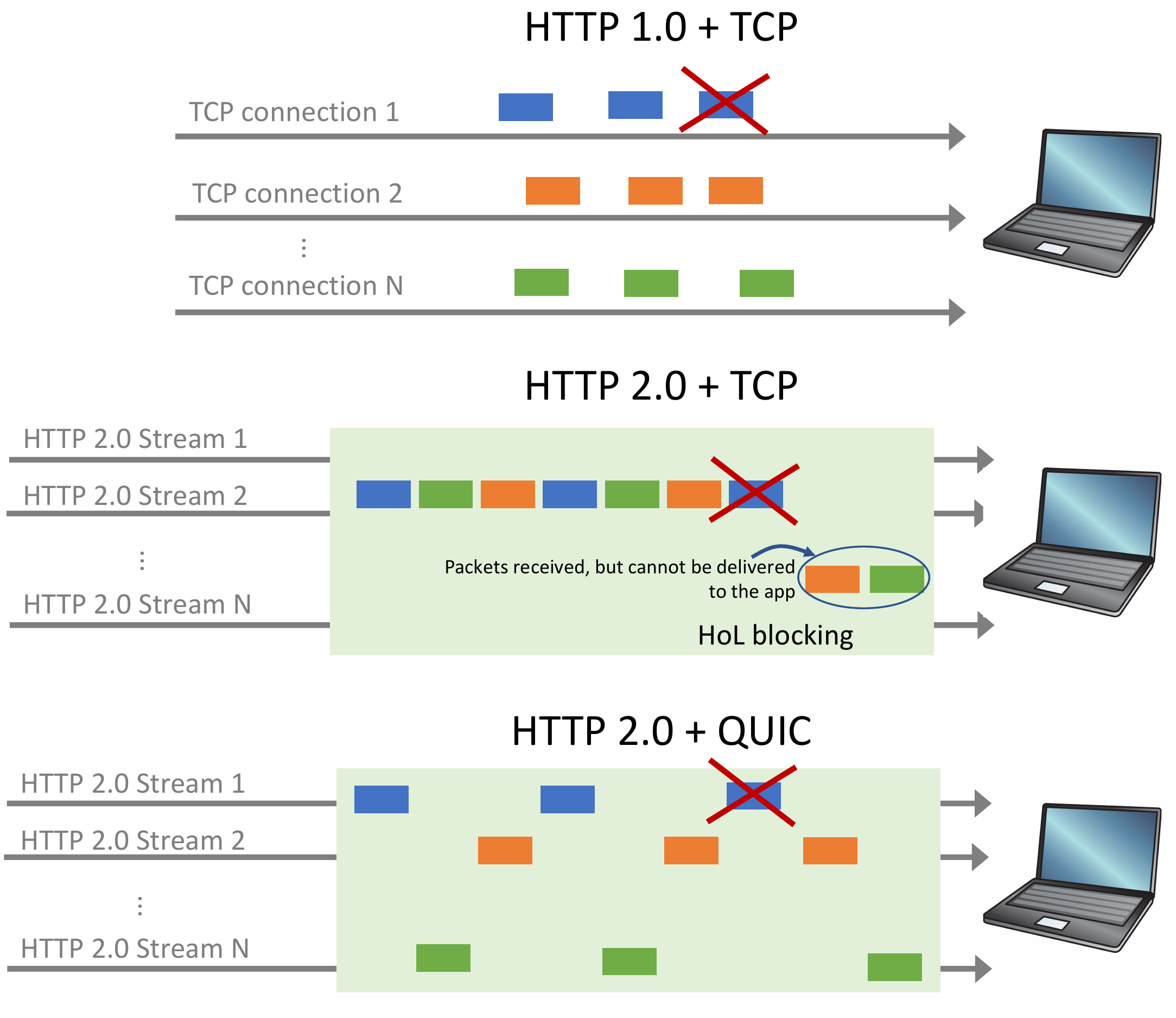}
\caption{The \gls{hol} problem with \gls{tcp} and \gls{http}/2, solved by \gls{quic} thanks to the support of multiple independent streams per connection.}
\label{fig:Multiplexing}
\end{figure}

Web traffic is a textbook example of the \gls{hol} problem: usually, web pages contain several objects, such as text, images, media and third-party scripts; when a client requests a page to the server, each of these objects is downloaded with a single \gls{http} GET request, but they do not need to be displayed at the same time. \gls{http}/1.1 did not allow multiplexing, so the client was forced to open one \gls{tcp} connection for every object, with \rev{the issues described above}. Version 2 of the protocol, introduced in 2015 as RFC 7540~\cite{rfc7540}, was supposed to solve this issue by using a single \gls{tcp} connection to handle all the requests, with significant page load time reductions. However, this advantage is nullified by \gls{hol} in lossy networks: given that all the packets for multiple HTTP 2.0 streams are multiplexed over the same \gls{tcp} connection, which requires in-order delivery, then the loss of a single packet would halt the reception on all streams, and not just on the one interested by the loss, as reported in Fig.~\ref{fig:Multiplexing}. For example, measurements in cellular networks show that HTTP/2 does not have a significant performance advantage over HTTP/1.1~\cite{4150963}.

In principle, in-order delivery is not necessary for reliability, but \rev{transport protocols such as} \gls{tcp} require both, and although a proposal exists to extend \gls{tcp} to allow out-of-order transmission for multimedia services to avoid the \gls{hol} problem~\cite{mcquistin2016tcp}, it has not been widely adopted.
Fig.~\ref{fig:Multiplexing} shows that protocols that support multiple data streams such as \rev{\gls{sctp} or \gls{quic}, which will be described in Sec.~\ref{sec:transport},} can overcome the \gls{hol} problem because there is no single line to block: each stream has its own buffer and in-order delivery is ensured on a per stream basis~\cite{7867726}. 

The multipath scenario is another case in which \gls{tcp} suffers heavily from the \gls{hol} problem; we will describe the issue in detail in Sec.~\ref{sec:mp}.

\subsection{Performance on Wireless Channels}
\label{sec:wireless}
The performance of \gls{tcp} on wireless channels has been investigated since the introduction of the first commercial wireless services in the 1990s~\cite{balakrishnan1997comparison}. \rev{Traditional congestion control schemes} generally react to packet loss by reducing the sending rate, assuming that it is caused by congestion. In wireless links, however, packets may be lost because of degraded channel quality. Therefore, the abstraction of the overall connection is poor, and the end-to-end performance suffers. Common solutions include the provisioning of retransmission and protection mechanisms on the wireless link~\cite{917504} (even though this increases the delay variability), and in-network performance-enhancing proxies (e.g., to split the connection and use a different congestion control algorithm on the wireless link)~\cite{rfc3135}. Several techniques have been proposed throughout the years for different technologies, e.g., LTE~\cite{liu2016improving} and Wi-Fi~\cite{7430293}. We refer the reader to~\cite{liu2016improving,4020606} for a more extensive discussion on the performance of \gls{tcp} on wireless links.

Recently, the adoption of the mmWave frequencies for the next generation of cellular networks~\cite{rappaport1} has sparked a renewed interest in the performance of \gls{tcp} on wireless links. In particular, in this wireless medium the variability of the channel is much higher than at sub-6 GHz frequencies, given the sensitivity to blockage from common materials, and the directionality of the communications. These limitations not only impact the design of the lower layers of the protocol stack, but also affect the transport layer performance, as discussed in~\cite{mmNet,polese2017tcp}. \gls{tcp}'s control loop is indeed too slow to properly react to the dynamics of the channel quality and offered data rate that affect mmWave links. For example, even though bufferbloat is also present in sub-6 GHz cellular networks~\cite{jiang2012understanding}, this phenomenon is much more disruptive at mmWaves~\cite{zhang2018will}, where large buffers are needed to protect from temporary but sudden variations of the link capacity due to, e.g., \gls{los} to \gls{nlos} transitions~\cite{mmNet}. Moreover, it has been shown that \gls{tcp} makes a suboptimal use of the very high mmWave data rates, since it takes a long time to reach full capacity after a loss or at the beginning of a connection~\cite{mmNet,zhang2018will}. Since the issue is inherent in the congestion control logic, any protocol implementing congestion control will need to work efficiently also in a wireless scenario.

\section{\rev{Recent Transport Protocols}}
\label{sec:transport}

In this section, we will focus on the description of three recently proposed transport protocols: \gls{quic}, \gls{sctp} and \gls{dccp}. In particular, we will describe the main features of each protocol, and the novelties introduced with respect to \gls{tcp}. Table~\ref{tab:tpsum} summarizes the main characteristics of these protocols, together with the relevant \gls{ietf} documents and some details of their implementations. The discussion of the congestion control mechanisms that can be used by these protocols and the recent multipath extensions will be presented in later subsections.

\subsection{\acrfull{quic}}
\label{sec:quic}

\gls{quic} is an \textit{application layer} transport mechanism~\cite{kakhki2017taking} introduced by Google in 2013 and currently considered for standardization by the \gls{ietf}~\cite{draftquicktr,draftquickrec}.\footnote{The set of \gls{ietf} drafts related to \gls{quic} can be found at \url{https://datatracker.ietf.org/wg/quic/documents/}.} \gls{quic} is not implemented in the kernel, but distributed either as a user-space library or in the application itself, and relies on \gls{udp} as the underlying transport protocol.  
The main motivations related to the introduction of \gls{quic} are discussed in~\cite{Langley-2017-QTP}. While several alternatives to \gls{tcp} have been recently proposed at the transport layer, their adoption is not widespread, due to the ossification caused by the presence of middleboxes in the network that drop packets from unknown protocols, i.e., protocols different from \gls{tcp} and/or \gls{udp}.\footnote{Notice that some firewalls and middleboxes also drop \gls{udp} traffic~\cite{kakhki2017taking}.} \gls{quic} encapsulates its packets into \gls{udp} datagrams, thus avoiding most of the incompatibilities with middleboxes. Secondly, as we mentioned in Sec.~\ref{sec:intro}, \gls{tcp} is implemented in the kernel, and any change or new congestion control algorithm requires an operating system update, which is not always feasible. Finally, in terms of protocol design, \gls{quic} aims at solving the \gls{hol} issue of \gls{tcp}, described in Sec.~\ref{ssec:hol}, and at reducing the handshake delay at the beginning of a connection.


\begin{figure}[t]
\centering
\includegraphics[width=.85\columnwidth]{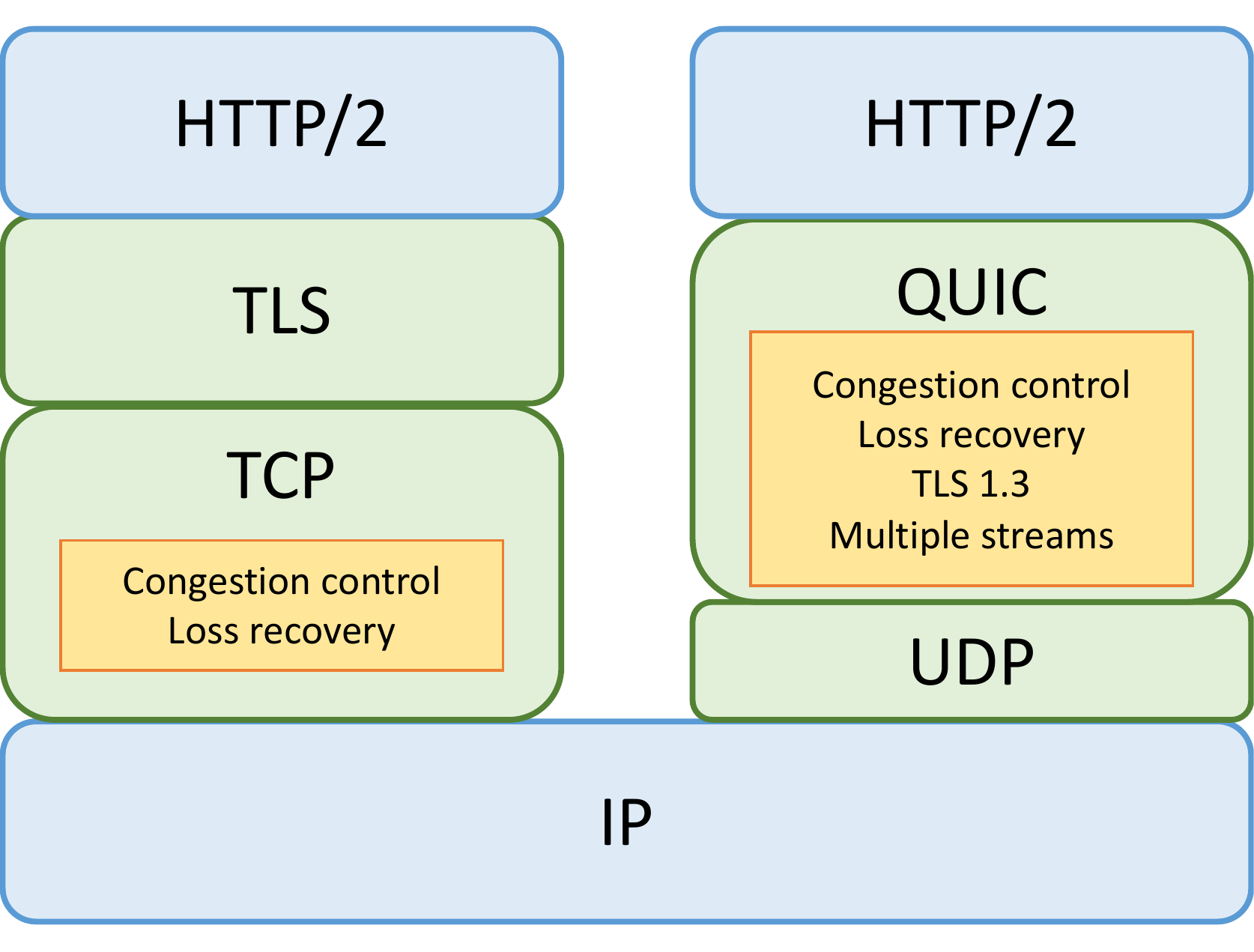}
\caption{Protocol stack with \gls{quic} and \gls{tcp}, \rev{as described in }~\cite{7867726}.}
\label{fig:QUICvsTCP}
\end{figure}


\gls{quic} incorporates (i) some of the \gls{tcp} features, including the acknowledgment mechanism, congestion control and loss recovery; (ii) the key negotiation of \gls{tls} 1.3, requiring an all-encrypted connection; and (iii) features of HTTP/2, like multi-streaming~\cite{7867726}. Fig.~\ref{fig:QUICvsTCP} shows the difference in the protocol stack for a combination of HTTP/2 and \gls{tcp} and HTTP/2 and \gls{quic}. 
The combination of these elements allows \gls{quic} to optimize several key areas. For example, the integration of transport- and cryptographic-related handshakes in the same messages makes it possible to reduce the time for the initial handshake. Furthermore, packet header encryption makes packet inspection and tracking harder for middleboxes. Moreover, it improves \gls{tcp}'s loss recovery by explicitly distinguishing in the ACKs between lost and out-of-order packets. A single connection is composed of multiple independent streams (which can be mapped to HTTP/2 streams) so that the loss of a packet of a single stream does not block the others, effectively mitigating the \gls{hol} issue. Finally, additional identifiers are introduced to label a connection, which can then be maintained even in case of changes in the \gls{ip} addresses of the endpoints.

\begin{figure}[!t]
 \centering
 \scriptsize{
\input{img/quic-conn-setup.tex}
\caption{Connection establishment procedures; the blue arrows represent the \gls{tcp} or \gls{quic} handshake, the black ones the cryptographic handshake. The red arrow is the first data transfer.}
\label{fig:connection}
\end{figure}
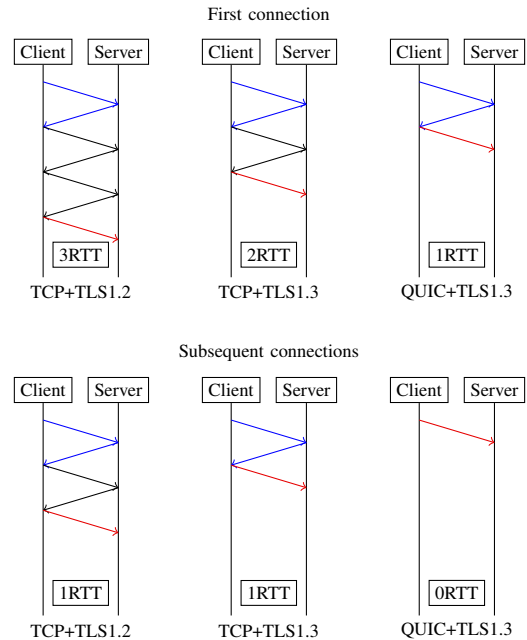

\subsubsection{\textbf{Connection Establishment}}
The connection establishment is one of the main novelties introduced in \gls{quic}, especially with the 0-\gls{rtt} option. \gls{tcp} requires at least one \gls{rtt} for its handshake when using \gls{tls} 1.3~\cite{rfc8446}, and one or two additional \glspl{rtt} for the cryptographic handshake when using the older 1.2 version~\cite{7867726}.

\gls{quic} enforces the use of at least \gls{tls} 1.3 and, in case of a first-time establishment, it carries over all the relevant \gls{quic} transport setup parameters in the first packet, a \textit{Client Hello} message, so that a single \gls{rtt} is sufficient.
The parameters negotiated during the first connection and the server's Diffie-Hellman value, used to calculate the encryption key, are stored at the client. In case of subsequent connections, this information is sent by the client to the server together with the first data packet, and the server uses it to authenticate the client and decrypt the data, thus realizing the 0-\gls{rtt} handshake shown in Fig.~\ref{fig:connection}.


\subsubsection{\textbf{\gls{quic} Packets}}
\begin{figure}[t]
\centering
\includegraphics[width=.9\columnwidth]{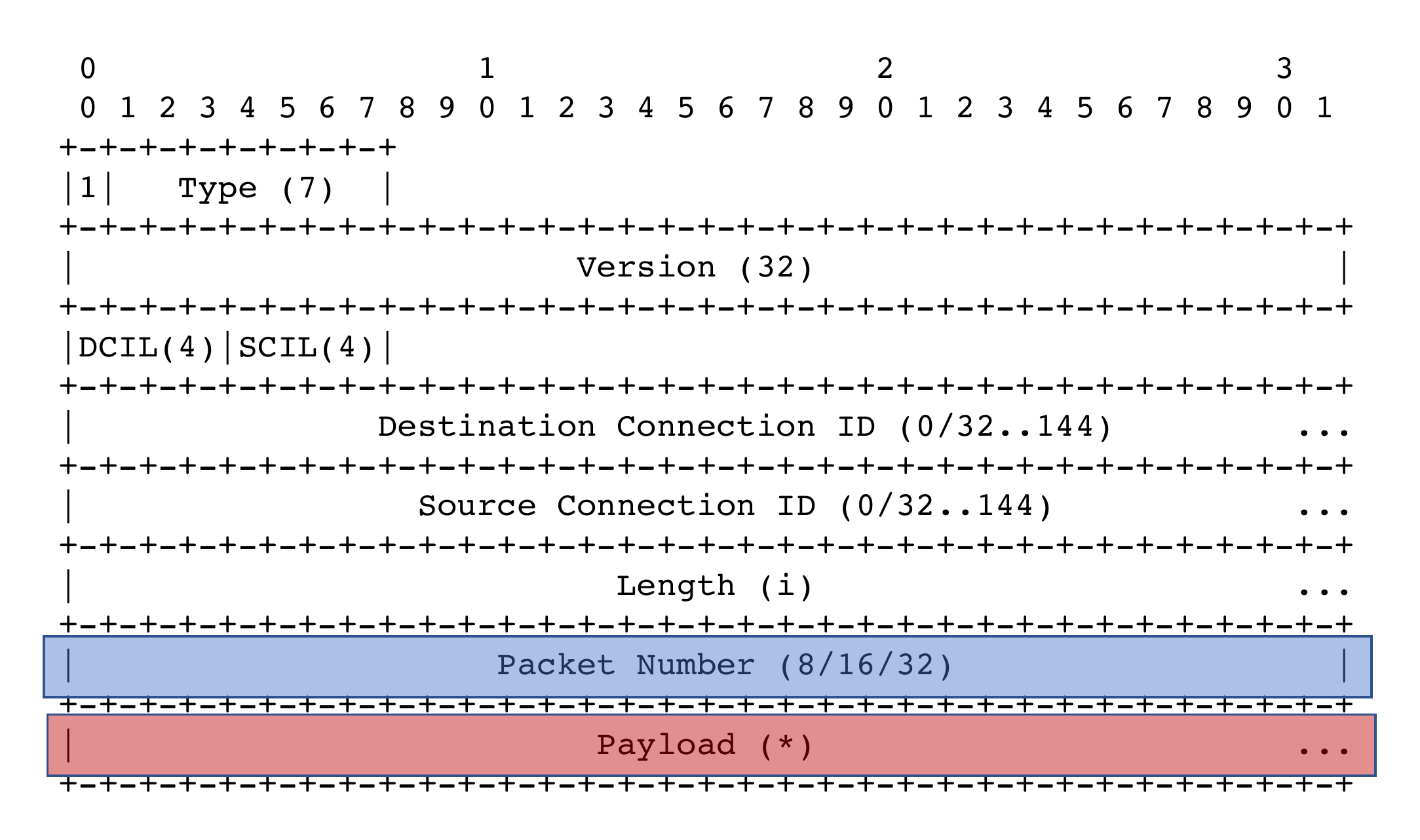}
\caption{Example of structure of a \gls{quic} packet with \textit{long header} for \gls{ietf} \gls{quic} (version 14), adapted from \cite{draftquicktr}. The red fields are encrypted, while the blue ones (i.e., the packet number field in the header) are protected against casual observation.}
\label{fig:QUICpkt}
\end{figure}

The use of \gls{tls} 1.3 also allows \gls{quic} packets to traverse middleboxes without having their inside information tampered with.
In fact, the only bytes that are not encrypted are the \gls{udp} header and a portion of the \gls{quic} header, as shown in Fig.~\ref{fig:QUICpkt}: this hides key information on \gls{quic} flows from network equipment and operators.
After the connection establishment, the only values that are sent in cleartext in the \gls{quic} header are the source and destination connection IDs (to allow routing, identification and IP address changes), and the packet type, version number and packet length. The packet number is not protected with the same encryption that is applied to the payload, but has a confidentiality protection against casual observation (e.g., middleboxes on the path).
Packet numbers, as in \gls{tcp}, are different for sending and receiving flows, with the constraint that they must not be reused within the same connection, not even for retransmissions~\cite{draftquicktr}.

The packet payload consists in one or more frames.
Each frame has a type, like \texttt{ack}, \texttt{ping} or \texttt{stream}, and a series of fields that are type-dependent, such as \textit{stream ID} and \textit{data length}~\cite{draftquicktr}. Therefore, a packet can multiplex multiple streams.

\subsubsection{\textbf{\gls{quic} Performance}}

\begin{figure}[t]
\centering
\includegraphics[width=\columnwidth]{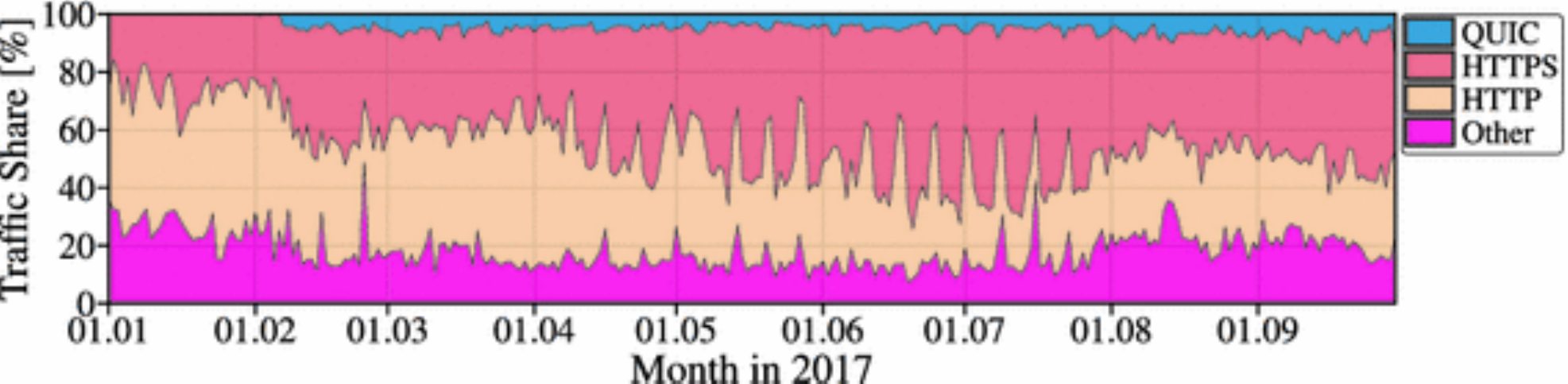}
\caption{Percentage of traffic over \gls{quic} vs other protocols on the MAWI backbone during 2017. Reprinted, with permission, from~\cite{DBLP:journals/corr/abs-1801-05168}.}
\label{fig:QUICtraffic}
\end{figure}

Given \gls{quic}'s promising design goals, its performance has been recently under the spotlight~\cite{kakhki2017taking,Carlucci-2015-HOU,Langley-2017-QTP,7510788}, even though the protocol has not yet been completely specified by the IETF. As of today, the \gls{quic} protocol is implemented in Google servers, and, client-side, in Google Chrome and in some mobile applications like YouTube or Google Search. According to~\cite{Langley-2017-QTP}, in 2017 it accounted for 7\% of all Internet traffic, and up to 30\% of Google's egress traffic. A more detailed analysis of \gls{quic} deployment is provided in \cite{DBLP:journals/corr/abs-1801-05168}, which shows that the number of \gls{quic}-capable endpoints is increasing, mainly thanks to Google and Akamai deployments, while the traffic is almost exclusively covered by Google and its services, reaching a $6.7\%$ share with respect to \gls{tcp}/HTTPS in September 2017 along the MAWI backbone, as shown in Fig.~\ref{fig:QUICtraffic}, and a $9.1\%$ share in a mobile ISP.
The authors also point out a possible problem in long-term stability due to the frequent and possibly not backward-compatible updates of the protocol.

\begin{figure}[t]
\centering
\includegraphics[width=\columnwidth]{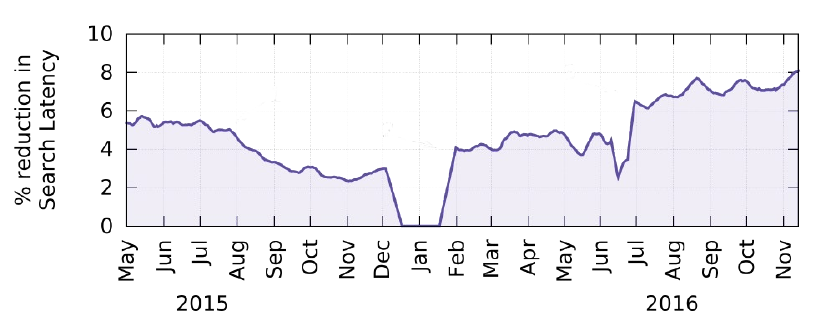}
\caption{Percentage of reduction of \textit{Search Latency}. Reprinted, with permission, from~\cite{Langley-2017-QTP}.}
\label{fig:QUICsrclat}
\end{figure}

Another work~\cite{Langley-2017-QTP} discusses some metrics measured from Google's live traffic. Fig.~\ref{fig:QUICsrclat} reports the reduction in search latency\footnote{\textit{Search Latency} is defined as the delay from the time when a user enters a search term to the time all the result content is delivered to the client, including embedded content~\cite{Langley-2017-QTP}.} experienced by clients using \gls{quic} with respect to clients using \gls{tcp}/\gls{tls}. The same paper claims that a reduction in the re-buffering rates in the order of 18\% has been observed for YouTube traffic.  Finally, the authors highlight that, with current implementation at the time of writing, the CPU load is on average 3.5 times higher with \gls{quic} than with \gls{tcp}/\gls{tls} \cite{Langley-2017-QTP}.

More recently, another work~\cite{kakhki2017taking} has tested the performance of 13 different versions of \gls{quic}, in desktop and mobile scenarios. With respect to previous studies~\cite{Carlucci-2015-HOU,7510788}, this paper provides a more thorough investigation, using a more advanced testbed and more recent versions of the protocol. The main findings (using CUBIC as congestion control) confirm that \gls{quic} performs better than \gls{tcp}/\gls{tls} in a desktop environment, but suffers from packet re-ordering issues (which may arise in a mobile environment). With respect to variations in the underlying data rate, \gls{quic} outperforms \gls{tcp} thanks to the more advanced ACK mechanism. Finally, \gls{quic} is not fair with \gls{tcp}, consuming more bottleneck bandwidth than the fair share, and has performance issues in older hardware, given that packet processing and cryptographic operations are not as optimized in the user-space as they are in the kernel. An implementation of \gls{quic} for the open source network simulator ns-3 has also been recently released~\cite{debiasio2019quic}.

\subsection{\acrfull{sctp}}
\label{sec:sctp}

\gls{sctp}, which has been defined in~\cite{rfc4960}, is a standardized signaling protocol developed by the \gls{ietf} Signaling Transport working group that later evolved into a general-purpose protocol at the transport layer. It is now maintained within IETF's Transport Area (TSVWG) working group~\cite{5233615-SCTPartWWW}.
Like \gls{tcp}, it is implemented in the kernel of the operating system, and offers a reliable, point-to-point, and connection-oriented data transport service over IP networks~\cite{968833-SCTPartIntComputing}.
\gls{sctp} was specifically designed to be \gls{tcp}-friendly, and shares its window-based congestion control, error detection and retransmission mechanisms. Nonetheless, \gls{sctp} also incorporates several new features~\cite{1284931-SCTPartState2004}, such as:
\begin{itemize}
	\item \textbf{multihoming} - while a \gls{tcp} connection is established between two sockets (identified by an IP address and a port number each) at the two end hosts, an \gls{sctp} \textit{association} can span across multiple IP addresses and possibly comprise several types of links in the network. For each client, a primary address is used to exchange data, and the others serve as backup in case of link failure or network fault, in order to ensure higher availability without interrupting the ongoing data transfers. Periodic \textit{heartbeat} packets are sent to the backup addresses to verify the link state, and the retransmission of lost packets can also occur using backup addresses.
	\item \textbf{multistreaming} - like \gls{quic}, \gls{sctp} introduces multistreaming capabilities. Using the terminology from~\cite{rfc4960}, 
	it is possible to define
	an \gls{sctp} stream as a unidirectional logical data flow within an \gls{sctp} association~\cite{1244536-SCTPartProposedStd}. \gls{sctp} then allows 
	to split the application layer data into multiple substreams, according to the specific application that generates the packets. The sequencing and delivery are performed independently within each substream, thus overcoming the \gls{hol} blocking issue, previously described in Sec.~\ref{ssec:hol}. Moreover, the delivery within each stream is strictly in order by default, as in \gls{tcp}, but it is possible to enable unordered delivery~\cite{968833-SCTPartIntComputing}.
\end{itemize}

The connection establishment and shutdown slightly differ from \gls{tcp} to improve operations and security. A cookie mechanism during the initial 4-way handshake prevents SYN flooding\footnote{SYN flooding is a denial-of-service attack where an attacker sends a succession of \gls{tcp} SYN requests to a server but never replies to SYN-ACK responses. These half-open connections allocate memory that is freed only later, but in the meantime the system is not able to reply to legitimate queries.} and masquerade attacks; moreover half-closed connections where one endpoint is still allowed to send data are not allowed.

The performance of \gls{sctp} has been evaluated in several scenarios:~\cite{985806-SCTPbattlefield} reports that partial order and partial reliability increase the performance over traditional transfer protocols such as \gls{tcp}. Jungmaier \textit{et al.}~\cite{856658-SCTPperfEval} proved that \gls{sctp} traffic has the same impact of standard \gls{tcp} traffic, that is, the two protocols can coexist in the same network.
Moreover, the multihoming scheme can achieve better throughput and faster network accident recovery when applied to wireless LANs~\cite{1209105-SCTPperfWAN}.

Finally, some papers have tested the performance of \gls{sctp} with respect to \gls{hol} blocking issues. Evaluations based on the \gls{sip} protocol at the application layer, and \gls{sctp}, \gls{tcp} and \gls{udp} as transport, showed how the \gls{hol} blocking avoidance mechanism of \gls{sctp} does not provide better performance in normal scenarios of concurrent traffic on the network, and only slightly decreases the end-to-end latency with a high level of packet losses~\cite{1233916-SCTPartEvalSIP}.
In addition, when the Maximum Transmission Unit (MTU) value suddenly decreases, \gls{sctp} is unable to avoid IP fragmentation and NATs may not be able to correctly route fragmented datagrams.

\subsection{\acrfull{dccp}}
\label{sec:dccp}
\gls{dccp}, introduced in~\cite{rfc4340,rfc5595,rfc5596}, is a protocol that augments \gls{udp} with connection-oriented functionalities such as congestion control, while maintaining its message-oriented structure. Like normal \gls{udp}, it does not guarantee reliable delivery, and packets can be out of order. Therefore, it presents a viable alternative as transport for applications which do not need reliability, but at the same time do not wish to implement their own congestion control mechanism~\cite{Kohler-dccp}.
The main reason why this protocol was introduced was the rapid growth of latency-oriented applications relying on \gls{udp} connections, like online games or IP telephony. These real-time applications cannot afford the additional latency introduced by reliability mechanisms, but still need to perform congestion control to conform with~\cite{rfc8085} and avoid choking \gls{tcp} flows when sharing a bottleneck with them~\cite{rfc4336}. \gls{dccp} provides a common framework that can perform congestion control for these applications, but also allows new mechanisms to be implemented in a straightforward manner.

\gls{dccp} establishes a bidirectional connection, which is logically composed by two half-duplex unidirectional connections~\cite{rfc4340}. Given that it is an unreliable transport protocol, \gls{dccp} does not perform retransmissions. Nonetheless, it needs to detect losses to perform congestion control~\cite{Kohler-dccp}. Therefore, the protocol does not feature cumulative ACKs, but a per-packet sequence number which allows the detection of missing datagrams. The ACKs are also acknowledged, using the same mechanism, and this makes it possible to perform congestion control on these feedback packets as well. The features of the connection (such as the security mechanisms, or congestion control algorithms), must be negotiated when establishing the connection, using control fields embedded in the \gls{dccp} headers~\cite{rfc4340}. Therefore, it is possible to tune the congestion control in each single end-to-end connection.
\gls{dccp} also natively provides \gls{ecn}~\cite{rfc3168}, thus supporting the advertisement of network congestion without necessarily recording packet losses. 

\input{img/cc-table.tex}

\rev{
\subsection{Open Challenges and Research Directions}
\label{sec:tp-ch}

The protocols presented in this section have been proposed in the attempt to solve a number of open issues and challenges related to TCP. Nonetheless, there still exists a main problem related to their wide-scale adoption in the Internet, which, as we discussed, is throttled by the ossification of the networking equipment~\cite{7738442}. \gls{quic} has adopted a promising direction, thanks to a user-space implementation, and to the support of major web content providers, such as Google.

Another important trend in \gls{quic} and \gls{sctp} is the support for updates at the IP layer and/or multi-homing, so that the end-to-end connection does not need to be broken whenever one of the endpoints changes its IP address. This is a useful feature in highly mobile wireless networks, where the user may frequently roam across different networks. Additionally, these protocols natively support or can be extended to support multiple paths, as we will discuss in detail in Sec.~\ref{sec:mp}.

Moreover, \gls{quic}, \gls{sctp} and \gls{dccp} share flexibility as a main design characteristic, which represents a novelty with respect to \gls{tcp}. For example, in \gls{quic} and \gls{sctp} it is possible to disable reliable transmissions for certain streams, and \gls{quic} and \gls{dccp} have mechanisms for feature and parameter negotiation during the connection handshake. This reconfigurability should be however matched by a proper evolution of the APIs to the application layer, so that developers can benefit from it~\cite{tran2019beyond}. 
Explicit \gls{qos} signaling represents a step further in this direction: in order to fully support new applications and services, endpoints will need to be able to specify their \gls{qos} requirements~\cite{ando2011characteristics}, as will \gls{qos}-aware schemes exploiting tools such as \gls{sdn}~\cite{mohan2013performance}.
}

\section{Congestion control algorithms}\label{sec:cc}
\rev{Congestion control is an essential function in modern networks~\cite{rfc6582}, since the high volumes of traffic that needs to be delivered reliably can only be supported if senders limit their rate before flooding their connections and affecting other flows. It can be performed directly by transport layer protocols such as \gls{dccp}, \gls{sctp}, \gls{quic}, and the omnipresent \gls{tcp}. Applications that run directly over \gls{udp} usually perform their own congestion control, as recommended by the \gls{ietf}~\cite{rfc8085}.}

\rev{The rationale behind congestion control is the following: since links with limited capacity need to be shared by multiple flows, they can get overloaded by the aggregate flow generated by multiple (or even single) sources, becoming the bottleneck of the connection. If no limit is set to the sending rate, senders will start retransmitting packets when they do not get a positive reception acknowledgment (ACK) from the destination within an interval called \gls{rto}. This, in turn, increases the load on the network, triggering a destructive spiral known as \emph{congestion collapse}~\cite{rfc896}. Congestion control is aimed at avoiding this condition: senders react to signs of congestion by backing off and reducing their own rate proactively. There are several congestion control design philosophies, with different ways to detect congestion and react to it, but they all share this common core.}

The issues we mentioned in Sec.~\ref{sec:issues} make the design of good congestion control mechanisms extremely difficult; specific features of the connection and of each individual link can significantly affect the performance, often in unpredictable ways~\cite{10gComparison}. Mathematical models of protocols and their rigorous assumptions are rarely realistic, and often theoretically optimal algorithms do not work as expected in real networks~\cite{grieco2004performance}. \rev{Moreover, in most cases (e.g., for TCP), the congestion control algorithm for a certain transport protocol is implemented in the same kernel code base of that protocol and, consequently, it is the same for every end-to-end connection. Therefore, it is not possible to customize the response of the congestion algorithm to the characteristics of each connection.}

Furthermore, detecting when congestion is taking place and how the congestion window size should be maintained are not trivial problems. Most early solutions adopted packet loss as a sign of congestion and used an \gls{aimd} strategy, which ensured asymptotic fairness and stability~\cite{NewVegasHL}.
\rev{The \gls{aimd} scheme is simple, as it only needs two commands to update the congestion window (often referred to as CWND for short):}
\begin{equation}
 \text{CWND} \leftarrow \text{CWND} + \frac{\alpha}{\text{CWND}}
\end{equation}
 on received ACKs and
 \begin{equation}
  \text{CWND} \leftarrow \frac{\text{CWND}}{\beta}
 \end{equation}
on packet losses, where $\alpha$ and $\beta$ are mechanism-set parameters. Newer protocols often include delay information as a potential congestion signal, and their adaptation of the congestion window is more tailored to their operating requirements. \gls{sack} is another option that improves the information available to the congestion control mechanism~\cite{rfc2018}.
We now list the main congestion control design philosophies and their advantages and shortcomings, \rev{quickly reviewing older algorithms in order to provide the necessary background information to better appreciate the new developments}; the main mechanisms presented in this section are summarized in Table~\ref{tab:ccsum}. 
\rev{We also highlight that all the \gls{tcp} congestion control mechanisms discussed below can be implemented for \gls{quic} and \gls{sctp} senders with no modifications. \gls{dccp} requires more effort, since it does not include a retransmission mechanism, but the basic principles and design philosophies are the same. \gls{dccp} provides a framework for congestion control~\cite{Kohler-dccp} mechanisms which can be selected by the application through the \glspl{ccid}. The congestion control mechanisms are separated from the core of the protocol enabling modularity and possible expansions. Each one of the two half-connections can even choose a different \gls{ccid} since they are logically separated. In the following, we will refer to the standardized \gls{dccp} congestion control schemes by their \gls{ccid}.}
Another taxonomy of congestion control mechanisms for \gls{tcp} was presented in~\cite{LAR2013126}, but lacks the most recent developments (e.g., those related to machine-learning-based algorithms) that we survey in this paper.

\subsection{Loss-based mechanisms}
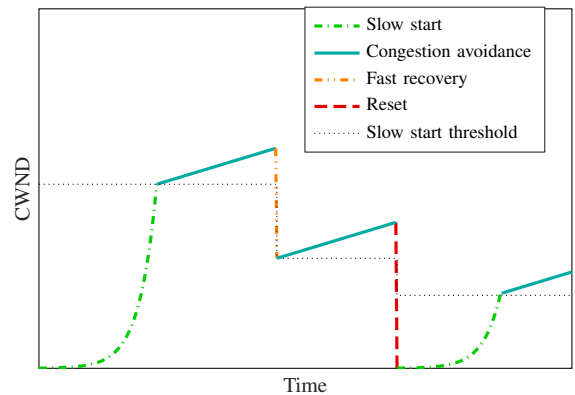
\begin{figure}[t]
\centering
\input{img/newreno.tex}
\caption{Evolution of \gls{tcp} New Reno's CWND over time}
\label{fig:nrgrowth}
\end{figure}

Classic \gls{tcp} congestion control algorithms such as Tahoe~\cite{rfc793}, Reno~\cite{rfc5681} and New Reno~\cite{rfc6582} use packet losses to detect congestion. If a packet is determined to be lost, either when the \gls{rto} timer expires or after three consecutive duplicate ACKs, the \gls{aimd} mechanism sharply decreases the congestion window. While Tahoe is very conservative, starting back from a CWND of 1 packet and entering the Slow Start phase, Reno and New Reno only reduce CWND by half in case of three duplicate ACKs, as shown in Fig.~\ref{fig:nrgrowth}. \rev{The default \gls{quic} congestion control is also based on New Reno~\cite{rfc6582}, even though other congestion control algorithms may be used, as in \gls{tcp}~\cite{draftquickrec}. A number of refinements that have been proposed for New Reno are directly included in the \gls{ietf} \gls{quic} protocol drafts. For example, the packet number is monotonically increasing, so that the \gls{quic} sender knows if a received ACK is for the original packet or its retransmission. Moreover, while with the \gls{tcp} SACK feature only three ACK ranges can be specified~\cite{rfc6675}, \gls{quic} does not have a limit for the number of ACK ranges in a feedback packet. The draft~\cite{draftquickrec} also recommends the use of pacing, to avoid the bursty transmission of packets, and a tail loss probe mechanism to efficiently detect packet loss at the end of a data transfer.}

\glsreset{lfn}
The \gls{bdp} is the product of the bottleneck capacity and the minimum \gls{rtt}; networks with a \gls{bdp} over 100 kb are considered \glspl{lfn}~\cite{rfc7323}. The classic loss-based mechanisms are extremely inefficient in \glspl{lfn}, since the recovery phase will take slightly less than a minute for a connection with a minimum \gls{rtt} of 100 ms and a capacity of 100 Mb/s, and about 10 minutes if the capacity is increased to 1 Gb/s~\cite{BIC}.

\begin{figure}[t]
\centering
\input{img/cubic.tex}
\caption{CWND growth of CUBIC algorithm}
\label{fig:cubicgrowth}
\end{figure}
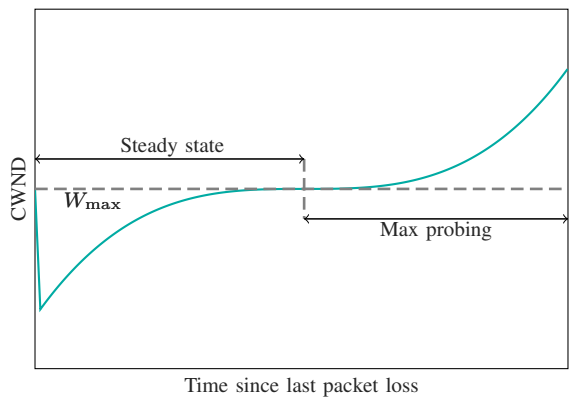

\gls{bic}~\cite{BIC} and CUBIC~\cite{CUBIC} are two congestion control algorithms developed to solve earlier mechanisms' issues with \glspl{lfn}. They abandon pure \gls{aimd} for a more complex function (a binary search in \gls{bic}'s case, and a cubic function of the time since the last loss for CUBIC, as shown in Fig.~\ref{fig:cubicgrowth}) in order to achieve fairness between flows with different \glspl{rtt}. However, \gls{bic} was deemed too aggressive and unfair to legacy flows~\cite{BicCubicHybla}, while CUBIC is now widely deployed~\cite{CubicPerf} (it is the default congestion control algorithm in the Linux kernel since version 2.6.19) and is currently an IETF Internet-Draft \cite{draftcubic} because it can achieve high performance without being unfair to flows using older congestion control mechanisms. \rev{The \gls{quic} implementation in the Chromium code base also features CUBIC congestion control~\cite{CUBIC,Langley-2017-QTP}. The main difference between CUBIC applied to \gls{tcp} and \gls{quic} is that $\beta_{\text{TCP}}=0.3$ (i.e., the decrease factor for the congestion window after loss events), while $\beta_{\text{\gls{quic}}} = \beta_{\text{TCP}}/2$~\cite{Carlucci-2015-HOU}.}

\rev{\gls{dccp} implements three loss-based mechanisms: \gls{ccid} 2~\cite{rfc4341} is a New Reno-like mechanism that uses the same \gls{aimd} principle, but using packets as a data unit instead of bytes because of the datagram-oriented nature of the underlying \gls{udp} socket. \gls{ccid} 2 also applies congestion control to acknowledgments, and has no retransmission mechanism since it is meant for applications that do not require reliable delivery and might even be hampered by it. \gls{ccid} 3, or \gls{tfrc}, uses the \gls{tcp} throughput equation explicitly~\cite{rfc5348} to calculate the sending rate in a way that is both fair to competing \gls{tcp} flows and relatively stable. It is recommended for applications that need to maintain a more stable throughput while remaining \gls{tcp}-friendly. \gls{ccid} 4 is a variant of \gls{tfrc} designed for applications that send small packets, achieving roughly the same bandwidth as an equivalent \gls{tcp} flows sending full-sized packets~\cite{rfc5622}. Another extension of \gls{tfrc}, which behaves better when multiple \gls{dccp} flows share a bottleneck, was proposed in~\cite{schier2012using}.}

\rev{Nowadays, most of the research community is moving away from loss-based congestion estimation. Several simulation studies show that none of these mechanisms performs well in wireless networks~\cite{CubicPerf}, in \glspl{manet}~\cite{1367260-SCTPcompMANET} and in broadband, high-latency networks,~\cite{1243196-SCTPperfHighLatency}. \gls{tcp}, \gls{quic}, and \gls{sctp} all suffer from the same issues, since the problem is inherent in the loss-based congestion control philosophy. However, it can still be a useful design principles for some network scenarios, particularly when very high \glspl{rtt} are involved. One example is} \gls{tcp} Noordwijk~\cite{roseti2008tcp}, a cross-layer version of \gls{tcp} that exploits the peculiarities of satellite links. The main feature of \gls{tcp} Noordwijk is that it transmits data in bursts at the maximum capacity of the bottleneck, assuming a large buffer at the bottleneck (as is the case for satellite links) and not considering competing flows. Fairness is ensured by appropriately scheduling transmission bursts; a recent proposal, called \gls{tcp} Wave~\cite{abdelsalam2017tcp}, extends the protocol by removing the cross-layer aspects and adapting the algorithm to any kind of link. The protocol quickly achieves fairness with any other loss-based flow, and goes back to full bandwidth utilization just as quickly if the competing flows stop. However, \rev{if the protocol sends packets in bursts, packets will be queued at the sender until the next burst, causing a longer queuing time. Burst-based mechanisms are also highly vulnerable to jitter and capacity fluctuations because they need to commit to a sending rate in advance and have fewer opportunities to correct their mistakes than schemes which make smaller commitments: if the bottleneck link is time-varying, they take longer to react and can make bigger errorsMany common applications are sensitive to both delay and jitter, reducing the possible usefulness of Wave in standard networks.}

\subsection{Delay-based mechanisms}
Loss-based mechanisms such as CUBIC can efficiently make use of most connections; however, the only signal of congestion they perceive is, as the name suggests, packet loss. If the buffer at the bottleneck link is large, packets will accumulate until the buffer fills, while the sender will keep increasing its sending rate and the queueing delay will keep rising. Latency can even reach 10 seconds in the conditions of heavy bufferbloat we talked about in Sec.~\ref{ssec:bufferbloat}~\cite{jiang2012understanding}, and filling up the buffer completely can cause instability in the throughput observed by other flows.

One possible solution to this is delay-based congestion control: losses may be a signal of congestion, but congestion may occur long before the buffer is full, and its early detection can increase the reactivity of the congestion control, which can then back off as soon as the \gls{rtt} increases substantially and avoid sharp declines in the throughput and high latency.

\gls{tcp} Vegas~\cite{Vegas} was the first implementation to use delay as a signal of congestion: after a milder slow start phase~\cite{VegasCC}, it adjusts the congestion window to the expected throughput of the channel. The channel is assumed to be underused if the \gls{rtt} is close to the minimum measured one; as soon as the \gls{rtt} grows beyond a certain threshold, the protocol backs off. Vegas is fair and uses capacity better than Reno~\cite{Vegas} while maintaining a low latency. However, the first version of Vegas had severe issues with the Delayed ACK mechanism, which affected its \gls{rtt} estimation, and with links with high latency~\cite{NewVegasHL}.  These issues prompted the development of several improvements, like Vegas-A \cite{VegasA}, which addresses rerouting and bandwidth sharing fairness, Vegas-V \cite{VegasV}, which increases the aggressiveness of the original algorithm, while remaining fair to other flows, and Adaptive Vegas \cite{VegasAdaptive}, where \textit{adaptive} refers to the ability of the algorithm to change its parameters according to the statistics of throughput and \gls{rtt} variation. Some specialized versions were developed for wireless \cite{VegasW} and mobile \cite{VegasManet} ad hoc networks; three particular modifications became part of the so called New Vegas algorithm~\cite{NewVegasRev}. New Vegas~\cite{NewVegasPerf} uses packet pacing in the slow start phase to avoid issues caused by burstiness, and sends packets in pairs to avoid the Delayed ACK problem, while maintaining most of Vegas's main ideas.

\rev{A simple delay-based congestion control mechanism for \gls{dccp} has been proposed in~\cite{liu2009extended}, reducing the sending rate by the ratio between the minimum and current \gls{rtt}. A similar Vegas-like mechanism~\cite{ye2009qos} uses one-way delay to measure congestion, adapting the send rate to maintain \gls{qos}. Since these schemes have not been standardized, they were not assigned a specific \gls{ccid}.}

\rev{Over the last few years, the development of delay-based protocols has seen a new renaissance thanks to the rise of interactive applications with strict latency requirements.}
Verus \cite{Verus} is a recent delay-based end-to-end congestion control protocol that exploits delay measurements to quickly adapt the congestion window size to enhance the transport performance in cellular networks.
It retains the same \gls{aimd} scheme as traditional \gls{tcp}, but changes the additive increase part:
by constantly sensing the channel, the window size is increased or reduced at every step according to the observation of short-term packet delay variations, that the authors called ``learning a \textit{delay profile}.'' Basically, the protocol builds an empirical function by associating CWND and delay and observing when the rate exceeds the capacity of the channel. The multiplicative decrease and slow start steps remain unmodified.
This strategy can be more effective than using an explicit model of the connection, and is practically achieved by using a sliding window over a period equal to the estimated \gls{rtt}.

Real world tests were performed by the authors of \cite{Verus} on 3G and LTE networks considering multiple competing flows between some devices and a server on a high bandwidth and low delay network, whose results show that Verus can achieve a throughput similar to that of \gls{tcp} CUBIC while reducing the delay by an order of magnitude.
The authors also pointed out how Verus outperforms other protocols and quickly adapts to rapid network changes and to the arrival of new flows, with throughput being independent of the RTT, confirming the protocol's fairness. While the positive results were confirmed by independent measurements~\cite{LEAP}, these tests also highlighted the massive computational load that the Verus delay profile estimation imposes on the sender; its use in uplink or high-throughput scenarios might be impossible because of the protocol's complexity.

In general, delay-based mechanisms are more stable, retransmit fewer packets and have lower latencies than loss-based mechanisms, with a similar throughput in realistic conditions; however, their adoption has been blocked because of the higher aggressiveness of the existing protocols. If a delay-based and a loss-based flow share a bottleneck, and the buffer at the bottleneck node is large enough, the delay-based mechanism will sense congestion far before its loss-based competitor, which will keep increasing its sending rate~\cite{896302}. As a result, the throughput of the delay-based flow will basically drop to zero, while the loss-based flow will take up all the available capacity. Since the majority of the servers in the Internet run a loss-based congestion control mechanism, and separating loss-based and delay-based traffic was extremely expensive and required extensive changes, delay-based protocols have never been deployed on a wide scale. However, deploying better performing delay-based protocols, that might suffer from the aggressiveness of loss-based flows, has recently become possible thanks to techniques such as network slicing and \gls{nfv}~\cite{aijaz2017realizing}, although it still involves a significant effort on the network operator's part.

Two delay-based protocols that can switch to a more aggressive mode to avoid being outcompeted by loss-based flows have recently been proposed. Like Verus, Copa~\cite{arun2018copa} is a protocol that uses the delay profile to determine the sending rate. It uses Markov chains to explicitly model the bottleneck queue, and it dynamically adjusts its aggressiveness to compete fairly with loss-based protocols. Nimbus~\cite{goyal2018elasticity} is another enhancement that also models the elasticity of cross traffic, detecting how other flows will react to changes in the perceived bandwidth and adapting to the resulting scenario. The cross traffic modeling is performed by observing the Fourier transform of the capacity and detecting periodic behavior; this approach works well with CUBIC or Copa cross-traffic, but can fail for \gls{bbr} cross-traffic and is too aggressive in the presence of Vegas flows, which it senses as elastic flows and proceeds to outcompete.

Another interesting development is represented by \gls{tcp} \gls{ledbat}~\cite{rfc6817}, a delay-based algorithm developed for BitTorrent traffic: since background traffic should have a lower priority than user traffic, the low aggressiveness of delay-based mechanisms becomes a strength. \gls{ledbat} estimates the capacity left unused by foreground flows by measuring the extra delay it adds after sending a packet, and it only transmits if it deems its actions will not affect more important flows. It is currently deployed on Microsoft Windows machines as a solution for bulk transfer applications. TIMELY~\cite{2787510} is another delay-based protocol which adjusts the congestion window using the gradient of the \gls{rtt}, designed specifically for data centers. Finally, \gls{tcp} \gls{lola}~\cite{8109356} adapts the CUBIC mechanism to a delay-based detection method; when the queueing delay starts to increase, \gls{lola} switches to a Vegas-like holding mechanism to maintain capacity and fairness to other flows.

\subsection{Capacity-based mechanisms}

\gls{tcp} Westwood\cite{mascolo2001tcp} was a congestion control mechanism from 2001 that explicitly tracked the estimated bandwidth in order to adaptively change the congestion window after a loss, while maintaining the additive increase scheme of \gls{tcp} Reno. Westwood was designed for wireless and lossy links~\cite{zanella2001tcp}, \rev{which can have a fast-varying capacity, as well as packet losses caused by the physical layer. Both of these factors can affect transport layer performance, as traditional loss-based schemes will overreact to losses by halving the send rate even when there is no need for it, and fast-varying capacity can also be interpreted as a sign of congestion. Westwood adapts the congestion window to the measured connection capacity after losses} and achieves a higher performance\cite{mascolo2004performance} than loss-based schemes in those conditions, as well as in wired networks.

\rev{However, the idea of capacity-based congestion control was recently revived by} Sprout~\cite{Sprout}, an end-to-end protocol designed for cellular wireless networks that aims to be a compromise between achieving the highest possible throughput and preventing long packet delays in a network queue.
\rev{Sprout exploits some of the peculiarities of cellular networks, improving congestion control in that particular scenario. Since cellular networks often suffer from extreme bufferbloat, using packet loss to sense congestion leads to extremely high latency. Additionally, cellular network users are not subject to queues accumulated by other users, since carriers generally set up separate uplink and downlink queues for each device belonging to a cell. Sprout exploits that fact by periodically measuring the connection capacity and predicting its future distribution with a \gls{hmm}.}
Estimating the available capacity is instead done by counting the received bytes in a long interval and then dividing the result by its duration. This estimate is used to forecast the number of packets that is safe to send over the links, that is, the right number of packets so that they will not wait too long in a queue. It is a highly customizable protocol: the model can be more throughput or delay-oriented by changing an aggressiveness parameter.
However, this approach requires a dedicated buffer: if multiple flows share the same buffer, Sprout will leave almost all the capacity to the competing flows \cite{LEAP}.


\begin{figure}[!t]
 \centering
 \scriptsize{
\begin{tikzpicture}[auto]

 \node at (1,0)[name=cbeg1] {BDP};
 \node at (5.5,0)[name=sbeg1] {BDP + bottleneck buffer};
 \coordinate (cend1) at(1,-2); 
 \coordinate (send1) at(5,-2); 
 \coordinate (x1) at (-1,-2);
 \coordinate (x2) at (7,-2);
 \coordinate (x3) at (-1,-4.5);
 \coordinate (x4) at (7,-4.5);
 \node at(1,-1.5)[circle,line width=1pt,blue,draw,name=c11]{}; 
 \node at(5,-0.2)[circle,line width=1pt,red,draw,name=s11]{}; 
  \node at(1,-2.7)[circle,line width=1pt,blue,draw,name=c21]{}; 
 \node at(5,-2.7)[circle,line width=1pt,red,draw,name=s21]{}; 
 \coordinate (cend2) at(1,-4.5); 
 \coordinate (send2) at(5,-4.5); 
 \coordinate (cbeg2) at(1,-2.5); 
 \coordinate (sbeg2) at(5,-2.5); 
 \coordinate (c12) at(-1,-1.5);   
 \coordinate (s12) at(7,-0.2);
 \coordinate (c22) at(-1,-4.2);  
 \coordinate (s22) at(7,-2.7);  
 
  \node at (3,0.2)[rectangle,draw,name=label1] {Round-Trip Time};
  \node at (3,-2.3)[rectangle,draw,name=label2] {Delivery rate};
  \node at (3,-0.4)[rectangle,name=label3] {Bandwidth-limited};
  \node at (6,-0.4)[rectangle,name=label4] {Buffer-limited};
  \node at (0,-1.3)[rectangle,name=label5] {RTT$_{\min}$};
  \node at (0,-0.4)[rectangle,name=label6] {Application-limited};
  \node at (0.2,-2.25)[rectangle,name=label7] {Optimal operating point};
  \node at (5.8,-2.25)[rectangle,name=label8] {Loss-based operating point};

 \draw[-] (cend1) to (1,-0.15);
 \draw[-] (send1) to (5,-0.15); 
 \draw[-] (cend2) to (cbeg2);
  \draw[arrows={->}] (x1) -- (x2);
  \draw[arrows={->},line width=1pt] (label7) -- (c11);
  \draw[arrows={->},line width=1pt] (label7) -- (c21);  
  \draw[arrows={->},line width=1pt] (label8) -- (s11);
  \draw[arrows={->},line width=1pt] (label8) -- (s21);
  \draw[arrows={->}] (x3) -- (x4);
 \draw[-] (send2) to (sbeg2);
 \draw[orange,line width=1pt,-] (1,-1.5) -- (c12);
  \draw[orange,line width=1pt,-] (5,-0.2) -- (1,-1.5);
 \draw[orange,-,line width=1pt,dashed] (5,-0.2) -- (s12);
 \draw[cyan,line width=1pt,-] (1,-2.7) -- (5,-2.7);
 \draw[cyan,line width=1pt,-] (c22) -- (1,-2.7);
 \draw[cyan,-,line width=1pt,dashed] (5,-2.7) -- (s22);

\end{tikzpicture}
}
\caption{BBR basic principle: the mechanism tries to work at the optimal operating point instead of filling the bottleneck buffer}
\label{fig:vj1}
\end{figure}
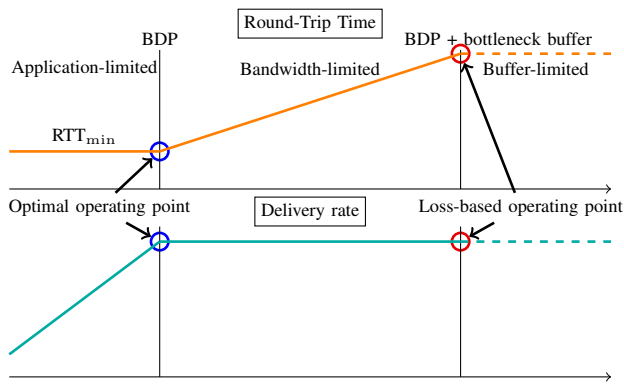

\subsection{Hybrid mechanisms}\label{ssec:bbr}

Some congestion control mechanisms try to combine two approaches to reap the benefits of both. Compound \gls{tcp}~\cite{4146841} is a well-known example, as it is available on all Microsoft Windows machines by default. Compound uses the sum of a delay-based window and a loss-based Reno window as its CWND, and as such it can deal with \glspl{lfn} better than pure loss-based mechanisms while maintaining a measure of fairness toward them, since its compound window will grow faster than a purely loss-based one, but the protocol's behavior will revert to loss-based congestion control when it senses congestion.
Illinois~\cite{liu2008tcp} is another algorithm that considers both loss and delay to infer congestion: while it uses losses to detect it, the variation in the delay is used to determine the rate of change in the congestion window. It is friendly to CUBIC, but, like Compound, its CWND grows significantly faster when it is far lower than the capacity; Veno~\cite{fu2003tcp}, a hybrid between Vegas and Reno, uses an explicit model of the bottleneck buffer occupancy to achieve the same kind of CWND growth.

\begin{figure*}[t]
 \centering
 \scriptsize{
\begin{tikzpicture}[auto]

\coordinate (o1) at (0,0);
\coordinate (x1) at (3.5,0);
\coordinate (y1) at (0,2);
  \draw[arrows={->},gray,line width=1pt] (o1) -- (x1);
  \draw[arrows={->},gray,line width=1pt] (o1) -- (y1);  
  \coordinate (o2) at (4,0);
\coordinate (x2) at (7.5,0);
\coordinate (y2) at (4,2);
  \draw[arrows={->},gray,line width=1pt] (o2) -- (x2);
  \draw[arrows={->},gray,line width=1pt] (o2) -- (y2);  
\coordinate (o3) at (8,0);
\coordinate (x3) at (11.5,0);
\coordinate (y3) at (8,2);
  \draw[arrows={->},gray,line width=1pt] (o3) -- (x3);
  \draw[arrows={->},gray,line width=1pt] (o3) -- (y3);  
\coordinate (o4) at (12,0);
\coordinate (x4) at (15.5,0);
\coordinate (y4) at (12,2);
  \draw[arrows={->},gray,line width=1pt] (o4) -- (x4);
  \draw[arrows={->},gray,line width=1pt] (o4) -- (y4);  
   \node at (1.75,-0.2)[rectangle,name=xlabel1] {Congestion window};
   \node at (5.75,-0.2)[rectangle,name=xlabel2] {Congestion window};
   \node at (9.75,-0.2)[rectangle,name=xlabel3] {Congestion window};
   \node at (13.75,-0.2)[rectangle,name=xlabel4] {Congestion window};
   \node at (-0.2,1)[rectangle,rotate=90,name=ylabel1] {Delivery rate};
   \node at (3.8,1)[rectangle,rotate=90,name=ylabel2] {Delivery rate};
   \node at (7.8,1)[rectangle,rotate=90,name=ylabel3] {Delivery rate};
   \node at (11.8,1)[rectangle,rotate=90,name=ylabel4] {Delivery rate};   
   
      \node at (1.75,2)[rectangle,draw,name=label1] {Startup};
   \node at (5.75,2)[rectangle,draw,name=label2] {Drain};
   \node at (9.75,2)[rectangle,draw,name=label3] {Bandwidth probe};
   \node at (13.75,2)[rectangle,draw,name=label4] {RTT probe};

   \coordinate (f1) at (1,1.5);
   \coordinate (g1) at (3.5,1.5);
  \draw[cyan,-,line width=1pt] (o1) -- (f1);
  \draw[green,-,line width=1pt] (f1) -- (g1);
  \coordinate (f2) at (5,1.5);
  \coordinate (g2) at (7.5,1.5);
  \draw[cyan,-,line width=1pt] (o2) -- (f2);
  \draw[green,-,line width=1pt] (f2) -- (g2);
  \coordinate (f3) at (9,1.5);
  \coordinate (g3) at (11.5,1.5);
  \draw[cyan,-,line width=1pt] (o3) -- (f3);
  \draw[green,-,line width=1pt] (f3) -- (g3);
  \coordinate (f4) at (13,1.5);
  \coordinate (g4) at (15.5,1.5);
  \draw[cyan,-,line width=1pt] (o4) -- (f4);
  \draw[green,-,line width=1pt] (f4) -- (g4);

  \node at (0.125,0.1875)[draw,shape=circle,red,scale=0.4,fill=red,name=p11]{};
  \node at (0.25,0.375)[draw,shape=circle,red,scale=0.4,fill=red,name=p12]{};
  \node at (0.5,0.75)[draw,shape=circle,red,scale=0.4,fill=red,name=p13]{};
  \node at (1,1.5)[draw,shape=circle,red,scale=0.4,fill=red,name=p14]{};
  \node at (1.75,1.5)[draw,shape=circle,red,scale=0.4,fill=red,name=p15]{};
  \node at (2.125,1.5)[draw,shape=circle,red,scale=0.4,fill=red,name=p16]{};
  \node at (2.5,1.5)[draw,shape=circle,red,scale=0.4,fill=red,name=p17]{};
   \draw [->,red] (p11) to [out=86,in=206] (p12);
   \draw [->,red] (p12) to [out=86,in=206] (p13);
   \draw [->,red] (p13) to [out=86,in=206] (p14);
   \draw [->,red] (p14) to [out=30,in=150] (p15);
   \draw [->,red] (p15) to [out=30,in=150] (p16);
   \draw [->,red] (p16) to [out=30,in=150] (p17);

    \node at (5,1.5)[draw,shape=circle,red,scale=0.4,fill=red,name=p21]{};
  \node at (5.2,1.5)[draw,shape=circle,red,scale=0.4,fill=red,name=p22]{};
  \node at (5.4,1.5)[draw,shape=circle,red,scale=0.4,fill=red,name=p23]{};
  \node at (6.5,1.5)[draw,shape=circle,red,scale=0.4,fill=red,name=p24]{};
     \draw [->,green] (p24) to [out=150,in=30] (p23);
   \draw [->,green] (p23) to [out=150,in=30] (p22);
   \draw [->,green] (p22) to [out=150,in=30] (p21);
  
    \coordinate (ss3) at (8.8,1.5);
    \node at (9,1.5)[draw,shape=circle,red,scale=0.4,fill=red,name=p31]{};
  \node at (9.5,1.5)[draw,shape=circle,red,scale=0.4,fill=red,name=p32]{};
     \draw [->,green] (p32) to [out=150,in=30] (p31);
   \draw [->,red] (p31) to [out=330,in=210] (p32);
   \draw [-,blue] (p31) to [out=90,in=120] (ss3);
   \draw [->,blue] (ss3) to [out=270,in=240] (p31);

  \node at (9,1.5)[draw,shape=circle,red,scale=0.4,fill=red,name=p31]{};
  \node at (9.5,1.5)[draw,shape=circle,red,scale=0.4,fill=red,name=p32]{};
  
    \node at (12.125,0.1875)[draw,shape=circle,red,scale=0.4,fill=red,name=p41]{};
  \node at (13,1.5)[draw,shape=circle,red,scale=0.4,fill=red,name=p42]{};
    \draw [->,red] (p41) to [out=86,in=206] (p42);
   \draw [->,green] (p42) to [out=266,in=26] (p41);

\end{tikzpicture}
}
\caption{Overview of the operating points in the states of the BBR algorithm}
\label{fig:bbrstates}
\end{figure*}
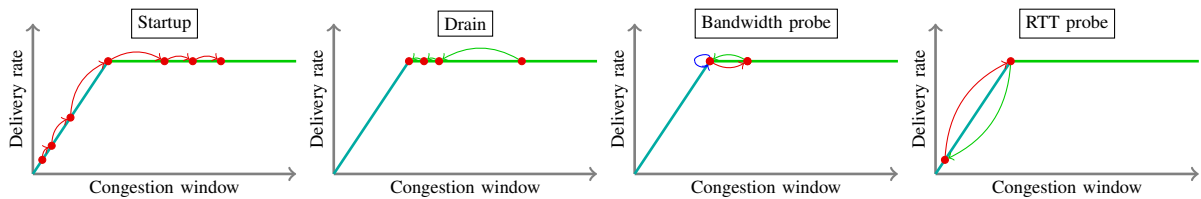

\rev{The most notable recent development in hybrid congestion control schemes is undoubtedly }Google's \gls{bbr} algorithm, which takes a different approach  by trying to estimate both the bandwidth and the \gls{rtt}~\cite{BBR}. It tries to reach the optimal operating point~\cite{Kleinrock79} by keeping a CWND equal to the \gls{bdp}; any further increase in the CWND just increases the delay without any throughput benefits, as shown in Fig.~\ref{fig:vj1}. While converging to the \gls{bdp} with traditional mechanisms is impossible~\cite{Jaffe81}, \gls{bbr} alternately estimates the minimum \gls{rtt} and the capacity by periodically draining the buffer, then measuring the \gls{rtt}, and finally returning to the initial delivery rate. This mechanism is shown in Fig.~\ref{fig:bbrstates}.

\gls{bbr} can share the bandwidth fairly with other \gls{bbr} or loss-based flows~\cite{BBR}, but its latency is low only when the connection is used by other delay-conscious protocols, as there is no way to seize a fair share of the capacity from loss-based flows without having significant queueing delays. \gls{bbr} also works well with high-capacity connections; however, the presence of short buffers causes the \gls{bbr} operating point to increase too much, thus leading to massive packet losses and unfairness towards loss-based flows~\cite{BBRexp}.

Fairness and stability when sharing a connection with other congestion control algorithms are two of \gls{bbr}'s biggest open issues. Competition with other algorithms can cause severe throughput fluctuations~\cite{BBRCyclic}, and \gls{bbr}'s aggressiveness against loss-based flows strongly depends on the buffer size: as previously mentioned, the mechanism becomes too aggressive when buffers are very small~\cite{BBRexp} and too conservative when the network suffers from bufferbloat.
The same problem was noticed by Farrow in his comparative study of various congestion control algorithms, namely CUBIC, NewReno and BBR, in heterogeneous network environments~\cite{PerfAnalysisHetEnv}:
virtual tests with several competing flows both with the same and with different algorithms lead the author to conclude that NewReno and CUBIC are sufficiently fair, with the latter offering more predictable and stable throughput, while asserting that \gls{bbr} is unfair and still not ready for public use.

Doubts have also been raised about \gls{bbr}'s performance in cellular networks~\cite{BBRPerfMobile} and under mobility~\cite{BBRMobile,BBRCubicHighway}, but work to adapt it to this kind of environment is currently ongoing. \rev{\gls{bbr} has also been proposed for \gls{quic}~\cite{BBRIetf100}.}

\subsection{Machine learning approaches}
\label{sec:ml}

Over the past few years, machine learning has become an important tool for network and protocol designers~\cite{8121867}; research on machine-learning based congestion control is ongoing, and there already are several working mechanisms in the literature.

\gls{tcp} Remy~\cite{2486020} is the first example of machine learning-based congestion control: the authors define a Markov model of the channel and an objective function with a parameter $\alpha$ which can be tuned to set the aggressiveness of the protocol ($\alpha=1$ corresponds to proportional fairness, while $\alpha=0$ does not consider fairness at all and $\alpha=\infty$ achieves max-min fairness), and use a machine learning algorithm to define the behavior of the congestion control mechanism. The inputs given to the mechanism are the ratio between the most recent \gls{rtt} and the lowest measured \gls{rtt} during the connection, an \gls{ewma} of the interarrival times of the latest ACKs, and an \gls{ewma} of the sending times of those same packets. The mechanism itself is essentially a Monte Carlo simulation, and has some limitations, since it requires some prior knowledge about the network scenario and operates on the basic assumption that all competing flows use it. A more advanced version of Remy, called \gls{tao}\cite{2626324}, solves the problem of \gls{tcp} awareness and performs well with heterogeneous competing flows, but still requires extensive prior knowledge about the network to function. 

\gls{pcc}~\cite{189004} is another learning-oriented mechanism that uses online experiments instead of offline pre-training: it uses \glspl{sack} to measure the utility of an action, and adjusts the sending rate accordingly. \gls{pcc} is more aggressive than even loss-based \gls{tcp} versions, and performs well with short buffers and high \glspl{rtt}, but its performance in wireless networks or links affected by bufferbloat has not been tested. \gls{pcc} Vivace~\cite{211245} is a new version of the congestion control mechanism with improved \gls{tcp} friendliness and a more advanced learning mechanism using linear regression techniques.

An attempt at using more advanced supervised learning techniques to design congestion control mechanisms was made in~\cite{3229550}, and Q-learning was used in~\cite{7536338} and\cite{7733018}. However, all these algorithms are optimized for very simple network topologies and situations and strongly depend on the considered scenario, so to the best of our knowledge there is no fully implemented and well-tested congestion control mechanism using these techniques. \rev{Along this line, the authors of~\cite{8668690} investigate a reinforcement learning approach to improve the performance of both short and long TCP flows. In particular, the proposed approach, TCP-RL, exploits two different learning processes to adapt the initial window of the connection (i.e., the number of packets that can be sent when the connection starts) and the congestion control algorithm used for each single flow. The performance evaluation is based on a real implementation, and either real traffic or synthetic traces generated from a major web service provider. Finally, also QTCP~\cite{8357943} adopts a reinforcement learning approach, but the actions that are available are directly related to the dimensioning of the congestion window. The authors, however, compare the performance of their scheme with TCP NewReno.}

A more detailed discussion on the design considerations for new machine learning-based congestion control mechanisms can be found in~\cite{3152446}.

\subsection{Cross-layer approaches}
\label{sec:cl}

As previously mentioned, \gls{tcp} relies on an abstraction of the multiple hops between the two end-points of the connection to perform its congestion control operations. This, however, is sometimes sub-optimal, given that \gls{tcp} may fail to capture medium-specific characteristics (e.g., mmWave link variations, as discussed in Sec.~\ref{sec:wireless}). Therefore, cross-layer approaches aim at designing congestion control algorithms that exploit additional and more precise information provided by the lower layers of the protocol stack, and are popular especially in the wireless domain. 

Several papers have recently proposed cross-layer solutions to cope with cellular networks operating at mmWave frequencies. In~\cite{zhang2017tcp,azzino2017x} the authors tune the \gls{tcp} congestion window to the \gls{bdp} of the connection, tracked with physical and \gls{mac} layer information in the mobile terminal, for downlink and uplink \gls{tcp} connections, respectively. The paper \cite{polese2017milliproxy} proposes a proxy-based mechanism at the base station that intercepts ACKs and changes the advertised window value to force the unmodified \gls{tcp} sender to track the \gls{bdp} of the end-to-end connection with a mmWave link.

The \gls{sdn} paradigm has also opened new possibilities for cross-layer approaches, giving the transport layer the possibility to reserve resources for low-latency applications~\cite{naman2018responsive} by communicating directly with the network controller.

Another possible cross-layer approach is to consider the needs of the application: several \gls{tcp} versions that consider rate distortion and content type in multimedia streaming have been proposed~\cite{6146438,habachi2013online,aghdam2014wccp}.

\rev{In~\cite{lu2015cqic}, the authors revisited a cross-layer approach for 3G and 4G cellular networks applying it to \gls{quic}'s congestion control, with modifications both at the sender (which alternates between a rate-limited approach, when a valid rate estimation is available from the lower layers, and a congestion-window-limited state) and at the receiver (which feeds back the rate estimation to the sender).}

\rev{
\subsection{Datacenter networks and the Incast issue}
\label{sec:cc-datacenter}
As we described in Sec.~\ref{ssec:incast}, data centers are a very peculiar networking environment, which requires several adjustments on the transport layer to avoid issues like the Incast problem. Over the past few years, there have been several attempts to mitigate these issues.
\gls{dctcp}~\cite{DCTCP} is a congestion control algorithm based on \gls{ecn}, which reduces the congestion window based on the amount of experienced congestion, leaving enough buffer space to avoid the Incast issue. 

Another option is to abandon standard congestion control altogether and use explicit rate control, as in the \gls{d3} scheme~\cite{wilson2011better}, which can solve the problem if full network support is available. This can be implemented by using \gls{sdn} and \gls{nfv} solutions~\cite{drutskoy2013scalable}, but scalability remains an issue, and \gls{d3} is vulnerable to bursts. \gls{d2tcp}~\cite{vamanan2012deadline} is a distributed scheme that considers application-level deadlines for data blocks like \gls{d3}, while avoiding its need for full network support and vulnerability to bursty traffic. It is similar to \gls{dctcp}, but it has full awareness of the deadlines when performing congestion avoidance.

The pFabric scheme~\cite{alizadeh2013pfabric} is another network support technique based on early dropping: switches implement priority dropping with very short buffers, enforcing flow scheduling even if the flows are aggressive and do not back off until the loss rate is consistently high. In this kind of scenario, flows need to be aggressive, or the early dropping mechanism would reduce their throughput far below the achievable limit.
The pHost~\cite{gao2015phost} protocol aims at replicating pFabric's performance without any network support, using a token-based scheme at the receivers to assign priorities to senders. Receivers prioritize the flow with fewest remaining bytes when assigning tokens, achieving near optimal performance. ExpressPass~\cite{cho2017credit} is another technique that uses receiver-side credit packets to control sender-side rate; credit packets can be lost without consequence, and the loss rate is used to gauge the connection capacity. Homa~\cite{montazeri2018homa} is a new connectionless protocol inspired by pHost and ExpressPass, which can significantly reduce latency with no network support. It aggressively uses priority queues on switches, using receiver-driven flow control to assign the priorities and explicitly limiting the Incast issue by counting outstanding requests and marking new packets to reduce their priority if they are over a certain length. Its connectionless nature limits \gls{hol} blocking and avoids explicit acknowledgments, further contributing to the protocol's efficiency.

\gls{pase}~\cite{munir2015friends} is a strategy that synthesizes the approaches of \gls{dctcp}, \gls{d3}, and pFabric. It uses all three mechanisms at different scales: explicit rate control is performed at coarse time-scales and with low precision, since the endpoints themselves can find an efficient allocation by probing according to their assigned priority. Finally, pFabric-like prioritization and dropping are performed at very short timescales. This hybrid approach can outperform each adaptation by itself.

For a more complete discussion of data center networks and their transport layer issues, we refer the reader to~\cite{noormohammadpour2017datacenter,7096919,zhang2018load,hafeez2018detection}.

}

\rev{
\subsection{Open Challenges and Research Directions}
\label{sec:cc-ch}

Congestion control is a critical component in the operation of most transport protocols. \gls{tcp}, \gls{sctp}, and \gls{quic} implement it natively, and \gls{dccp} represents an evolution in the design of unreliable transport protocols that also accounts for congestion. The research in congestion control has seen a fast evolution in several directions: while traditional loss-, delay-, and capacity-based approaches have all been used as stepping stones towards more efficient protocols, entirely new machine learning-based and cross-layer protocols are beginning to attract significant interest.

The main trend in this area is related to overcoming a one-size-fits-all approach to congestion control (e.g., the widely adopted NewReno and CUBIC) towards algorithms that target a particular use case. For example, several delay-based protocols target low queuing delay (e.g., \gls{tcp} Copa, LoLa), which translates into low end-to-end latency. However, updates in the \gls{tcp} congestion control mechanism require modifications in the kernel, and thus an Operating System update, which is a cumbersome operation. The development of user-space protocols, such as \gls{quic}, allows researchers to experiment more, and network support through \gls{sdn} and slicing makes it easier to develop mechanisms that are not as aggressive as CUBIC, since they can be isolated in separate buffers and thus do not need to compete with it.
}

\section{Multipath Transport Protocols}
\label{sec:mp}

Nowadays, most devices can use multiple communication technologies at the same time: for example, modern smartphones can connect to both Wi-Fi and LTE. For this reason, multipath communications have become the subject of considerable interest over the past few years. At the transport layer, multipath-capable protocols need to be designed to successfully exploit the advantages of multipath diversity, but this is not always simple.

\subsection{MPTCP}
\label{sec:mptcp}

\gls{mptcp} is a fully backward-compatible extension of \gls{tcp}, published as an experimental standard in 2011~\cite{rfc6824,rfc6356,rfc6182} and now widely deployed~\cite{2798088}. It allows applications to use multiple connections at the same time without any changes to the socket API. Using multiple connections can improve the total capacity, provide redundancy against link failures, and reduce the load on congested paths. In order to ensure backward compatibility and transparency to applications~\cite{rfc6182}, \gls{mptcp} needs to be implemented within the operating system's kernel. 

\begin{figure}[!t]
\centering
\includegraphics[width=.9\columnwidth]{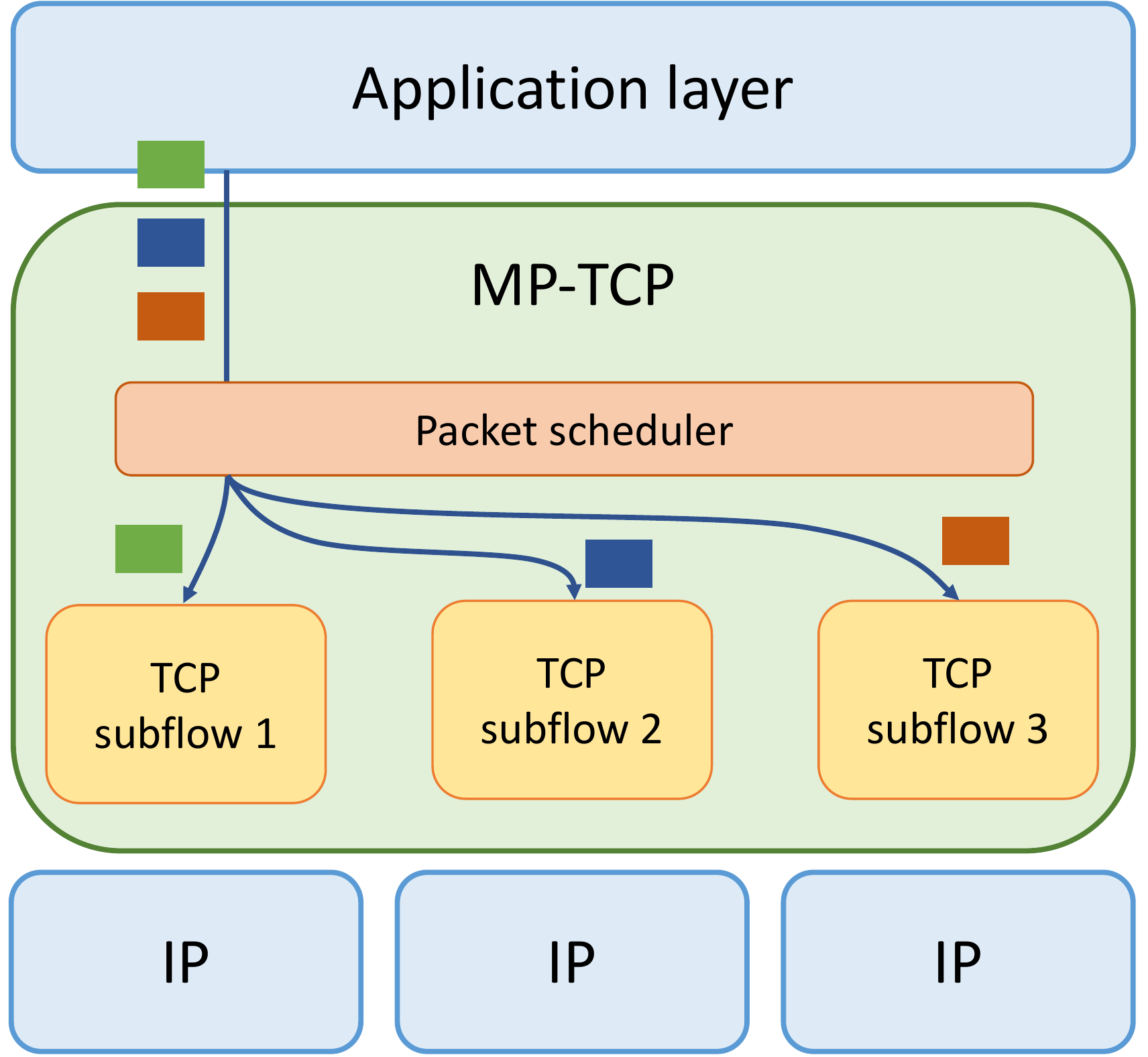}
\caption{Protocol stack with \gls{mptcp}.}
\label{fig:MPTCP}
\end{figure}

Fig.~\ref{fig:MPTCP} shows the basic architecture of an \gls{mptcp} host: the connection is composed of two separate \gls{tcp} flows on different paths~\cite{228338}, each with its own congestion control and ACKing mechanism. Using single-path \gls{tcp} flows is necessary because many middleboxes inspect \gls{tcp} traffic and discard packets with behaviors inconsistent with the normal operation of the protocol (e.g., missing ACKs, out-of-order or with-gaps sequence numbers, or badly formed options) for security reasons.

An \gls{mptcp} session starts with a single \gls{tcp} sub-flow; \gls{mptcp}-capable hosts can recognize each other by setting the \texttt{MP_CAPABLE} option in the first packets of the connection handshake. The two hosts then exchange cryptographic keys to add new sub-flows securely; these can be established by setting the \texttt{MP_JOIN} option and using a hash of the connection's keys during the standard \gls{tcp} handshake. \gls{mptcp} supports also the addition and removal of addresses on a host, both implicitly and explicitly.

\gls{mptcp} assigns packets two sequence numbers: the standard sequence number for each sub-flow and a connection-level \gls{dsn}. Since congestion control and retransmission need to happen at the sub-flow level, as each sub-flow is a full-fledged \gls{tcp} connection in its own right, the flow-level sequence numbers need to be consecutive and data on each sub-flow is delivered in-order; the connection-level \gls{dsn} is needed to join the two split data streams and reconstruct the original data~\cite{rfc6897}. Retransmissions of the same data over multiple sub-flows are also possible, since the connection-level \gls{dsn} can identify the packet at the connection level and mitigate \gls{hol} if a sub-flow is blocked. However, data transmitted on a sub-flow must also be retransmitted and delivered on it, even if received correctly on another, since doing otherwise would break the \gls{tcp}-compliance of the sub-flow. 

As with sequence numbers, there are two types of ACKs: regular acknowledgments on each sub-flow, and connection-level ACKs, which act as a cumulative acknowledgment for the whole data flow. 

The support of multiple IP addresses for the same \gls{mptcp} connection may result in security vulnerabilities~\cite{rfc6181}, but the design of the protocol minimizes this risk, especially if we compare it to single-path \gls{tcp}, which is vulnerable for example to the man-in-the-middle attack as well.
Multiple IP support can also be a problem for applications, since the application cannot implicitly rely on the one-to-one mapping between host and address and has to adapt its code to deal with multiple IP addresses~\cite{rfc6897,BPB11}.

\gls{mptcp}'s main rationale is that using multiple flows will reduce delay and increase throughput and reliability; the protocol is now implemented in most major operating systems and used in data centers and wireless networks~\cite{rfc8041}. However, several issues need to be taken into consideration, and the multipath scheduler and congestion control mechanism need to be carefully tuned, particularly in wireless networks~\cite{han2016should,polese2017tcp}.

The \gls{hol} problem we described in Sec.~\ref{ssec:hol} is particularly severe in \gls{mptcp}: since each sub-flow forwards data to the multipath level only in-order, a loss on one sub-flow can block the whole connection until the packet is retransmitted on that same sub-flow. If the loss happens on the sub-flow with the longest \gls{rtt}, this can take a long time, particularly in networks affected by bufferbloat. \gls{hol} can even reduce the throughput of other sub-flows, since the receiver-side buffer is limited and once it is filled other flows will have to wait for the head of line packet without transmitting anything~\cite{6193502}. In some cases, \gls{mptcp} is clearly outperformed by a single-path \gls{tcp} flow on the best path~\cite{polese2017tcp}; it has even been proposed to completely disable it and automatically fall back to single-path transmission in lossy scenarios~\cite{7856155}.

\begin{figure}[!t]
\centering
\includegraphics[width=\columnwidth]{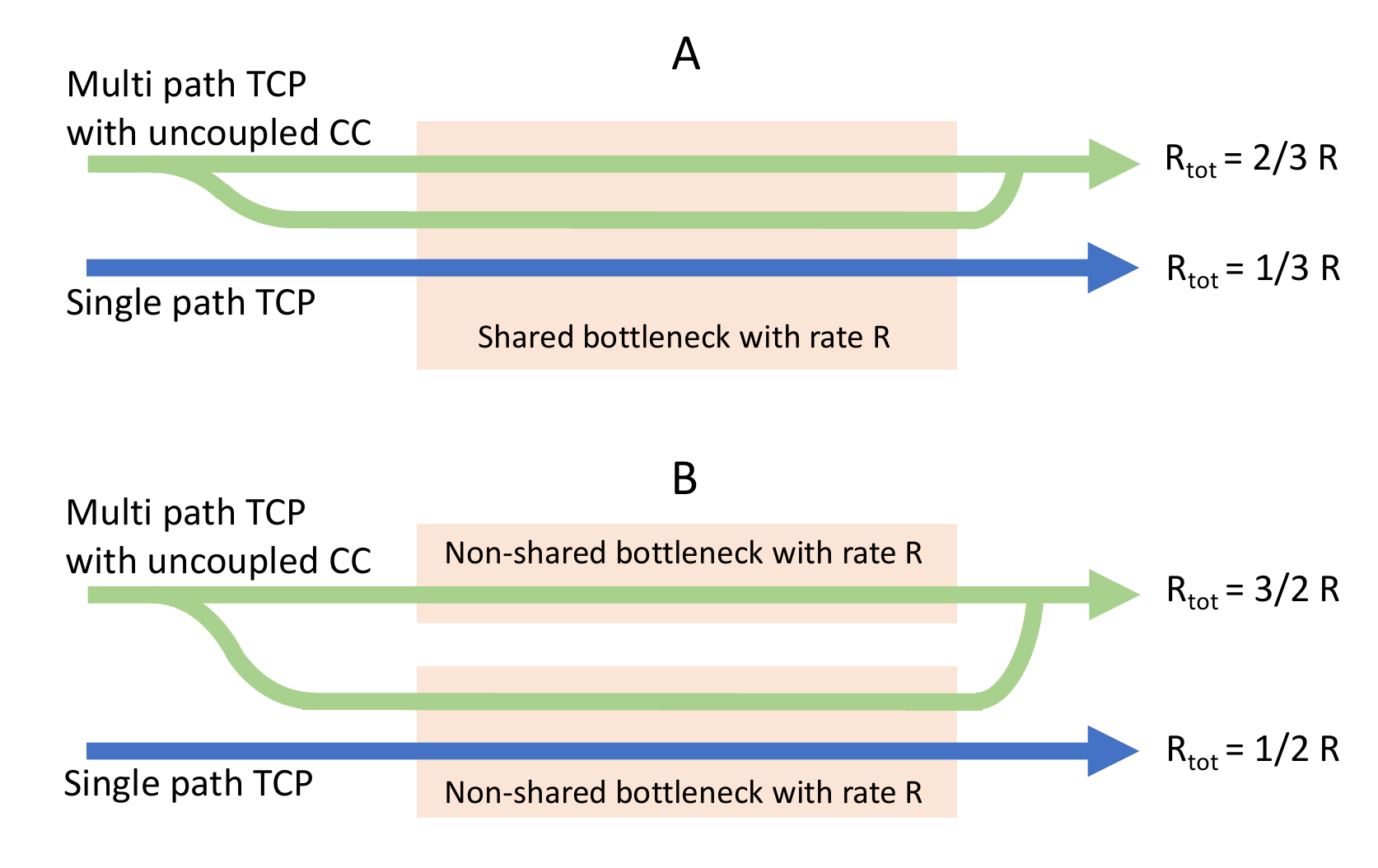}
\caption{Scenarios with unfairness between \gls{mptcp} (with uncoupled congestion control) and \gls{tcp}, adapted from~\cite{7460213}.}
\label{fig:MPTCP_unfair}
\end{figure}

\begin{table*}[ht]
	\centering 
	\caption{Summary of the main presented congestion control schemes for multipath, divided by protocol}
	\label{tab:mpcc}
	\scriptsize
	\begin{tabular}{cc|cccc}
		\toprule
		Protocol & Algorithm &   Mechanism               &      Type & Pros         &              Cons              \\
		\midrule
		MPTCP & CUBIC~\cite{6363695} & Independent CUBIC & Uncoupled & Direct adaptation of TCP & Unfair to TCP, \gls{hol}\\
		& Coupled~\cite{han2006multi} & Joint CUBIC CWND & Fully coupled & Fair to TCP & Does not use most congested path \\
		& LIA~\cite{Wischik,rfc6356} & Weighted CWND increase & Semi-coupled &Fair to TCP, uses all paths& \gls{hol}, unfairness to other LIA flows\\
		& OLIA~\cite{2578934} & Resource pooling & Semi-coupled & Pareto optimal& \gls{hol}, unstable behavior \\
		& BALIA~\cite{walid2014balia} & Mix of LIA and OLIA & Semi-coupled & Mix of OLIA and LIA's pros& \gls{hol}, wireless networks\\ 
		& BELIA~\cite{zhu2017belia} & Capacity-based & Semi-coupled & Good in wireless networks& \gls{hol}, unfairness\\ 
		& DRL-CC~\cite{xu2019experience}& Reinforcement learning & Fully coupled & High fairness & Untested in wireless networks\\ \midrule
		SCTP & CMT~\cite{iyengar2006concurrent} & SACK-based CWND growth & Semi-coupled&No \gls{hol} &Retransmissions, wireless networks \\
		& MPTCP-like~\cite{becke2013comparison}& Adaptation of LIA & Semi-coupled & No \gls{hol}, fair to TCP &Retransmissions, unfairness\\ \midrule
		MP-DCCP & FSCC~\cite{huang2015packet} & Adaptation of CCID 2& Semi-coupled & Free paths help congested ones & Untested, high reordering \\ \midrule
		LEAP & CKF~\cite{LEAP}& Capacity-based Kalman filtering & Uncoupled & No \gls{hol}, good in wireless networks & Vegas-like conservativeness\\
		\bottomrule
	\end{tabular}
\end{table*}

\subsubsection{\textbf{Congestion control}}\label{ssec:mptcpcc} The first versions of \gls{mptcp} used \emph{uncoupled} congestion control: each path had an independent congestion window, updated with one of the single-path mechanisms we described in Sec.~\ref{sec:cc}. However, measurements show that uncoupled \gls{mptcp} can be unfair to single-path \gls{tcp} users sharing one of the paths~\cite{6363695}, and can often lead to increased resource usage without a corresponding performance benefit~\cite{2578934}. 

In general, a congestion control algorithm for multipath flows should~\cite{rfc6356}:
\begin{enumerate}
\item Obtain at least the same throughput as a single path connection on the best sub-path;
\item Not take up more capacity than what a single path on the best sub-path would;
\item Avoid to push traffic on the most congested sub-paths.
\end{enumerate}

\rev{Fig.~\ref{fig:MPTCP_unfair} shows two different scenarios which lead to unfairness when \gls{mptcp} applies uncoupled congestion control~\cite{7460213}: in the first one, the two paths of the \gls{mptcp} share a bottleneck. In this case, goal 2 is violated, since \gls{mptcp} behaves like two separate \gls{tcp} flows, taking up two thirds of capacity and squeezing out the competing single-path flow.
In the second scenario, one of the two paths shares its bottleneck with a standard \gls{tcp} flow. In this case, \gls{mptcp} violates goals 2 and 3: since it has a non-congested path, \gls{mptcp} should steer most of its traffic towards the free path, avoiding the congested one (goal 3) and taking up a total capacity of $R$ (which would respect goal 2).
}

In order to achieve fairness with single-path flows, a \emph{semicoupled} algorithm achieving good performance while being fair to single-path flows, called \gls{lia}, was proposed~\cite{Wischik} and subsequently standardized~\cite{rfc6356}; the fully \emph{coupled} version~\cite{han2006multi} was considered too conservative, as it did not use the most congested path at all and could not then discover any changes in the traffic on that path.

The \gls{lia} congestion mechanism \rev{maintains the uncoupled single-path slow start, fast recovery and fast retransmit mechanisms of single-path \gls{tcp}}, but increases the congestion window after each ACK received on sub-flow $i$ by
\begin{equation}
 \min\left(
\frac{\alpha \times \text{bytes}_{\text{acked}} \times MSS_i}{CWND_{\text{total}}},\frac{\text{bytes}_{\text{acked}} \times MSS_i}{CWND_i}\right),
\end{equation}
\rev{where the second argument in the minimum is the increase that an uncoupled \gls{tcp} flow would have, while the first one is a coupled value that takes the overall congestion window into account. The parameter $\alpha$ describes the aggressiveness of the algorithm.}
Theoretically, we get
\begin{equation}
\alpha = CWND_{\text{total}}\,\,\frac{ \max_i (CWND_i / RTT_i^2)}{ \sum_i(CWND_i/RTT_i)^2}, 
\end{equation}
\rev{which corresponds to the congestion window increase that satisfies goal 1: by setting that value of $\alpha$, the \gls{mptcp} flow can theoretically obtain the same throughput as a single-path flow on the best path.}
The complexity of calculating the congestion window or $\alpha$ for each ACK scales linearly with the number of flows.

The \gls{olia}~\cite{2578934} is a modification to the \gls{lia} algorithm that achieves optimal resource pooling and avoids \gls{lia}'s fairness issues.
\gls{balia}~\cite{walid2014balia} is another proposal, now widely adopted, that tries to balance \gls{tcp}-friendliness and throughput stability, striking a balance between \gls{lia}'s and \gls{olia}'s strengths and weaknesses.

\begin{table*}[ht]
	\centering 
	\caption{Summary of the main presented scheduling algorithms}
	\label{tab:mpsch}
	\scriptsize
	\begin{tabular}{cc|ccc}
		\toprule
		Protocol & Algorithm &   Mechanism & Pros         &              Cons              \\
		\midrule
		MPTCP & LowRTT~\cite{8012365} & Lowest RTT first &Simple, widely deployed & Often inefficient \\
		&Loss-aware~\cite{ni2014fine} & Lowest RTT, weighted by loss rate & Works in wireless networks &Bad with asymmetric paths\\
		&Weighted round robin~\cite{choi2017optimal} & Round robin, weighted by path performance & Optimal load balancing & Unknown fairness \\
		&ECF~\cite{3143376} &Lowest completion time&Avoids idle flows&Can be hurt by \gls{hol} \\
		&DEMS~\cite{guo2017dems} &Forward on one path, backwards on the other& Object level delay& Requires data in blocks\\
		&STMS~\cite{216043}, DAPS~\cite{kuhn2014daps} & Delay modeling& Packets arrive in the correct order& RTT error sensitivity\\
		&BLEST~\cite{7497206} &Explicit \gls{hol} minimization& Limits the \gls{hol} issue&RTT error sensitivity\\
		&Application-aware~\cite{fahmi2018low}& Layer 7 delay minimization & Improved QoE& Requires application-level objects\\
		&Q-aware~\cite{shreedhar2018qaware} & Direct estimation of buffer occupancy & Limits the \gls{hol} issue& Requires network assistance\\ \midrule
		MPQUIC & Stream-aware~\cite{8422951}& Scheduling by stream & Improved webpage download times & Limited applicability outside HTTP/3 \\ \midrule
		MP-DCCP&AOPS~\cite{huang2015packet} & Delivery time and reliability modeling & Limits reordering & RTT error sensitivity\\ \midrule
		LEAP & DKF~\cite{LEAP} & Deadline-based Kalman filter & Reliable latency with FEC & FEC \\
		\bottomrule
	\end{tabular}
\end{table*}

There are also proposals for delay-based multipath congestion control, either by using delay-based algorithms on each sub-flow or by using specifically designed solutions~\cite{xu-mptcp-congestion-control-05}. The same goes for capacity-based~\cite{zhu2017belia,le2013improving} and cross-layer schemes~\cite{shreedhar2017more}, but all these proposals are untested and still under development.

However, these algorithms have been shown to be inefficient in wireless networks~\cite{8229265,7398634}: the \gls{hol} problem is exacerbated by the volatility of wireless links, and errors and retransmissions negatively affect both the average throughput and its stability, resulting in an oscillatory behavior~\cite{chen2013measurement}. \rev{A promising approach to multipath congestion control is to use reinforcement learning tools: \gls{drlcc}~\cite{xu2019experience}, an application of the actor-critic method which jointly sets the congestion window for all active flows and all paths, achieves high fairness in a wired network scenario with multiple active flows. The performance of these types of algorithms in more volatile wireless environments is still untested.}

\gls{fec} has been proposed as a solution for the \gls{hol} problem: by adding some redundancy packets to the flow, errors can be recovered by the receiver by decoding them and avoiding blocking other sub-flows unnecessarily. However, each sub-flow still needs to retransmit lost packets and deliver everything in order, even if the data has already been received on other flows. \gls{scmptcp} is a hybrid solution~\cite{li2013tolerating} that uses \gls{fec} to reconstruct missing packets, and other recent solutions~\cite{garcia2017low, 7998267, cui2015fmtcp,6193502,ferlin2018mptcp} use different coding schemes to achieve the same objective. These schemes can mitigate the \gls{hol} problem, especially when the receiver buffer size is limited, but they often require specifically designed schedulers and do not fix other congestion control issues in multipath, such as throughput volatility~\cite{ferlin2014multi}. Another workaround for the \gls{hol} problem is to retransmit lost packets on a faster path while avoiding transmitting new packets on the congested path~\cite{8330650}.

The main multipath congestion control algorithms we presented are summarized in Table~\ref{tab:mpcc}; for a more thorough survey of the research on multipath congestion control and \gls{mptcp}, we refer the reader to~\cite{7460213}.

\subsubsection{\textbf{Scheduling}} Congestion control is not the only factor affecting performance, as scheduling is also extremely important in \gls{mptcp}: sending data on the wrong path can exacerbate the \gls{hol} problem, increase latency and lower throughput. 

The scheduler in the Linux kernel currently follows the Lowest-\gls{rtt}-First (LowRTT) policy, so the first packet in the send buffer will always be sent through the lowest \gls{rtt} available path, but this simple heuristic is not always efficient. As Hwang \textit{et al.} point out in \cite{7182529}, waiting until the fastest path becomes free could be more efficient than immediately sending a packet on the slowest one if the difference between their \glspl{rtt} is large enough. Even simple heuristics based on delay, transmission rate, and loss rate often perform better than the Lowest-\gls{rtt} scheduler~\cite{8012365,ni2014fine}.

In \cite{choi2017optimal}, the authors use a weighted round robin scheduler in combination with load balancing to overcome this difficulty; loss-aware scheduling is also a possibility~\cite{8109323}. \gls{ecf}~\cite{3143376} is another scheduler that considers completion time as its main objective; it tries to reduce underutilization of flows by avoiding long idle periods, as they would cause CWND resets and consequent inefficiency in the capacity utilization.
\gls{dems}~\cite{guo2017dems} uses the two paths to transmit the data out-of-order: the data is sent from the first packet of a chunk on the first path, and from the last packet backwards, until the two flows meet and the chunk is fully downloaded.

In general, the optimal scheduling might not always send packets in-order, as sending future packets so that they arrive at the same time as the first (which is then sent later on a much faster flow) can be advantageous. The \gls{stms}~\cite{216043} and \gls{daps}~\cite{kuhn2014daps} schedulers explicitly model this aspect, interleaving packets so that successive packets sent over different paths arrive at the same time. The \gls{fec}-based congestion control scheme in~\cite{cui2015fmtcp} also uses a similar strategy with good results. 

The \gls{blest} scheduler~\cite{7497206} is the first one in the literature to explicitly consider \gls{hol}; it tries to estimate which sub-flows are likely to cause it and dynamically adapts the scheduling to prevent it. 
Cross-layer scheduling is also a possibility; in~\cite{fahmi2018low}, the authors schedule application-level objects on different paths to minimize webpage-loading times.
The Q-aware scheduler~\cite{shreedhar2018qaware} uses information from the network in a cross-layer fashion, directly considering buffer occupancy.
An experimental evaluation of several of the scheduling algorithms above is presented in~\cite{paasch2014experimental}.

We summarize the main features, advantages and drawbacks of most of the scheduling algorithms described above in Table~\ref{tab:mpsch}.

\rev{
\subsubsection{Multipath TCP in data center networks}\label{ssec:mptcp_datacenter}
One of the original proposed use cases for \gls{mptcp} was the data center scenario~\cite{raiciu2010data}: since data centers often have complex topologies with multiple available paths between hosts, \gls{mptcp} is a natural solution. Large-scale simulations show that using \gls{mptcp} can improve load balancing, leading to fewer underutilized links and higher overall throughput, particularly in optical networks~\cite{tariq2014performance}. However, it is not immune to the Incast issue described in Sec.~\ref{ssec:incast}~\cite{li2014mptcp}: whenever a client requests data from multiple servers at once, throughput collapses, even if \gls{mptcp} can actively relieve congested links. While the subflows from an \gls{mptcp} connection are not more aggressive than a single-path \gls{tcp} flow, multiple \gls{mptcp} connections are not aware of each other. The solution proposed in~\cite{li2014mptcp} is to consider the number of existing flows when updating the congestion window of each subflow, combining an equally weighted coupling between flows to the subflow-level coupling described in Sec.~\ref{ssec:mptcpcc}.

The possibility of quickly retransmitting lost packets on less congested flows is another enhancement that has been proposed for the data center scenario~\cite{hwang2018fast}: this kind of technique is not needed for long-lived flows, but it can compensate for \gls{mptcp}'s inability to steer short flows towards less congested paths and alleviate the Incast issue for this kind of frequent, short-lived traffic.

The benefits of \gls{mptcp} in data center networks can be increased when combined with \gls{sdn}: network support can improve routing~\cite{zannettou2016exploiting}, leading to fewer congestion events. The performance benefit is even larger when \gls{mptcp} senders themselves are aware of the network situation and can dynamically add and remove subflows to avoid congestion~\cite{duan2015responsive}.
}

\subsection{Multipath in other protocols}
\label{sec:mpother}
The first alternative to \gls{tcp} to have a multipath capability was \gls{sctp}~\cite{iyengar2006concurrent}: the \gls{cmt} extension increases throughput, but it has \gls{tcp} friendliness issues. Several \gls{cmt}-\gls{sctp} congestion control algorithms have been proposed~\cite{dreibholz2011impact}, both using resource pooling concepts and following in \gls{mptcp}'s footsteps; a full Markov model of congestion control is presented in~\cite{wallace2014concurrent}.
An experimental comparison between \gls{mptcp} and \gls{cmt}-\gls{sctp} was performed in~\cite{becke2013comparison}; the results showed that \gls{mptcp} has a slight performance advantage, but both protocols have issues with high-delay paths. For a more thorough analysis of \gls{cmt}-\gls{sctp} congestion control algorithms, we refer the reader to the multipath congestion control survey we mentioned above~\cite{7460213}, which examines it in detail.

More recent proposals add multipath capabilities to the \gls{quic} protocol, paralleling the development of \gls{mptcp}: since \gls{quic} is implemented in the user-space and not in the kernel, it is much easier to extend. Proposals for \gls{mpquic} were recently advanced in \cite{DeConinck:2017:MQD:3143361.3143370,deconinck-multipath-quic-00} and \cite{8422951}.

Both works develop similar aspects of the protocol, principally taking advantage of \gls{quic}'s features, like the 0-RTT connection establishment or its transparency to middleboxes.
The great advantage of \gls{quic} over \gls{tcp} in the multipath domain is that it does not require in-order delivery on each sub-flow, sidestepping the \gls{hol} problem. For this reason, the design parameters and constraints for multipath congestion control in \gls{quic} are not equivalent to those of \gls{mptcp}, and future work comparing them should take this into account. At the moment, research on \gls{mpquic} is still very limited, as the multipath extension was only proposed in 2017. A first example of \gls{quic}-specific work in multipath is given by~\cite{8422951}, which proposes a stream-aware scheduler.

\rev{
Finally, \gls{mpdccp} is a recent draft~\cite{draft-amend-tsvwg-multipath-framework-mpdccp-00} that extends \gls{dccp} to provide multipath capabilities. The peculiarities of \gls{dccp} are also present in its multipath version: data transfer is unreliable and out of order. Contrary to \gls{mptcp}, the communication establishment does not rely on a set initial path: as long as at least one functional path is available, the \gls{mpdccp} connection can be established. The standard is still unfinished, and parts of the signaling and receiver-side reassembly procedures are yet to be defined.

Reordering is a major problem in \gls{mpdccp}: even though single-path \gls{dccp} is also unreliable and does not guarantee in-order delivery, reordering is rare on single-path connections. In a multipath scenarios, reorderings can be much more common, and the use of a scheduler that prevents them as much as possible is recommended to prevent an excessive degradation of the application performance.

A possible multipath congestion control scheme is \gls{fscc}~\cite{huang2015packet}, based on \gls{ccid} 2. It adds a ``helping state'' to each flow's congestion control state, which is triggered when another path is congested. After the congested path's CWND is reduced, the helping path's is increased accordingly to increase the overall throughput stability. The authors of \gls{fscc} also propose a predictive scheduler called \gls{aops} as a way to deal with the reordering issue: the scheme considers the predicted delivery time of packets, as well as the reliability of each path, to optimize the schedule for in-order delivery. Reliability can also be adapted to the type of data being sent, protecting more important data so that it is delivered on the most reliable path, even if this increases the latency.}

\begin{figure}[t]
 \centering
  \input{img/multipath_wifilte.tex}
 \caption{Performance of well-known multipath protocols in the throughput/reliability plane in a multipath scenario [Wi-Fi+LTE]. Reprinted, with permission, from~\cite{LEAP}.}
\label{fig:multi-path_wifilte}
\end{figure}
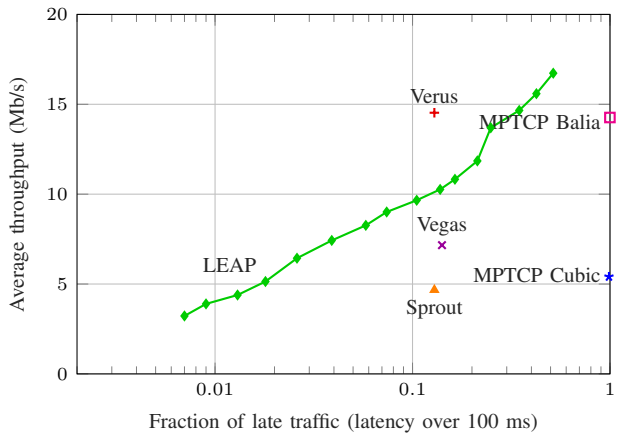

\rev{There are also some natively multipath protocols: }\gls{leap}~\cite{LEAP} is a protocol developed specifically to exploit the multipath scenario in order to guarantee low latency to applications. It uses a capacity-based congestion control mechanism on each sub-flow, and adaptively sends \gls{fec} to compensate for capacity estimation errors, exploiting path diversity to guarantee that packets will reach the receiver with a strictly bounded latency. Since it runs in user-space over \gls{udp}, it does not require retransmission on each sub-flow: lost or late packets that are recovered using the \gls{fec} protection on the other paths do not need to be retransmitted at all.

Fig.~\ref{fig:multi-path_wifilte} shows the relative performance of several congestion control mechanisms in a wireless scenario with two paths, one over LTE and one over an office Wi-Fi: while \gls{mptcp} \gls{balia} performs far better than the uncoupled CUBIC because it handles the \gls{hol} problem far better, it does not provide any latency guarantees. Less aggressive protocols such as Verus, Vegas, and Sprout do better in terms of latency, but they still do not manage to go below a lateness rate of 10\%. \gls{leap} can control the trade-off between the strictness of the latency constraint and the throughput by dynamically adjusting the coding rate, and it is the only protocol that manages to  violate the latency requirement less than 1\% of the time.

\rev{
\subsection{Open Challenges and Research Directions}
\label{sec:mp-ch}

As discussed in Sec.~\ref{sec:transport} and in the previous paragraphs, the exploitation of multiple paths at the transport layer is a promising research trend, made possible by the advanced capabilities of recent communication devices, such as smartphones capable of connecting to the network over multiple wireless technologies (usually cellular and Wi-Fi). The field is still in its infancy, and there are unsolved issues such as fairness and \gls{hol} blocking: some of these issues are caused by the backward compatibility requirement and can be mitigated by using flexible protocols such as \gls{mpquic}, but other problems are more fundamental.

The close coupling of scheduling and congestion control, which have significant interactions with hard-to-model effects, is one of the biggest issues for multipath protocols, and it affects all protocols: all the issues described in Sec.~\ref{sec:issues} are exacerbated by the additional layer of adaptation. Designing an efficient integrated multipath scheduling and congestion control scheme is very challenging, and will probably require the use of \gls{fec} and extensive cross-layer support.
Finally, the use of multipath protocols in wireless and data center networks requires specific adaptations, and there are several research challenges that still have to be solved.

}

\section{Conclusions}\label{sec:conclusion}

\gls{tcp} has been the \textit{de facto} standard transport protocol for years, but, despite its wide adoption, it presents sub-optimal performance in a number of use cases and scenarios. Moreover, new emerging technologies (e.g., mmWave communications) and requirements (e.g., those for ultra-low latency \gls{vr} streaming) are overtaking the performance that \gls{tcp} can achieve. 
Therefore, the research related to the transport layer has seen a renewed interest in the last few years. In this paper, we reviewed the main results related to these efforts.\rev{
In particular, we analyzed three main research areas: transport protocols in general, congestion control, and multipath transport. 

First, we discussed the main features of three proposed protocols, namely \gls{quic}, \gls{sctp} and \gls{dccp}, which represent significant evolutions or alternatives to the widely used \gls{tcp}. These protocols have several improvements related to the acknowledgment mechanism, the connection establishment, the management of the connection life cycle, and the embedding of the cryptographic stack in the protocol itself. \gls{sctp} and \gls{dccp} have been standardized by the \gls{ietf}, but have not reached a wide adoption in the Internet. \gls{quic}, instead, has not been fully standardized yet, but is implemented in the user space, thus can be deployed without the need to upgrade the operating system. Besides, some early results show its promising performance, at the cost of a higher computational load with respect to kernel-based solutions. \gls{quic} can be seen as a possible enabler of a flexible deployment of congestion control to target different scenarios and use cases.

Then, we reviewed a selection of recently proposed congestion control algorithms for \gls{tcp} and, in general, for congestion-aware transport protocols. We analyzed different approaches, from evolutions of traditional loss-based mechanisms to new delay-based or hybrid proposals and machine learning strategies.  }

Finally, another promising trend is represented by the usage of multiple paths at the transport layer, to provide macro diversity and, possibly, increase the throughput and reliability. A first solution is \gls{mptcp}, an extension of \gls{tcp} that dispatches packets over multiple subflows when multiple network interfaces are available. However, while maintaining \gls{tcp} compatibility with respect to middleboxes by design, it suffers from \gls{hol} issues (depending on the scheduler implementation) and fairness with respect to single path \gls{tcp}. Multi-path extensions of other protocols (e.g., \gls{sctp}, \gls{quic}, and \gls{dccp}) have also been proposed but are still open research areas.

The research related to several of the topics presented in this survey is still ongoing, and, given the fast update rate of communication technologies, the interest in innovation at the transport layer will hardly fade away. A promising research direction is related to the coupling of network slicing and \gls{tcp} fairness: slices can be instantiated by network operators to protect delay-based flows against less fair algorithms. Similarly, adaptive congestion control algorithms are still being studied, with the aim of targeting the optimal operating point that minimizes the latency while maximizing the data rate. Finally, in the multipath domain, standardization efforts are still ongoing, and new efficient and fair multi-link scheduling and congestion control algorithms will need to be identified.

	\renewcommand{\arraystretch}{1}
	\footnotesize
	\setlength{\glsdescwidth}{0.75\columnwidth}
	\printglossary[style=index]
\normalsize

\bibliographystyle{IEEEtran}
\bibliography{ref-ieee,ref-rfc,ref-site}

\begin{IEEEbiography}
    [{\includegraphics[width=0.9in,clip,keepaspectratio]{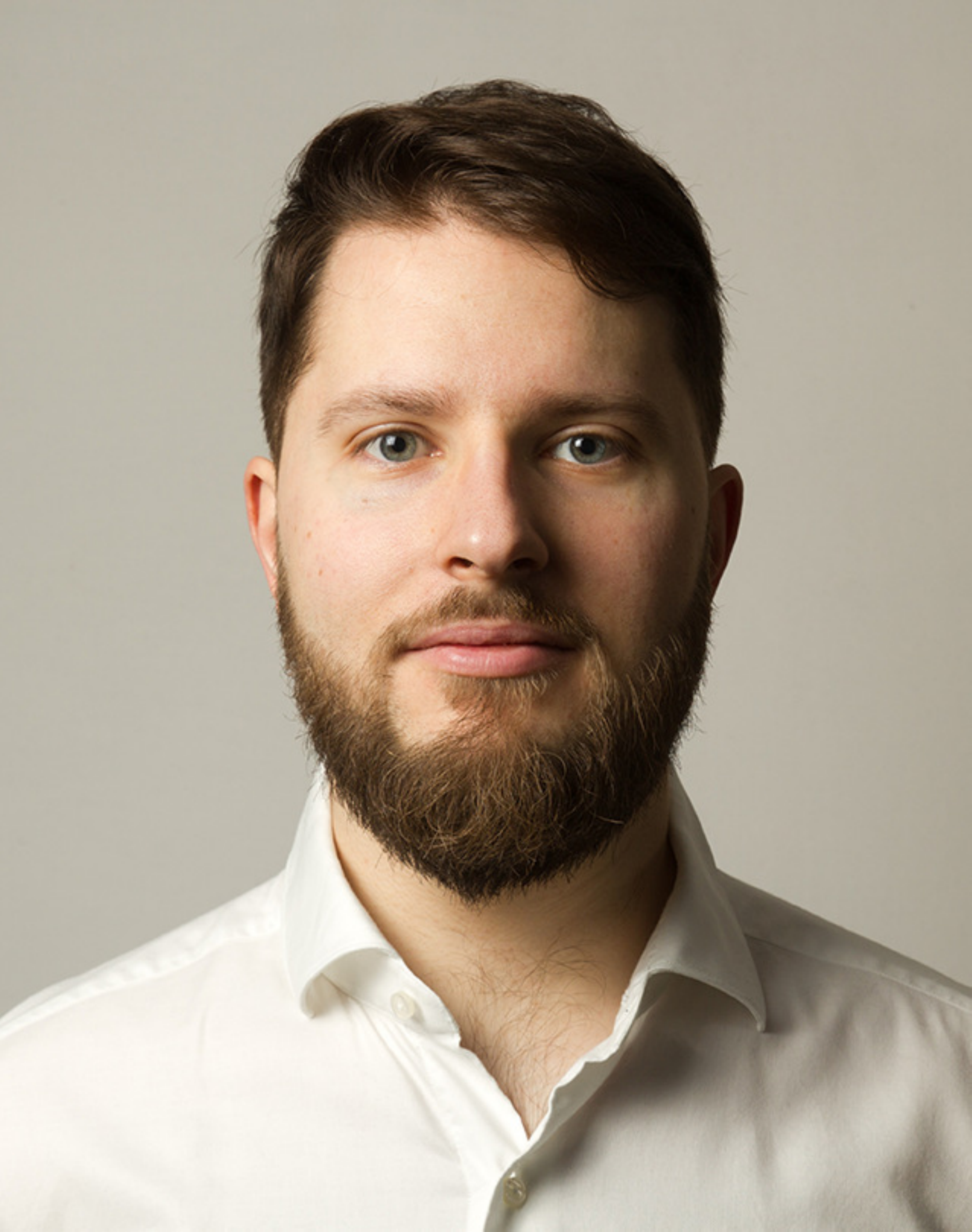}}]{Michele Polese}
[S'17] received his B.Sc. in Information Engineering in 2014 and M.Sc. in Telecommunication Engineering in 2016 from the University of Padova, Italy. Since October 2016 he has been a Ph.D. student at the Department of Information Engineering of the University of Padova, under the supervision of Prof. Michele Zorzi. He visited New York University (NYU), NY in 2017, AT\&T Labs in Bedminster, NJ in 2018, and Northeastern University, Boston, MA in 2019. He is collaborating with several academic and industrial research partners, including Intel, InterDigital, NYU, AT\&T Labs, University of Aalborg, King's College, Northeastern University and NIST. He was awarded with the Best Journal Paper Award of the IEEE ComSoc Technical Committee on Communications Systems Integration and Modeling (CSIM) 2019 and the Best Paper Award at WNS3 2019. His research interests are in the analysis and development of protocols and architectures for the next generation of cellular networks (5G), in particular for millimeter-wave communication, and in the performance evaluation of complex networks.
\end{IEEEbiography}

\begin{IEEEbiography}
    [{\includegraphics[width=0.9in,clip,keepaspectratio]{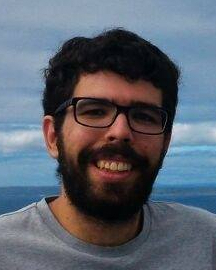}}]{Federico Chiariotti}
[S'15 M'19] is a post-doctoral researcher at the University of Padova, in Italy, where he received his Ph.D. in information engineering in 2019. He received the bachelor's and master's degrees in telecommunication engineering (both cum laude) from the University of Padova, in 2013 and 2015, respectively. In 2017 and 2018, he was a Research Intern with Nokia Bell Labs, Dublin. He has authored over 20 published papers on wireless networks and the use of artificial intelligence techniques to improve their performance. He was a recipient of the Best Paper Award at the Workshop on ns-3 in 2019 and the Best Student Paper Award at the International Astronautical Congress in 2015. His current research interests include network applications of machine learning, transport layer protocols, Smart Cities, bike sharing system optimization, and adaptive video streaming.
\end{IEEEbiography}

\begin{IEEEbiography}
    [{\includegraphics[width=0.9in,clip,keepaspectratio]{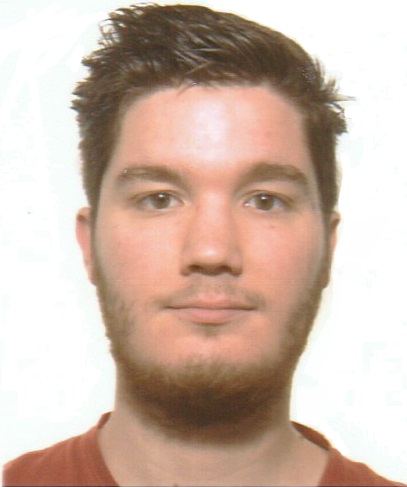}}]{Elia Bonetto}
Elia Bonetto received his B.Sc. degree in computer engineering from the University of Padova, Padova, Italy, in 2017, where he is currently pursuing the M.Sc. degree in ICT for Internet and multimedia.
His main interests are in computer vision and robotics besides computer networks and network protocols.
\end{IEEEbiography}

\begin{IEEEbiography}
    [{\includegraphics[width=0.9in,clip,keepaspectratio]{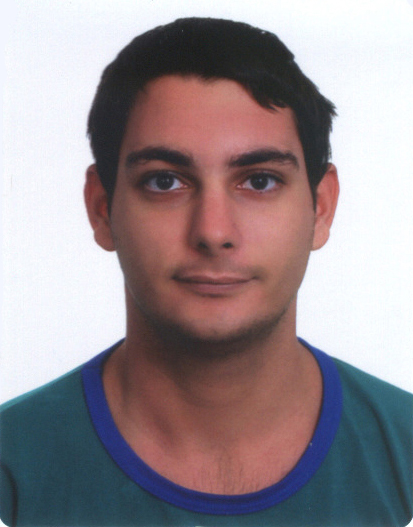}}]{Filippo Rigotto}
received his B.Sc. degree in computer engineering from the University of Padova, Italy, in 2017, where he is currently pursuing the M.Sc. degree in ICT for Internet and multimedia.
His present interests are in computer networks and network protocols, machine learning, computer vision and robotics areas.
\end{IEEEbiography}

\begin{IEEEbiography}
    [{\includegraphics[width=0.9in,clip,keepaspectratio]{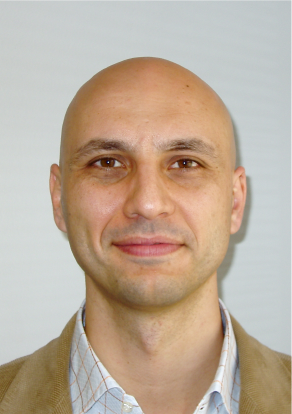}}]{Andrea Zanella}
[S'98-M'01-SM'13] is Associate Professor at the University of Padova, in Italy. He received the master degree in Computer Engineering and the Ph.D. degree in Electronic and Telecommunications Engineering from the same University, in 1998 and 2000, respectively. He has (co)authored more than 130 papers, five books chapters and three international patents in multiple subjects related to wireless networking and Internet of Things, Vehicular Networks, cognitive networks, and microfluidic networking. He serves as Technical Area Editor for the \textsc{IEEE Internet of Things Journal}, and as Associate Editor for the \textsc{IEEE Communications Surveys \& Tutorials}, and the \textsc{IEEE Transactions on Cognitive Communications and Networking}.
\end{IEEEbiography}

    \begin{IEEEbiography}
    [{\includegraphics[width=0.99in,clip,keepaspectratio]{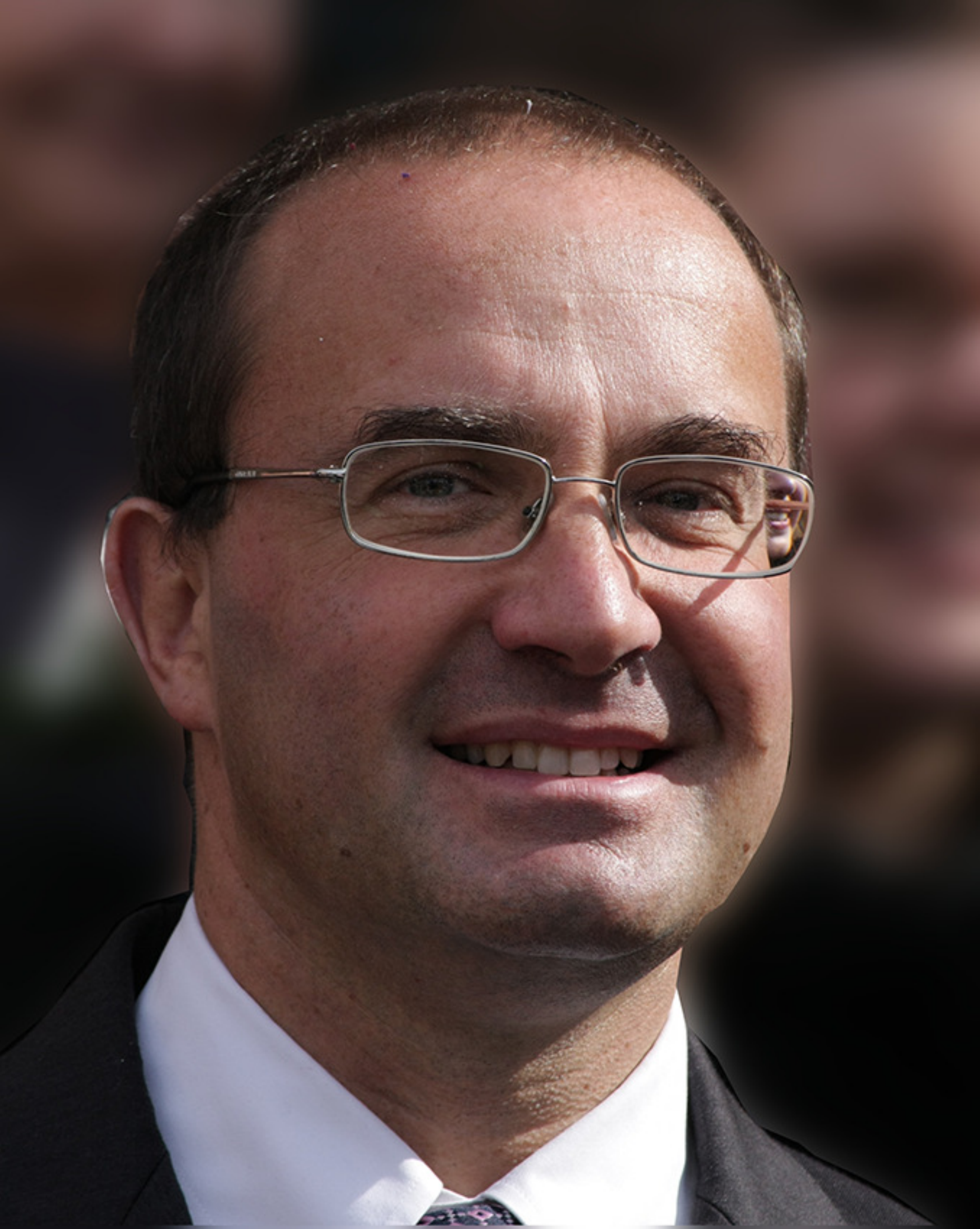}}]{Michele Zorzi}
        [F'07] received his Laurea and PhD degrees in electrical engineering from the University of Padova in 1990 and 1994, respectively. During academic year 1992-1993 he was on leave at the University of California at San Diego (UCSD). In 1993 he joined the faculty of the Dipartimento di Elettronica e Informazione, Politecnico di Milano, Italy. After spending three years with the Center for Wireless Communications at UCSD, in 1998 he joined the School of Engineering of the University of Ferrara, Italy, where he became a professor in 2000. Since November 2003 he has been on the faculty of the Information Engineering Department at the University of Padova. His present research interests include performance evaluation in mobile communications systems, WSN and Internet of Things, cognitive communications and networking, 5G mmWave cellular systems, vehicular networks, and underwater communications and networks. 
        He is the recipient of several awards 
from the IEEE Communications Society, including the Best Tutorial Paper 
Award (2008, 2019), the Education Award (2016), and the Stephen O. Rice Best 
Paper Award (2018).
        He was the Editor in Chief of the IEEE Transactions on Cognitive Communications and Networking from 2014 to 2018, of IEEE Wireless Communications from 2003 to 2005 and of the IEEE Transactions on Communications from 2008 to 2011. He served as a Member-at-Large of the Board of Governors of the IEEE Communications Society from 2009 to 2011, and as its Director of Education in 2014-15.
    \end{IEEEbiography}
\end{document}

%% file: acronyms.tex
\newacronym{ai}{AI}{Artificial Intelligence}
\newacronym{ecf}{ECF}{Earliest Completion First}
\newacronym{leap}{LEAP}{Latency-controlled End-to-End Aggregation Protocol}
\newacronym{qoe}{QoE}{Quality of Experience}
\newacronym{rl}{RL}{Reinforcement Learning}
\newacronym{tcp}{TCP}{Transmission Control Protocol}
\newacronym{dash}{DASH}{Dynamic Adaptive Streaming over HTTP}
\newacronym{dems}{DEMS}{Decoupled Multipath Scheduler}
\newacronym{http}{HTTP}{HyperText Transfer Protocol}
\newacronym{udp}{UDP}{User Datagram Protocol}
\newacronym{mdp}{MDP}{Markov Decision Process}
\newacronym{mpeg}{MPEG}{Moving Picture Experts Group}
\newacronym{lia}{LIA}{Linked Increases Algorithm}
\newacronym{olia}{OLIA}{Opportunistic Linked Increases Algorithm}
\newacronym{vi}{VI}{Value Iteration}
\newacronym{cdn}{CDN}{Content Delivery Network}
\newacronym{html}{HTML}{HyperText Markup Language}
\newacronym{mpd}{MPD}{Media Presentation Description}
\newacronym{url}{URL}{Uniform Resource Locator}
\newacronym{td}{TD}{Temporal Difference}
\newacronym{qos}{QoS}{Quality of Service}
\newacronym{mos}{MOS}{Mean Opinion Score}
\newacronym{lola}{LoLa}{Low Latency}
\newacronym{bic}{BIC}{Binary Increase Control}
\newacronym{psnr}{PSNR}{Peak Signal to Noise Ratio}
\newacronym{daps}{DAPS}{Delay Aware Packet Scheduling}
\newacronym{mse}{MSE}{Mean Square Error}
\newacronym{rto}{RTO}{Retransmission Timeout}
\newacronym{rmse}{RMSE}{Root Mean Square Error}
\newacronym{ssim}{SSIM}{Structural Similarity Index}
\newacronym{psqa}{PSQA}{Pseudo-Subjective Quality Assessment}
\newacronym{cac}{CAC}{Call Admission Control}
\newacronym{mss}{MSS}{Microsoft Smooth Streaming}
\newacronym{pomdp}{POMDP}{Partially Observable MDP}
\newacronym{svc}{SVC}{Scalable Video Coding}
\newacronym{sarsa}{SARSA}{State-Action-Reward-State-Action}
\newacronym{stms}{STMS}{Slide Together Multipath Scheduler}
\newacronym{dqn}{DQN}{Deep Q-network}
\newacronym{mlp}{MLP}{Multilayer Perceptron}
\newacronym{lstm}{LSTM}{Long-Short Term Memory}
\newacronym{festive}{FESTIVE}{Fair, Efficient, and Stable adapTIVE algorithm}
\newacronym{panda}{PANDA}{Probe and Adapt}
\newacronym{rahas}{QoE-RAHAS}{QoE-driven Rate Adaptation Heuristic for Adaptive video Streaming}
\newacronym{rnn}{RNN}{Recurrent Neural Network}
\newacronym{btt}{BTT}{Backpropagation Through Time}
\newacronym{bdp}{BDP}{Bandwidth-Delay Product}
\newacronym{mpc}{MPC}{Model Predictive Control}
\newacronym{cbr}{CBR}{Constant Bit Rate}
\newacronym{vbr}{VBR}{Variable Bit Rate}
\newacronym{ckf}{CKF}{Congestion Kalman Filter}
\newacronym{dkf}{DKF}{Delivery Kalman Filter}
\newacronym{rtt}{RTT}{Round Trip Time}
\newacronym{fec}{FEC}{Forward Error Correction}
\newacronym{cdf}{CDF}{Cumulative Distribution Function}
\newacronym{pdf}{PDF}{Probability Distribution Function}
\newacronym{mptcp}{MPTCP}{Multipath TCP}
\newacronym{mpquic}{MPQUIC}{Multipath QUIC}
\newacronym{cmt}{CMT}{Concurrent Multipath Transport}
\newacronym{aqm}{AQM}{Active Queue Management}
\newacronym{hmm}{HMM}{Hidden Markov Model}
\newacronym{balia}{BALIA}{Balanced Link Adaptation}
\newacronym{scmptcp}{SC-MPTCP}{MPTCP with Systematic Coding}
\newacronym{fmtcp}{FMTCP}{Fountain-based MPTCP}
\newacronym{bbr}{BBR}{Bottleneck Bandwidth and Round-trip propagation time}
\newacronym{lan}{LAN}{Local Area Network}
\newacronym{jfi}{JFI}{Jain Fairness Index}
\newacronym{vr}{VR}{Virtual Reality}
\newacronym{ar}{AR}{Augmented Reality}
\newacronym{ue}{UE}{User Equipment}
\newacronym{sc}{SC}{Smart City}
\newacronym{kpi}{KPI}{Key Performance Indicator}
\newacronym{m2m}{M2M}{Machine to Machine}
\newacronym{v2v}{V2V}{Vehicle to Vehicle}
\newacronym{v2i}{V2I}{Vehicle to Infrastructure}
\newacronym{i2v}{I2V}{Infrastructure to Vehicle}
\newacronym{d2d}{D2D}{Device to Device}
\newacronym{son}{SON}{Self-Organized Networking}
\newacronym{ran}{RAN}{Radio Access Network}
\newacronym{cn}{CN}{Core Network}
\newacronym{cloudran}{CloudRAN}{Cloud Radio Access Network}
\newacronym{nfv}{NFV}{Network Function Virtualization}
\newacronym{vm}{VM}{Virtual Machine}
\newacronym{rss}{RSS}{Received Signal Strength}
\newacronym{rssi}{RSSI}{Received Signal Strength Indicator}
\newacronym{sdn}{SDN}{Software-Defined Networking}
\newacronym{ttt}{TTT}{Time-to-Trigger}
\newacronym{enb}{eNB}{evolved Node Base}
\newacronym{scoot}{SCOOT}{Split Cycle Offset Optimization Technique}
\newacronym{utc}{UTC}{Urban Traffic Control}
\newacronym{tfl}{TfL}{Transport for London}
\newacronym{hetnet}{HetNet}{Heterogeneous Network}
\newacronym{snr}{SNR}{Signal to Noise Ratio}
\newacronym[plural=MMEs,firstplural=Mobility Management Entities (MMEs)]{mme}{MME}{Mobility Management Entity}
\newacronym{bs}{BS}{Base Station}
\newacronym{svr}{SVR}{Support Vector Regression}
\newacronym{svm}{SVM}{Support Vector Machine}
\newacronym{rf}{RF}{Random Forest}
\newacronym{nn}{NN}{Neural Network}
\newacronym{mimo}{MIMO}{Multi-Input Multi-Output}
\newacronym{gb}{GB}{Graphical Bayesian}
\newacronym{aimd}{AIMD}{Additive Increase Multiplicative Decrease}
\newacronym{csi}{CSI}{Channel State Information}
\newacronym{cr}{CR}{Cognitive Radio}
\newacronym{mcs}{MCS}{Modulation and Coding System}
\newacronym{lfn}{LFN}{Long Fat Network}
\newacronym{ofdm}{OFDM}{Orthogonal Frequency Division Modulation}
\newacronym{gps}{GPS}{Global Positioning System}
\newacronym{sv}{SV}{Support Vector}
\newacronym{lte}{LTE}{Long Term Evolution}
\newacronym{rbf}{RBF}{Radial Basis Function}
\newacronym{nr}{NR}{New Radio}
\newacronym{blest}{BLEST}{Blocking Estimation}
\newacronym{iot}{IoT}{Internet of Things}
\newacronym{mip}{MIP}{Mixed Integer Programming}
\newacronym{mmpp}{MMPP}{Markov-Modulated Poisson Process}
\newacronym{brp}{BRP}{Bicycle Routing Problem}
\newacronym{comp}{CoMP}{Coordinated Multi-Point}
\newacronym{dude}{DUDe}{Downlink and Uplink Decoupling}
\newacronym{knn}{k-NN}{k-Nearest Neighbors}
\newacronym{quic}{QUIC}{Quick UDP Internet Connections}
\newacronym{nbiot}{NB-IoT}{Narrowband IoT}
\newacronym{lpwa}{LPWA}{Low-Power Wide-Area}
\newacronym{hol}{HoL}{Head of Line}
\newacronym{ledbat}{LEDBAT}{Low Extra Delay Background Transport}
\newacronym{ewma}{EWMA}{Exponentially Weighted Moving Average}
\newacronym{pcc}{PCC}{Performance-oriented Congestion Control}
\newacronym{tao}{TAO}{Tractable Attempt at Optimal}
\newacronym{red}{RED}{Random Early Dropping}
\newacronym{sack}{SACK}{Selective ACK}
\newacronym{dsn}{DSN}{Data Sequence Number}
\newacronym{ecn}{ECN}{Explicit Congestion Notification}
\newacronym{sctp}{SCTP}{Stream Control Transmission Protocol}
\newacronym{dccp}{DCCP}{Datagram Congestion Control Protocol}
\newacronym{mpdccp}{MP-DCCP}{Multipath Datagram Congestion Control Protocol}
\newacronym{los}{LOS}{Line of Sight}
\newacronym{nlos}{NLOS}{Non Line of Sight}
\newacronym{mtc}{MTC}{Machine-type Communication}
\newacronym{mac}{MAC}{Medium Access Control}
\newacronym{tls}{TLS}{Transport Layer Security}
\newacronym{codel}{CoDel}{Controlled Delay Management}
\newacronym{sip}{SIP}{Session Initiation Protocol}
\newacronym{ietf}{IETF}{Internet Engineering Task Force}
\newacronym{ccid}{CCID}{Congestion Control ID}
\newacronym{manet}{MANET}{Mobile Ad Hoc Network}
\newacronym{tfrc}{TFRC}{TCP-Friendly Rate Control}
\newacronym{aops}{AOPS}{Adaptive Order Prediction Scheduling}
\newacronym{fscc}{FSCC}{Flow Sharing Congestion Control}
\newacronym{drlcc}{DRL-CC}{Deep Reinforcement Learning Congestion Control}
\newacronym{dctcp}{DCTCP}{Data Center TCP}
\newacronym{pase}{PASE}{Prioritization, Arbitration, and Self-adjusting Endpoints}
\newacronym{d2tcp}{D$^2$TCP}{Deadline-aware Data center TCP}
\newacronym{d3}{D$^3$}{Deadline-Driven Delivery}
\newacronym{ip}{IP}{Internet Protocol}
\newacronym{api}{API}{Application Programming Interface}

%% file: img/tp-table.tex
\begin{table*}[b]
	\centering 
	\caption{Summary of the main features of the three recent transport protocols reviewed in this paper (\gls{sctp}, \gls{quic} and \gls{dccp}). We also include a review of the legacy TCP and UDP protocols as a comparison.
	}
	\label{tab:tpsum}
	\footnotesize
  	\begin{tabularx}{\textwidth}{>{\hsize=\hsize}l|>{\hsize=\hsize}l>{\hsize=\hsize}X>{\hsize=\hsize}X>{\hsize=\hsize}X}
		\toprule
		Protocol & RFC or Drafts &   Features & Implementations & Multipath extensions \\
		\midrule
		\gls{quic}$^{\dagger}$ & \cite{draftquicktr,draftquickrec} & Byte stream, flow and congestion control, reliability (configurable), multi-streaming, integration with \gls{tls}, zero-\gls{rtt} connection establishment & User space, multiple libraries are available~\cite{QGo,LSQ,libquic,proto-quic} & Multipath \gls{quic}, described in Sec.~\ref{sec:mpother} \\
		\gls{sctp} & \cite{rfc4960} & Byte stream, flow and congestion control, reliability (configurable), multi-streaming, multi-homing & Operating system kernel & \gls{cmt}-\gls{sctp} described in Sec.~\ref{sec:mpother} \\
		\gls{dccp} & \cite{rfc4340,rfc5595,rfc5596} & Datagram based, congestion control & Operating system kernel & \gls{mpdccp}, described in Sec.~\ref{sec:mpother}\\
		\rowcolor{gray!25} TCP & \cite{rfc793,rfc2018} & Byte stream, flow and congestion control, reliability & Operating system kernel & \gls{mptcp}, described in Sec.~\ref{sec:mptcp} \\
		\rowcolor{gray!25} UDP & \cite{rfc768} & Datagram based, best effort & Operating system kernel & No \\
		\bottomrule
	\end{tabularx}
	\raggedright \scriptsize$^{\dagger}$\gls{quic} is implemented on top of UDP, but provides transport layer functionalities.
\end{table*}

%% file: img/quic-conn-setup.tex
\begin{tikzpicture}[auto]

 \node at (0,0)[rectangle,draw,name=cbeg1] {Client};
 \node at (1,0)[rectangle,draw,name=sbeg1] {Server};
 \node at (0.5,-3.2)[rectangle,name=lab1] {TCP+TLS1.2};
 \node at (0.5,-2.7)[rectangle,draw,name=time1] {3RTT};
 \coordinate (send1) at(1,-3); 
 \coordinate (cend1) at(0,-3); 
 \coordinate (c11) at(0,-0.4); 
 \coordinate (s11) at(1,-0.7);
 \coordinate (c12) at(0,-1);   
 \coordinate (s12) at(1,-1.3); 
 \coordinate (c13) at(0,-1.6);    
 \coordinate (s13) at(1,-1.9); 
 \coordinate (c14) at(0,-2.2);  
 \coordinate (s14) at(1,-2.5);  
 
  \node at (3,0.5)[rectangle,name=label] {First connection};
  \node at (3,-4)[rectangle,name=label] {Subsequent connections};

  \node at (2.5,0)[rectangle,draw,name=cbeg2] {Client};
 \node at (3.5,0)[rectangle,draw,name=sbeg2] {Server};
 \node at (3,-3.2)[rectangle,name=lab2] {TCP+TLS1.3};
 \node at (3,-2.7)[rectangle,draw,name=time2] {2RTT};
 \coordinate (send2) at(3.5,-3); 
 \coordinate (cend2) at(2.5,-3); 
 \coordinate (c21) at(2.5,-0.4); 
 \coordinate (s21) at(3.5,-0.7);
 \coordinate (c22) at(2.5,-1);   
 \coordinate (s22) at(3.5,-1.3); 
 \coordinate (c23) at(2.5,-1.6);    
 \coordinate (s23) at(3.5,-1.9); 
 \coordinate (c24) at(2.5,-2.2);  
 \coordinate (s24) at(3.5,-2.5); 
 
   \node at (5,0)[rectangle,draw,name=cbeg3] {Client};
 \node at (6,0)[rectangle,draw,name=sbeg3] {Server};
 \node at (5.5,-3.2)[rectangle,name=lab3] {QUIC+TLS1.3};
 \node at (5.5,-2.7)[rectangle,draw,name=time3] {1RTT};
 \coordinate (send3) at(6,-3); 
 \coordinate (cend3) at(5,-3); 
 \coordinate (c31) at(5,-0.4); 
 \coordinate (s31) at(6,-0.7);
 \coordinate (s32) at(6,-1.3); 
 \coordinate (c32) at(5,-1); 
 \coordinate (s33) at(6,-1.3); 
 
   \node at (0,-4.5)[rectangle,draw,name=cbeg4] {Client};
 \node at (1,-4.5)[rectangle,draw,name=sbeg4] {Server};
 \node at (0.5,-7.7)[rectangle,name=lab4] {TCP+TLS1.2};
 \node at (0.5,-7.2)[rectangle,draw,name=time4] {1RTT};
 \coordinate (send4) at(1,-7.5); 
 \coordinate (cend4) at(0,-7.5); 
 \coordinate (c41) at(0,-4.9); 
 \coordinate (s41) at(1,-5.2);
 \coordinate (c42) at(0,-5.5);   
 \coordinate (s42) at(1,-5.8); 
 \coordinate (c43) at(0,-6.1);    
 \coordinate (s43) at(1,-6.4); 
 \coordinate (c44) at(0,-6.7);  
 \coordinate (s44) at(1,-7); 
 
   \node at (2.5,-4.5)[rectangle,draw,name=cbeg5] {Client};
 \node at (3.5,-4.5)[rectangle,draw,name=sbeg5] {Server};
 \node at (3,-7.7)[rectangle,name=lab5] {TCP+TLS1.3};
 \node at (3,-7.2)[rectangle,draw,name=time5] {1RTT};
 \coordinate (send5) at(3.5,-7.5); 
 \coordinate (cend5) at(2.5,-7.5); 
 \coordinate (c51) at(2.5,-4.9); 
 \coordinate (s51) at(3.5,-5.2);
 \coordinate (c52) at(2.5,-5.5);   
 \coordinate (s52) at(3.5,-5.8); 
 \coordinate (c53) at(2.5,-6.1);    
 \coordinate (s53) at(3.5,-6.4); 
 \coordinate (c54) at(2.5,-6.7);  
 \coordinate (s54) at(3.5,-7);

   \node at (5,-4.5)[rectangle,draw,name=cbeg6] {Client};
 \node at (6,-4.5)[rectangle,draw,name=sbeg6] {Server};
 \node at (5.5,-7.7)[rectangle,name=lab6] {QUIC+TLS1.3};
 \node at (5.5,-7.2)[rectangle,draw,name=time6] {0RTT};
 \coordinate (send6) at(6,-7.5); 
 \coordinate (cend6) at(5,-7.5); 
 \coordinate (c61) at(5,-4.9); 
 \coordinate (s61) at(6,-5.2);
 
 \draw[-] (cend1) to (cbeg1.south);
 \draw[-] (send1) to (sbeg1.south);
 \draw[blue,arrows={->}] (c11) -- (s11);
 \draw[blue,arrows={->}] (s11) -- (c12);
 \draw[arrows={->}] (c12) -- (s12);
 \draw[arrows={->}] (s12) -- (c13);
 \draw[arrows={->}] (c13) -- (s13);
 \draw[arrows={->}] (s13) -- (c14);
 \draw[red,arrows={->}] (c14) -- (s14);
 
  \draw[-] (cend2) to (cbeg2.south);
 \draw[-] (send2) to (sbeg2.south);
 \draw[blue,arrows={->}] (c21) -- (s21);
 \draw[blue,arrows={->}] (s21) -- (c22);
 \draw[arrows={->}] (c22) -- (s22);
 \draw[arrows={->}] (s22) -- (c23);
 \draw[red,arrows={->}] (c23) -- (s23);

  \draw[-] (cend3) to (cbeg3.south);
 \draw[-] (send3) to (sbeg3.south);
 \draw[blue,arrows={->}] (c31) -- (s31);
 \draw[blue,arrows={->}] (s31) -- (c32);
  \draw[red,arrows={->}] (c32) -- (s33);
  
    \draw[-] (cend4) to (cbeg4.south);
 \draw[-] (send4) to (sbeg4.south);
 \draw[blue,arrows={->}] (c41) -- (s41);
 \draw[blue,arrows={->}] (s41) -- (c42);
 \draw[arrows={->}] (c42) -- (s42);
 \draw[arrows={->}] (s42) -- (c43);
 \draw[red,arrows={->}] (c43) -- (s43);
 
  \draw[-] (cend5) to (cbeg5.south);
 \draw[-] (send5) to (sbeg5.south);
  \draw[blue,arrows={->}] (c51) -- (s51);
 \draw[blue,arrows={->}] (s51) -- (c52);
 \draw[red,arrows={->}] (c52) -- (s52);  
 
 \draw[-] (cend6) to (cbeg6.south);
 \draw[-] (send6) to (sbeg6.south);
 \draw[red,arrows={->}] (c61) -- (s61);

\end{tikzpicture}
}

%% file: img/cc-table.tex
\begin{table*}[ht]
	\centering 
	\caption{Summary of the main presented congestion control schemes, divided by design philosophy}
	\label{tab:ccsum}
	\scriptsize
	\renewcommand\tabularxcolumn[1]{m{#1}}
	\newcolumntype{C}{>{\centering\arraybackslash}X}

	\begin{tabularx}{\textwidth}{lC|CCC}
		\toprule
		Type & Algorithm &   Congestion control mechanism               &      Pros                &              Cons              \\
		\midrule
		Loss-based & NewReno~\cite{rfc6582}&  \gls{aimd}  & Proven convergence, fairness &          Inefficient in LFNs           \\
		  & BIC~\cite{BIC} &   Binary search increase function  &   Higher efficiency       &       Too aggressive       \\
		 & CUBIC~\cite{CUBIC}   & Cubic CWND function & \gls{rtt}-independent fairness  & High number of retransmissions       \\ 
		 & Wave~\cite{abdelsalam2017tcp} & Burst-based adaptation & Fairness and efficiency& Highly volatile \gls{rtt}\\
		 \midrule
		Delay-based & Vegas~\cite{Vegas,NewVegasPerf}   &  RTT changes as congestion signal  &    Fewer retransmissions, lower latency        & Suppressed by loss-based flows    \\
		 & Verus~\cite{Verus}   & Delay profile-based \gls{aimd}  &  Adaptability to volatile channels   &    High sender-side CPU load     \\ 
		 & Nimbus~\cite{goyal2018elasticity} & Explicit queue and cross traffic modeling & Scalable aggressiveness to deal with CUBIC & Unfair to Vegas and \gls{bbr}\\
		 & LEDBAT~\cite{rfc6817} & Extra delay to high-priority flows & Does not affect other flows & Limited to low-priority traffic\\ \midrule
		Capacity-based & Westwood~\cite{mascolo2001tcp} & Bandwidth estimation to decrease CWND& Good in wireless and lossy links & No latency control\\
		& Sprout~\cite{Sprout}   &  HMM capacity model   & Low delay, customizable &   Needs one buffer per flow    \\ \midrule
		Hybrid & Compound~\cite{4146841} & Sum of Reno and Vegas windows & Fast in \glspl{lfn}, fair to CUBIC & No latency control\\
		& Illinois~\cite{liu2008tcp} & Delay used to determine CWND change & Fast in \glspl{lfn}, fair to CUBIC & No latency control\\
		& Veno~\cite{fu2003tcp} & Explicit model of the buffer & Fast in \glspl{lfn}, fair to CUBIC & No latency control\\
		 &  BBR~\cite{BBR}    & Capacity and \gls{rtt} measurement & High throughput, low delay & Fairness and mobility issues\\  \midrule
		 Learning-based & Remy~\cite{2486020} & Monte Carlo-based policy & Reaches capacity with low delay & Fairness issues with other TCPs\\
		 &TAO~\cite{2626324}&Advancement on Remy&Fairness issues solved&Requires knowledge of the network\\
		 & PCC~\cite{189004} & Online experiments to determine CWND& Good in high \gls{rtt} networks& Untested with bufferbloat\\
		 & TCP-RL~\cite{8668690} & Reinforcement learning to determine CC algorithm & Self-organizing capabilities & Untested in highly dynamic environments \\
		 & QTCP~\cite{8357943} & Reinforcement learning to select CWND& Higher throughput than NewReno & Limited performance evaluation\\
		\bottomrule
	\end{tabularx}
\end{table*}

%% file: img/newreno.tex
\begin{tikzpicture}
\pgfplotsset{every tick label/.append style={font=\scriptsize}}
\tikzstyle{dotted}= [dash pattern=on \pgflinewidth off 0.5mm] 
\tikzstyle{dashed}= [dash pattern=on 7.5*0.8*0.8pt off 7.5*0.4*0.8pt]
\tikzstyle{dashdotted} = [dash pattern=on 7.5*0.8*0.6pt off 7.5*0.8*0.3pt on \the\pgflinewidth off 7.5*0.8*0.3pt]
\tikzstyle{dotted2} = [dash pattern=on 7.5*0.8*0.3pt off 7.5*0.8*0.2pt]

\begin{axis}[%
width=\fwidth,
height=\fheight,
at={(0\fwidth,0\fheight)},
scale only axis,
xmin=0,
xmax=1,
ticks=none,
xlabel near ticks,
xlabel style={font=\footnotesize\color{white!15!black}},
xlabel={Time},
ylabel near ticks,
ymin=0,
ymax=2000,
ylabel style={font=\footnotesize\color{white!15!black}},
ylabel={CWND},
axis background/.style={fill=white},
legend style={font=\scriptsize, at={(0.95,0.6)}, anchor=south east, legend cell align=left, align=left, draw=white!15!black}
]

\addplot [color=green,dash dot, line width=1.2pt]
  table[row sep=crcr]{%
0	1.07177346253629\\
0.00223214285714286	1.14869835499704\\
0.00446428571428571	1.23114441334492\\
0.00669642857142857	1.31950791077289\\
0.00892857142857143	1.41421356237310\\
0.0111607142857143	1.51571656651040\\
0.0133928571428571	1.62450479271247\\
0.0156250000000000	1.74110112659225\\
0.0178571428571429	1.86606598307361\\
0.0200892857142857	2\\
0.0223214285714286	2.14354692507259\\
0.0245535714285714	2.29739670999407\\
0.0267857142857143	2.46228882668983\\
0.0290178571428571	2.63901582154579\\
0.0312500000000000	2.82842712474619\\
0.0334821428571429	3.03143313302080\\
0.0357142857142857	3.24900958542494\\
0.0379464285714286	3.48220225318450\\
0.0401785714285714	3.73213196614723\\
0.0424107142857143	4\\
0.0446428571428571	4.28709385014517\\
0.0468750000000000	4.59479341998814\\
0.0491071428571429	4.92457765337967\\
0.0513392857142857	5.27803164309158\\
0.0535714285714286	5.65685424949238\\
0.0558035714285714	6.06286626604159\\
0.0580357142857143	6.49801917084989\\
0.0602678571428571	6.96440450636900\\
0.0625000000000000	7.46426393229446\\
0.0647321428571429	8.00000000000000\\
0.0669642857142857	8.57418770029035\\
0.0691964285714286	9.18958683997628\\
0.0714285714285714	9.84915530675933\\
0.0736607142857143	10.5560632861832\\
0.0758928571428571	11.3137084989848\\
0.0781250000000000	12.1257325320832\\
0.0803571428571429	12.9960383416998\\
0.0825892857142857	13.9288090127380\\
0.0848214285714286	14.9285278645889\\
0.0870535714285714	16\\
0.0892857142857143	17.1483754005807\\
0.0915178571428571	18.3791736799526\\
0.0937500000000000	19.6983106135187\\
0.0959821428571429	21.1121265723663\\
0.0982142857142857	22.6274169979695\\
0.100446428571429	24.2514650641664\\
0.102678571428571	25.9920766833995\\
0.104910714285714	27.8576180254760\\
0.107142857142857	29.8570557291778\\
0.109375000000000	32\\
0.111607142857143	34.2967508011614\\
0.113839285714286	36.7583473599051\\
0.116071428571429	39.3966212270373\\
0.118303571428571	42.2242531447326\\
0.120535714285714	45.2548339959390\\
0.122767857142857	48.5029301283327\\
0.125000000000000	51.9841533667991\\
0.127232142857143	55.7152360509519\\
0.129464285714286	59.7141114583557\\
0.131696428571429	64\\
0.133928571428571	68.5935016023228\\
0.136160714285714	73.5166947198102\\
0.138392857142857	78.7932424540746\\
0.140625000000000	84.4485062894653\\
0.142857142857143	90.5096679918781\\
0.145089285714286	97.0058602566655\\
0.147321428571429	103.968306733598\\
0.149553571428571	111.430472101904\\
0.151785714285714	119.428222916711\\
0.154017857142857	128\\
0.156250000000000	137.187003204646\\
0.158482142857143	147.033389439620\\
0.160714285714286	157.586484908149\\
0.162946428571429	168.897012578931\\
0.165178571428571	181.019335983756\\
0.167410714285714	194.011720513331\\
0.169642857142857	207.936613467196\\
0.171875000000000	222.860944203808\\
0.174107142857143	238.856445833423\\
0.176339285714286	256\\
0.178571428571429	274.374006409291\\
0.180803571428571	294.066778879241\\
0.183035714285714	315.172969816299\\
0.185267857142857	337.794025157861\\
0.187500000000000	362.038671967512\\
0.189732142857143	388.023441026662\\
0.191964285714286	415.873226934392\\
0.194196428571429	445.721888407616\\
0.196428571428571	477.712891666846\\
0.198660714285714	512\\
0.200892857142857	548.748012818582\\
0.203125000000000	588.133557758482\\
0.205357142857143	630.345939632598\\
0.207589285714286	675.588050315722\\
0.209821428571429	724.077343935025\\
0.212053571428571	776.046882053324\\
0.214285714285714	831.746453868785\\
0.216517857142857	891.443776815232\\
0.218750000000000	955.425783333691\\
0.220982142857143	1024\\};\addlegendentry{Slow start}

\addplot [color=cyan, line width=1.2pt]
  table[row sep=crcr]{%
0.220982142857143	1024\\
0.223214285714286	1026\\
0.225446428571429	1028\\
0.227678571428571	1030\\
0.229910714285714	1032\\
0.232142857142857	1034\\
0.234375000000000	1036\\
0.236607142857143	1038\\
0.238839285714286	1040\\
0.241071428571429	1042\\
0.243303571428571	1044\\
0.245535714285714	1046\\
0.247767857142857	1048\\
0.250000000000000	1050\\
0.252232142857143	1052\\
0.254464285714286	1054\\
0.256696428571429	1056\\
0.258928571428571	1058\\
0.261160714285714	1060\\
0.263392857142857	1062\\
0.265625000000000	1064\\
0.267857142857143	1066\\
0.270089285714286	1068\\
0.272321428571429	1070\\
0.274553571428571	1072\\
0.276785714285714	1074\\
0.279017857142857	1076\\
0.281250000000000	1078\\
0.283482142857143	1080\\
0.285714285714286	1082\\
0.287946428571429	1084\\
0.290178571428571	1086\\
0.292410714285714	1088\\
0.294642857142857	1090\\
0.296875000000000	1092\\
0.299107142857143	1094\\
0.301339285714286	1096\\
0.303571428571429	1098\\
0.305803571428571	1100\\
0.308035714285714	1102\\
0.310267857142857	1104\\
0.312500000000000	1106\\
0.314732142857143	1108\\
0.316964285714286	1110\\
0.319196428571429	1112\\
0.321428571428571	1114\\
0.323660714285714	1116\\
0.325892857142857	1118\\
0.328125000000000	1120\\
0.330357142857143	1122\\
0.332589285714286	1124\\
0.334821428571429	1126\\
0.337053571428571	1128\\
0.339285714285714	1130\\
0.341517857142857	1132\\
0.343750000000000	1134\\
0.345982142857143	1136\\
0.348214285714286	1138\\
0.350446428571429	1140\\
0.352678571428571	1142\\
0.354910714285714	1144\\
0.357142857142857	1146\\
0.359375000000000	1148\\
0.361607142857143	1150\\
0.363839285714286	1152\\
0.366071428571429	1154\\
0.368303571428571	1156\\
0.370535714285714	1158\\
0.372767857142857	1160\\
0.375000000000000	1162\\
0.377232142857143	1164\\
0.379464285714286	1166\\
0.381696428571429	1168\\
0.383928571428571	1170\\
0.386160714285714	1172\\
0.388392857142857	1174\\
0.390625000000000	1176\\
0.392857142857143	1178\\
0.395089285714286	1180\\
0.397321428571429	1182\\
0.399553571428571	1184\\
0.401785714285714	1186\\
0.404017857142857	1188\\
0.406250000000000	1190\\
0.408482142857143	1192\\
0.410714285714286	1194\\
0.412946428571429	1196\\
0.415178571428571	1198\\
0.417410714285714	1200\\
0.419642857142857	1202\\
0.421875000000000	1204\\
0.424107142857143	1206\\
0.426339285714286	1208\\
0.428571428571429	1210\\
0.430803571428571	1212\\
0.433035714285714	1214\\
0.435267857142857	1216\\
0.437500000000000	1218\\
0.439732142857143	1220\\
0.441964285714286	1222\\
0.444196428571429	1224\\};\addlegendentry{Congestion avoidance}
\addplot [color=orange,dash dot dot, line width=1.2pt]
  table[row sep=crcr]{%
0.444196428571429	1224\\
0.446428571428571	612\\};\addlegendentry{Fast recovery}
\addplot [color=red,dashed, line width=1.2pt]
  table[row sep=crcr]{%
  0.669642857142857	812\\
0.671875000000000	1\\};\addlegendentry{Reset}

\addplot [color=black,dotted, line width=0.4pt]
  table[row sep=crcr]{%
  0	1024\\
  0.446428571428571	1024\\
  0.446428571428571	612\\
  0.671875000000000	612\\
  0.671875000000000	406\\
  1			406\\
};\addlegendentry{Slow start threshold}

\addplot [color=cyan, line width=1.2pt]
  table[row sep=crcr]{%
0.446428571428571	612\\
0.448660714285714	614\\
0.450892857142857	616\\
0.453125000000000	618\\
0.455357142857143	620\\
0.457589285714286	622\\
0.459821428571429	624\\
0.462053571428571	626\\
0.464285714285714	628\\
0.466517857142857	630\\
0.468750000000000	632\\
0.470982142857143	634\\
0.473214285714286	636\\
0.475446428571429	638\\
0.477678571428571	640\\
0.479910714285714	642\\
0.482142857142857	644\\
0.484375000000000	646\\
0.486607142857143	648\\
0.488839285714286	650\\
0.491071428571429	652\\
0.493303571428571	654\\
0.495535714285714	656\\
0.497767857142857	658\\
0.500000000000000	660\\
0.502232142857143	662\\
0.504464285714286	664\\
0.506696428571429	666\\
0.508928571428571	668\\
0.511160714285714	670\\
0.513392857142857	672\\
0.515625000000000	674\\
0.517857142857143	676\\
0.520089285714286	678\\
0.522321428571429	680\\
0.524553571428571	682\\
0.526785714285714	684\\
0.529017857142857	686\\
0.531250000000000	688\\
0.533482142857143	690\\
0.535714285714286	692\\
0.537946428571429	694\\
0.540178571428571	696\\
0.542410714285714	698\\
0.544642857142857	700\\
0.546875000000000	702\\
0.549107142857143	704\\
0.551339285714286	706\\
0.553571428571429	708\\
0.555803571428571	710\\
0.558035714285714	712\\
0.560267857142857	714\\
0.562500000000000	716\\
0.564732142857143	718\\
0.566964285714286	720\\
0.569196428571429	722\\
0.571428571428571	724\\
0.573660714285714	726\\
0.575892857142857	728\\
0.578125000000000	730\\
0.580357142857143	732\\
0.582589285714286	734\\
0.584821428571429	736\\
0.587053571428571	738\\
0.589285714285714	740\\
0.591517857142857	742\\
0.593750000000000	744\\
0.595982142857143	746\\
0.598214285714286	748\\
0.600446428571429	750\\
0.602678571428571	752\\
0.604910714285714	754\\
0.607142857142857	756\\
0.609375000000000	758\\
0.611607142857143	760\\
0.613839285714286	762\\
0.616071428571429	764\\
0.618303571428571	766\\
0.620535714285714	768\\
0.622767857142857	770\\
0.625000000000000	772\\
0.627232142857143	774\\
0.629464285714286	776\\
0.631696428571429	778\\
0.633928571428571	780\\
0.636160714285714	782\\
0.638392857142857	784\\
0.640625000000000	786\\
0.642857142857143	788\\
0.645089285714286	790\\
0.647321428571429	792\\
0.649553571428571	794\\
0.651785714285714	796\\
0.654017857142857	798\\
0.656250000000000	800\\
0.658482142857143	802\\
0.660714285714286	804\\
0.662946428571429	806\\
0.665178571428571	808\\
0.667410714285714	810\\
0.669642857142857	812\\};
\addplot [color=green,dash dot, line width=1.2pt]
  table[row sep=crcr]{%
  0.671875000000000	1\\
0.674107142857143	1.07177346253629\\
0.676339285714286	1.14869835499704\\
0.678571428571429	1.23114441334492\\
0.680803571428571	1.31950791077289\\
0.683035714285714	1.41421356237310\\
0.685267857142857	1.51571656651040\\
0.687500000000000	1.62450479271247\\
0.689732142857143	1.74110112659225\\
0.691964285714286	1.86606598307361\\
0.694196428571429	2\\
0.696428571428571	2.14354692507259\\
0.698660714285714	2.29739670999407\\
0.700892857142857	2.46228882668983\\
0.703125000000000	2.63901582154579\\
0.705357142857143	2.82842712474619\\
0.707589285714286	3.03143313302080\\
0.709821428571429	3.24900958542494\\
0.712053571428571	3.48220225318450\\
0.714285714285714	3.73213196614723\\
0.716517857142857	4\\
0.718750000000000	4.28709385014517\\
0.720982142857143	4.59479341998814\\
0.723214285714286	4.92457765337967\\
0.725446428571429	5.27803164309158\\
0.727678571428571	5.65685424949238\\
0.729910714285714	6.06286626604159\\
0.732142857142857	6.49801917084989\\
0.734375000000000	6.96440450636900\\
0.736607142857143	7.46426393229446\\
0.738839285714286	8.00000000000000\\
0.741071428571429	8.57418770029035\\
0.743303571428571	9.18958683997628\\
0.745535714285714	9.84915530675933\\
0.747767857142857	10.5560632861832\\
0.750000000000000	11.3137084989848\\
0.752232142857143	12.1257325320832\\
0.754464285714286	12.9960383416998\\
0.756696428571429	13.9288090127380\\
0.758928571428571	14.9285278645889\\
0.761160714285714	16\\
0.763392857142857	17.1483754005807\\
0.765625000000000	18.3791736799526\\
0.767857142857143	19.6983106135187\\
0.770089285714286	21.1121265723663\\
0.772321428571429	22.6274169979695\\
0.774553571428571	24.2514650641664\\
0.776785714285714	25.9920766833995\\
0.779017857142857	27.8576180254760\\
0.781250000000000	29.8570557291778\\
0.783482142857143	32\\
0.785714285714286	34.2967508011614\\
0.787946428571429	36.7583473599051\\
0.790178571428571	39.3966212270373\\
0.792410714285714	42.2242531447326\\
0.794642857142857	45.2548339959390\\
0.796875000000000	48.5029301283327\\
0.799107142857143	51.9841533667991\\
0.801339285714286	55.7152360509519\\
0.803571428571429	59.7141114583557\\
0.805803571428571	64\\
0.808035714285714	68.5935016023228\\
0.810267857142857	73.5166947198103\\
0.812500000000000	78.7932424540746\\
0.814732142857143	84.4485062894653\\
0.816964285714286	90.5096679918781\\
0.819196428571429	97.0058602566655\\
0.821428571428571	103.968306733598\\
0.823660714285714	111.430472101904\\
0.825892857142857	119.428222916711\\
0.828125000000000	128\\
0.830357142857143	137.187003204646\\
0.832589285714286	147.033389439620\\
0.834821428571429	157.586484908149\\
0.837053571428571	168.897012578930\\
0.839285714285714	181.019335983756\\
0.841517857142857	194.011720513331\\
0.843750000000000	207.936613467196\\
0.845982142857143	222.860944203808\\
0.848214285714286	238.856445833423\\
0.850446428571429	256\\
0.852678571428571	274.374006409291\\
0.854910714285714	294.066778879241\\
0.857142857142857	315.172969816298\\
0.859375000000000	337.794025157861\\
0.861607142857143	362.038671967512\\
0.863839285714286	388.023441026662\\
0.866071428571429	415.873226934392\\
};

\addplot [color=cyan, line width=1.2pt]
  table[row sep=crcr]{%
0.868303571428571	417.873226934392\\
0.870535714285714	419.873226934392\\
0.872767857142857	421.873226934392\\
0.875000000000000	423.873226934392\\
0.877232142857143	425.873226934392\\
0.879464285714286	427.873226934392\\
0.881696428571429	429.873226934392\\
0.883928571428571	431.873226934392\\
0.886160714285714	433.873226934392\\
0.888392857142857	435.873226934392\\
0.890625000000000	437.873226934392\\
0.892857142857143	439.873226934392\\
0.895089285714286	441.873226934392\\
0.897321428571429	443.873226934392\\
0.899553571428571	445.873226934392\\
0.901785714285714	447.873226934392\\
0.904017857142857	449.873226934392\\
0.906250000000000	451.873226934392\\
0.908482142857143	453.873226934392\\
0.910714285714286	455.873226934392\\
0.912946428571429	457.873226934392\\
0.915178571428571	459.873226934392\\
0.917410714285714	461.873226934392\\
0.919642857142857	463.873226934392\\
0.921875000000000	465.873226934392\\
0.924107142857143	467.873226934392\\
0.926339285714286	469.873226934392\\
0.928571428571429	471.873226934392\\
0.930803571428571	473.873226934392\\
0.933035714285714	475.873226934392\\
0.935267857142857	477.873226934392\\
0.937500000000000	479.873226934392\\
0.939732142857143	481.873226934392\\
0.941964285714286	483.873226934392\\
0.944196428571429	485.873226934392\\
0.946428571428571	487.873226934392\\
0.948660714285714	489.873226934392\\
0.950892857142857	491.873226934392\\
0.953125000000000	493.873226934392\\
0.955357142857143	495.873226934392\\
0.957589285714286	497.873226934392\\
0.959821428571429	499.873226934392\\
0.962053571428571	501.873226934392\\
0.964285714285714	503.873226934392\\
0.966517857142857	505.873226934392\\
0.968750000000000	507.873226934392\\
0.970982142857143	509.873226934392\\
0.973214285714286	511.873226934392\\
0.975446428571429	513.873226934392\\
0.977678571428571	515.873226934392\\
0.979910714285714	517.873226934392\\
0.982142857142857	519.873226934392\\
0.984375000000000	521.873226934392\\
0.986607142857143	523.873226934392\\
0.988839285714286	525.873226934392\\
0.991071428571429	527.873226934392\\
0.993303571428571	529.873226934392\\
0.995535714285714	531.873226934392\\
0.997767857142857	533.873226934392\\
1	535.873226934392\\
};

%
%

\end{axis}
\end{tikzpicture}

%% file: img/cubic.tex
\begin{tikzpicture}
\pgfplotsset{every tick label/.append style={font=\scriptsize}}
\tikzstyle{dotted}= [dash pattern=on \pgflinewidth off 0.5mm] 
\tikzstyle{dashed}= [dash pattern=on 7.5*0.8*0.8pt off 7.5*0.4*0.8pt]
\tikzstyle{dashdotted} = [dash pattern=on 7.5*0.8*0.6pt off 7.5*0.8*0.3pt on \the\pgflinewidth off 7.5*0.8*0.3pt]
\tikzstyle{dotted2} = [dash pattern=on 7.5*0.8*0.3pt off 7.5*0.8*0.2pt]

\begin{axis}[%
width=\fwidth,
height=\fheight,
at={(0\fwidth,0\fheight)},
scale only axis,
xmin=-0.01,
xmax=1,
ticks=none,
xlabel near ticks,
xlabel style={font=\footnotesize\color{white!15!black}},
xlabel={Time since last packet loss},
ylabel near ticks,
ymin=-1.5,
ymax=1.5,
ylabel style={font=\footnotesize\color{white!15!black}},
ylabel={CWND},
axis background/.style={fill=white},
legend style={font=\scriptsize, at={(0.95,0.05)}, anchor=south east, legend cell align=left, align=left, draw=white!15!black}
]

\addplot [color=cyan, line width=0.8pt]
  table[row sep=crcr]{%
  -0.01 0\\
0.000100000000000000	-1.00481636742107\\
0.00110000000000000	-0.998798317773318\\
0.00210000000000000	-0.992804345107380\\
0.00310000000000000	-0.986834401163119\\
0.00410000000000000	-0.980888437680401\\
0.00510000000000000	-0.974966406399089\\
0.00610000000000000	-0.969068259059046\\
0.00710000000000000	-0.963193947400138\\
0.00810000000000000	-0.957343423162229\\
0.00910000000000000	-0.951516638085182\\
0.0101000000000000	-0.945713543908862\\
0.0111000000000000	-0.939934092373132\\
0.0121000000000000	-0.934178235217857\\
0.0131000000000000	-0.928445924182901\\
0.0141000000000000	-0.922737111008127\\
0.0151000000000000	-0.917051747433401\\
0.0161000000000000	-0.911389785198585\\
0.0171000000000000	-0.905751176043544\\
0.0181000000000000	-0.900135871708143\\
0.0191000000000000	-0.894543823932245\\
0.0201000000000000	-0.888974984455714\\
0.0211000000000000	-0.883429305018414\\
0.0221000000000000	-0.877906737360210\\
0.0231000000000000	-0.872407233220965\\
0.0241000000000000	-0.866930744340544\\
0.0251000000000000	-0.861477222458810\\
0.0261000000000000	-0.856046619315628\\
0.0271000000000000	-0.850638886650862\\
0.0281000000000000	-0.845253976204376\\
0.0291000000000000	-0.839891839716033\\
0.0301000000000000	-0.834552428925699\\
0.0311000000000000	-0.829235695573237\\
0.0321000000000000	-0.823941591398510\\
0.0331000000000000	-0.818670068141384\\
0.0341000000000000	-0.813421077541723\\
0.0351000000000000	-0.808194571339390\\
0.0361000000000000	-0.802990501274249\\
0.0371000000000000	-0.797808819086164\\
0.0381000000000000	-0.792649476515001\\
0.0391000000000000	-0.787512425300622\\
0.0401000000000000	-0.782397617182891\\
0.0411000000000000	-0.777305003901673\\
0.0421000000000000	-0.772234537196833\\
0.0431000000000000	-0.767186168808233\\
0.0441000000000000	-0.762159850475738\\
0.0451000000000000	-0.757155533939213\\
0.0461000000000000	-0.752173170938520\\
0.0471000000000000	-0.747212713213525\\
0.0481000000000000	-0.742274112504091\\
0.0491000000000000	-0.737357320550082\\
0.0501000000000000	-0.732462289091363\\
0.0511000000000000	-0.727588969867797\\
0.0521000000000000	-0.722737314619250\\
0.0531000000000000	-0.717907275085583\\
0.0541000000000000	-0.713098803006663\\
0.0551000000000000	-0.708311850122352\\
0.0561000000000000	-0.703546368172515\\
0.0571000000000000	-0.698802308897016\\
0.0581000000000000	-0.694079624035719\\
0.0591000000000000	-0.689378265328489\\
0.0601000000000000	-0.684698184515188\\
0.0611000000000000	-0.680039333335682\\
0.0621000000000000	-0.675401663529834\\
0.0631000000000000	-0.670785126837508\\
0.0641000000000000	-0.666189674998569\\
0.0651000000000000	-0.661615259752881\\
0.0661000000000000	-0.657061832840307\\
0.0671000000000000	-0.652529346000711\\
0.0681000000000000	-0.648017750973959\\
0.0691000000000000	-0.643526999499914\\
0.0701000000000000	-0.639057043318439\\
0.0711000000000000	-0.634607834169399\\
0.0721000000000000	-0.630179323792659\\
0.0731000000000000	-0.625771463928081\\
0.0741000000000000	-0.621384206315531\\
0.0751000000000000	-0.617017502694872\\
0.0761000000000000	-0.612671304805968\\
0.0771000000000000	-0.608345564388684\\
0.0781000000000000	-0.604040233182883\\
0.0791000000000000	-0.599755262928430\\
0.0801000000000000	-0.595490605365189\\
0.0811000000000000	-0.591246212233023\\
0.0821000000000000	-0.587022035271797\\
0.0831000000000000	-0.582818026221375\\
0.0841000000000000	-0.578634136821621\\
0.0851000000000000	-0.574470318812399\\
0.0861000000000000	-0.570326523933573\\
0.0871000000000000	-0.566202703925007\\
0.0881000000000000	-0.562098810526565\\
0.0891000000000000	-0.558014795478112\\
0.0901000000000000	-0.553950610519511\\
0.0911000000000000	-0.549906207390626\\
0.0921000000000000	-0.545881537831322\\
0.0931000000000000	-0.541876553581463\\
0.0941000000000000	-0.537891206380912\\
0.0951000000000000	-0.533925447969534\\
0.0961000000000000	-0.529979230087193\\
0.0971000000000000	-0.526052504473753\\
0.0981000000000000	-0.522145222869077\\
0.0991000000000000	-0.518257337013031\\
0.100100000000000	-0.514388798645478\\
0.101100000000000	-0.510539559506282\\
0.102100000000000	-0.506709571335307\\
0.103100000000000	-0.502898785872418\\
0.104100000000000	-0.499107154857478\\
0.105100000000000	-0.495334630030351\\
0.106100000000000	-0.491581163130902\\
0.107100000000000	-0.487846705898995\\
0.108100000000000	-0.484131210074493\\
0.109100000000000	-0.480434627397262\\
0.110100000000000	-0.476756909607163\\
0.111100000000000	-0.473098008444063\\
0.112100000000000	-0.469457875647825\\
0.113100000000000	-0.465836462958313\\
0.114100000000000	-0.462233722115391\\
0.115100000000000	-0.458649604858923\\
0.116100000000000	-0.455084062928774\\
0.117100000000000	-0.451537048064807\\
0.118100000000000	-0.448008512006886\\
0.119100000000000	-0.444498406494876\\
0.120100000000000	-0.441006683268640\\
0.121100000000000	-0.437533294068043\\
0.122100000000000	-0.434078190632949\\
0.123100000000000	-0.430641324703221\\
0.124100000000000	-0.427222648018725\\
0.125100000000000	-0.423822112319323\\
0.126100000000000	-0.420439669344880\\
0.127100000000000	-0.417075270835261\\
0.128100000000000	-0.413728868530328\\
0.129100000000000	-0.410400414169947\\
0.130100000000000	-0.407089859493981\\
0.131100000000000	-0.403797156242295\\
0.132100000000000	-0.400522256154751\\
0.133100000000000	-0.397265110971216\\
0.134100000000000	-0.394025672431552\\
0.135100000000000	-0.390803892275624\\
0.136100000000000	-0.387599722243295\\
0.137100000000000	-0.384413114074430\\
0.138100000000000	-0.381244019508893\\
0.139100000000000	-0.378092390286548\\
0.140100000000000	-0.374958178147260\\
0.141100000000000	-0.371841334830891\\
0.142100000000000	-0.368741812077306\\
0.143100000000000	-0.365659561626370\\
0.144100000000000	-0.362594535217946\\
0.145100000000000	-0.359546684591898\\
0.146100000000000	-0.356515961488091\\
0.147100000000000	-0.353502317646388\\
0.148100000000000	-0.350505704806654\\
0.149100000000000	-0.347526074708753\\
0.150100000000000	-0.344563379092548\\
0.151100000000000	-0.341617569697905\\
0.152100000000000	-0.338688598264686\\
0.153100000000000	-0.335776416532756\\
0.154100000000000	-0.332880976241979\\
0.155100000000000	-0.330002229132220\\
0.156100000000000	-0.327140126943341\\
0.157100000000000	-0.324294621415208\\
0.158100000000000	-0.321465664287684\\
0.159100000000000	-0.318653207300634\\
0.160100000000000	-0.315857202193921\\
0.161100000000000	-0.313077600707410\\
0.162100000000000	-0.310314354580964\\
0.163100000000000	-0.307567415554448\\
0.164100000000000	-0.304836735367726\\
0.165100000000000	-0.302122265760661\\
0.166100000000000	-0.299423958473119\\
0.167100000000000	-0.296741765244962\\
0.168100000000000	-0.294075637816056\\
0.169100000000000	-0.291425527926264\\
0.170100000000000	-0.288791387315450\\
0.171100000000000	-0.286173167723478\\
0.172100000000000	-0.283570820890213\\
0.173100000000000	-0.280984298555518\\
0.174100000000000	-0.278413552459258\\
0.175100000000000	-0.275858534341296\\
0.176100000000000	-0.273319195941497\\
0.177100000000000	-0.270795488999725\\
0.178100000000000	-0.268287365255843\\
0.179100000000000	-0.265794776449716\\
0.180100000000000	-0.263317674321209\\
0.181100000000000	-0.260856010610184\\
0.182100000000000	-0.258409737056506\\
0.183100000000000	-0.255978805400040\\
0.184100000000000	-0.253563167380649\\
0.185100000000000	-0.251162774738197\\
0.186100000000000	-0.248777579212548\\
0.187100000000000	-0.246407532543568\\
0.188100000000000	-0.244052586471118\\
0.189100000000000	-0.241712692735064\\
0.190100000000000	-0.239387803075270\\
0.191100000000000	-0.237077869231600\\
0.192100000000000	-0.234782842943917\\
0.193100000000000	-0.232502675952086\\
0.194100000000000	-0.230237319995972\\
0.195100000000000	-0.227986726815437\\
0.196100000000000	-0.225750848150347\\
0.197100000000000	-0.223529635740564\\
0.198100000000000	-0.221323041325954\\
0.199100000000000	-0.219131016646381\\
0.200100000000000	-0.216953513441707\\
0.201100000000000	-0.214790483451798\\
0.202100000000000	-0.212641878416518\\
0.203100000000000	-0.210507650075731\\
0.204100000000000	-0.208387750169300\\
0.205100000000000	-0.206282130437090\\
0.206100000000000	-0.204190742618964\\
0.207100000000000	-0.202113538454788\\
0.208100000000000	-0.200050469684425\\
0.209100000000000	-0.198001488047738\\
0.210100000000000	-0.195966545284593\\
0.211100000000000	-0.193945593134853\\
0.212100000000000	-0.191938583338383\\
0.213100000000000	-0.189945467635046\\
0.214100000000000	-0.187966197764706\\
0.215100000000000	-0.186000725467228\\
0.216100000000000	-0.184049002482475\\
0.217100000000000	-0.182110980550312\\
0.218100000000000	-0.180186611410603\\
0.219100000000000	-0.178275846803211\\
0.220100000000000	-0.176378638468001\\
0.221100000000000	-0.174494938144838\\
0.222100000000000	-0.172624697573584\\
0.223100000000000	-0.170767868494104\\
0.224100000000000	-0.168924402646263\\
0.225100000000000	-0.167094251769924\\
0.226100000000000	-0.165277367604951\\
0.227100000000000	-0.163473701891209\\
0.228100000000000	-0.161683206368561\\
0.229100000000000	-0.159905832776871\\
0.230100000000000	-0.158141532856004\\
0.231100000000000	-0.156390258345824\\
0.232100000000000	-0.154651960986195\\
0.233100000000000	-0.152926592516980\\
0.234100000000000	-0.151214104678044\\
0.235100000000000	-0.149514449209252\\
0.236100000000000	-0.147827577850466\\
0.237100000000000	-0.146153442341552\\
0.238100000000000	-0.144491994422372\\
0.239100000000000	-0.142843185832792\\
0.240100000000000	-0.141206968312675\\
0.241100000000000	-0.139583293601886\\
0.242100000000000	-0.137972113440288\\
0.243100000000000	-0.136373379567746\\
0.244100000000000	-0.134787043724123\\
0.245100000000000	-0.133213057649284\\
0.246100000000000	-0.131651373083093\\
0.247100000000000	-0.130101941765414\\
0.248100000000000	-0.128564715436110\\
0.249100000000000	-0.127039645835047\\
0.250100000000000	-0.125526684702088\\
0.251100000000000	-0.124025783777096\\
0.252100000000000	-0.122536894799937\\
0.253100000000000	-0.121059969510475\\
0.254100000000000	-0.119594959648572\\
0.255100000000000	-0.118141816954094\\
0.256100000000000	-0.116700493166905\\
0.257100000000000	-0.115270940026868\\
0.258100000000000	-0.113853109273848\\
0.259100000000000	-0.112446952647708\\
0.260100000000000	-0.111052421888314\\
0.261100000000000	-0.109669468735528\\
0.262100000000000	-0.108298044929215\\
0.263100000000000	-0.106938102209239\\
0.264100000000000	-0.105589592315465\\
0.265100000000000	-0.104252466987755\\
0.266100000000000	-0.102926677965975\\
0.267100000000000	-0.101612176989988\\
0.268100000000000	-0.100308915799658\\
0.269100000000000	-0.0990168461348499\\
0.270100000000000	-0.0977359197354273\\
0.271100000000000	-0.0964660883412542\\
0.272100000000000	-0.0952073036921949\\
0.273100000000000	-0.0939595175281133\\
0.274100000000000	-0.0927226815888734\\
0.275100000000000	-0.0914967476143396\\
0.276100000000000	-0.0902816673443756\\
0.277100000000000	-0.0890773925188457\\
0.278100000000000	-0.0878838748776140\\
0.279100000000000	-0.0867010661605444\\
0.280100000000000	-0.0855289181075011\\
0.281100000000000	-0.0843673824583481\\
0.282100000000000	-0.0832164109529495\\
0.283100000000000	-0.0820759553311694\\
0.284100000000000	-0.0809459673328718\\
0.285100000000000	-0.0798263986979209\\
0.286100000000000	-0.0787172011661807\\
0.287100000000000	-0.0776183264775152\\
0.288100000000000	-0.0765297263717886\\
0.289100000000000	-0.0754513525888649\\
0.290100000000000	-0.0743831568686082\\
0.291100000000000	-0.0733250909508826\\
0.292100000000000	-0.0722771065755521\\
0.293100000000000	-0.0712391554824808\\
0.294100000000000	-0.0702111894115328\\
0.295100000000000	-0.0691931601025722\\
0.296100000000000	-0.0681850192954629\\
0.297100000000000	-0.0671867187300692\\
0.298100000000000	-0.0661982101462550\\
0.299100000000000	-0.0652194452838845\\
0.300100000000000	-0.0642503758828218\\
0.301100000000000	-0.0632909536829308\\
0.302100000000000	-0.0623411304240757\\
0.303100000000000	-0.0614008578461205\\
0.304100000000000	-0.0604700876889294\\
0.305100000000000	-0.0595487716923663\\
0.306100000000000	-0.0586368615962954\\
0.307100000000000	-0.0577343091405807\\
0.308100000000000	-0.0568410660650864\\
0.309100000000000	-0.0559570841096764\\
0.310100000000000	-0.0550823150142148\\
0.311100000000000	-0.0542167105185658\\
0.312100000000000	-0.0533602223625934\\
0.313100000000000	-0.0525128022861617\\
0.314100000000000	-0.0516744020291347\\
0.315100000000000	-0.0508449733313765\\
0.316100000000000	-0.0500244679327512\\
0.317100000000000	-0.0492128375731229\\
0.318100000000000	-0.0484100339923556\\
0.319100000000000	-0.0476160089303134\\
0.320100000000000	-0.0468307141268605\\
0.321100000000000	-0.0460541013218607\\
0.322100000000000	-0.0452861222551783\\
0.323100000000000	-0.0445267286666773\\
0.324100000000000	-0.0437758722962218\\
0.325100000000000	-0.0430335048836758\\
0.326100000000000	-0.0422995781689034\\
0.327100000000000	-0.0415740438917687\\
0.328100000000000	-0.0408568537921358\\
0.329100000000000	-0.0401479596098688\\
0.330100000000000	-0.0394473130848316\\
0.331100000000000	-0.0387548659568885\\
0.332100000000000	-0.0380705699659033\\
0.333100000000000	-0.0373943768517404\\
0.334100000000000	-0.0367262383542636\\
0.335100000000000	-0.0360661062133372\\
0.336100000000000	-0.0354139321688250\\
0.337100000000000	-0.0347696679605913\\
0.338100000000000	-0.0341332653285001\\
0.339100000000000	-0.0335046760124155\\
0.340100000000000	-0.0328838517522015\\
0.341100000000000	-0.0322707442877222\\
0.342100000000000	-0.0316653053588417\\
0.343100000000000	-0.0310674867054241\\
0.344100000000000	-0.0304772400673334\\
0.345100000000000	-0.0298945171844337\\
0.346100000000000	-0.0293192697965891\\
0.347100000000000	-0.0287514496436637\\
0.348100000000000	-0.0281910084655215\\
0.349100000000000	-0.0276378980020266\\
0.350100000000000	-0.0270920699930430\\
0.351100000000000	-0.0265534761784350\\
0.352100000000000	-0.0260220682980664\\
0.353100000000000	-0.0254977980918014\\
0.354100000000000	-0.0249806172995041\\
0.355100000000000	-0.0244704776610386\\
0.356100000000000	-0.0239673309162688\\
0.357100000000000	-0.0234711288050589\\
0.358100000000000	-0.0229818230672730\\
0.359100000000000	-0.0224993654427752\\
0.360100000000000	-0.0220237076714294\\
0.361100000000000	-0.0215548014930998\\
0.362100000000000	-0.0210925986476505\\
0.363100000000000	-0.0206370508749455\\
0.364100000000000	-0.0201881099148489\\
0.365100000000000	-0.0197457275072247\\
0.366100000000000	-0.0193098553919371\\
0.367100000000000	-0.0188804453088501\\
0.368100000000000	-0.0184574489978278\\
0.369100000000000	-0.0180408181987343\\
0.370100000000000	-0.0176305046514336\\
0.371100000000000	-0.0172264600957898\\
0.372100000000000	-0.0168286362716670\\
0.373100000000000	-0.0164369849189293\\
0.374100000000000	-0.0160514577774407\\
0.375100000000000	-0.0156720065870653\\
0.376100000000000	-0.0152985830876671\\
0.377100000000000	-0.0149311390191103\\
0.378100000000000	-0.0145696261212589\\
0.379100000000000	-0.0142139961339771\\
0.380100000000000	-0.0138642007971288\\
0.381100000000000	-0.0135201918505781\\
0.382100000000000	-0.0131819210341891\\
0.383100000000000	-0.0128493400878259\\
0.384100000000000	-0.0125224007513526\\
0.385100000000000	-0.0122010547646333\\
0.386100000000000	-0.0118852538675319\\
0.387100000000000	-0.0115749497999126\\
0.388100000000000	-0.0112700943016394\\
0.389100000000000	-0.0109706391125765\\
0.390100000000000	-0.0106765359725879\\
0.391100000000000	-0.0103877366215376\\
0.392100000000000	-0.0101041927992898\\
0.393100000000000	-0.00982585624570850\\
0.394100000000000	-0.00955267870065779\\
0.395100000000000	-0.00928461190400176\\
0.396100000000000	-0.00902160759560445\\
0.397100000000000	-0.00876361751532994\\
0.398100000000000	-0.00851059340304232\\
0.399100000000000	-0.00826248699860565\\
0.400100000000000	-0.00801925004188403\\
0.401100000000000	-0.00778083427274150\\
0.402100000000000	-0.00754719143104213\\
0.403100000000000	-0.00731827325665002\\
0.404100000000000	-0.00709403148942922\\
0.405100000000000	-0.00687441786924383\\
0.406100000000000	-0.00665938413595790\\
0.407100000000000	-0.00644888202943550\\
0.408100000000000	-0.00624286328954072\\
0.409100000000000	-0.00604127965613762\\
0.410100000000000	-0.00584408286909029\\
0.411100000000000	-0.00565122466826278\\
0.412100000000000	-0.00546265679351918\\
0.413100000000000	-0.00527833098472355\\
0.414100000000000	-0.00509819898173998\\
0.415100000000000	-0.00492221252443254\\
0.416100000000000	-0.00475032335266529\\
0.417100000000000	-0.00458248320630231\\
0.418100000000000	-0.00441864382520766\\
0.419100000000000	-0.00425875694924544\\
0.420100000000000	-0.00410277431827972\\
0.421100000000000	-0.00395064767217454\\
0.422100000000000	-0.00380232875079401\\
0.423100000000000	-0.00365776929400218\\
0.424100000000000	-0.00351692104166313\\
0.425100000000000	-0.00337973573364095\\
0.426100000000000	-0.00324616510979968\\
0.427100000000000	-0.00311616091000342\\
0.428100000000000	-0.00298967487411623\\
0.429100000000000	-0.00286665874200219\\
0.430100000000000	-0.00274706425352537\\
0.431100000000000	-0.00263084314854984\\
0.432100000000000	-0.00251794716693968\\
0.433100000000000	-0.00240832804855895\\
0.434100000000000	-0.00230193753327174\\
0.435100000000000	-0.00219872736094212\\
0.436100000000000	-0.00209864927143415\\
0.437100000000000	-0.00200165500461191\\
0.438100000000000	-0.00190769630033948\\
0.439100000000000	-0.00181672489848092\\
0.440100000000000	-0.00172869253890032\\
0.441100000000000	-0.00164355096146174\\
0.442100000000000	-0.00156125190602925\\
0.443100000000000	-0.00148174711246693\\
0.444100000000000	-0.00140498832063886\\
0.445100000000000	-0.00133092727040911\\
0.446100000000000	-0.00125951570164174\\
0.447100000000000	-0.00119070535420083\\
0.448100000000000	-0.00112444796795046\\
0.449100000000000	-0.00106069528275470\\
0.450100000000000	-0.000999399038477621\\
0.451100000000000	-0.000940510974983293\\
0.452100000000000	-0.000883982832135792\\
0.453100000000000	-0.000829766349799191\\
0.454100000000000	-0.000777813267837564\\
0.455100000000000	-0.000728075326114987\\
0.456100000000000	-0.000680504264495526\\
0.457100000000000	-0.000635051822843257\\
0.458100000000000	-0.000591669741022254\\
0.459100000000000	-0.000550309758896590\\
0.460100000000000	-0.000510923616330340\\
0.461100000000000	-0.000473463053187572\\
0.462100000000000	-0.000437879809332363\\
0.463100000000000	-0.000404125624628784\\
0.464100000000000	-0.000372152238940909\\
0.465100000000000	-0.000341911392132813\\
0.466100000000000	-0.000313354824068566\\
0.467100000000000	-0.000286434274612241\\
0.468100000000000	-0.000261101483627913\\
0.469100000000000	-0.000237308190979654\\
0.470100000000000	-0.000215006136531539\\
0.471100000000000	-0.000194147060147638\\
0.472100000000000	-0.000174682701692026\\
0.473100000000000	-0.000156564801028775\\
0.474100000000000	-0.000139745098021959\\
0.475100000000000	-0.000124175332535651\\
0.476100000000000	-0.000109807244433923\\
0.477100000000000	-9.65925735808491e-05\\
0.478100000000000	-8.44830598405020e-05\\
0.479100000000000	-7.34304430769548e-05\\
0.480100000000000	-6.33864631542813e-05\\
0.481100000000000	-5.43028599365531e-05\\
0.482100000000000	-4.61313732878441e-05\\
0.483100000000000	-3.88237430722273e-05\\
0.484100000000000	-3.23317091537757e-05\\
0.485100000000000	-2.66070113965628e-05\\
0.486100000000000	-2.16013896646610e-05\\
0.487100000000000	-1.72665838221435e-05\\
0.488100000000000	-1.35543337330836e-05\\
0.489100000000000	-1.04163792615542e-05\\
0.490100000000000	-7.80446027162861e-06\\
0.491100000000000	-5.67031662737953e-06\\
0.492100000000000	-3.96568819288020e-06\\
0.493100000000000	-2.64231483220368e-06\\
0.494100000000000	-1.65193640942305e-06\\
0.495100000000000	-9.46292788611387e-07\\
0.496100000000000	-4.77123833841668e-07\\
0.497100000000000	-1.96169409187011e-07\\
0.498100000000000	-5.51693787204752e-08\\
0.499100000000000	-5.86360651512216e-09\\
0.500100000000000	8.04335598783523e-12\\
0.501100000000000	1.07057068197942e-08\\
0.502100000000000	7.44895198032363e-08\\
0.503100000000000	2.39619618233253e-07\\
0.504100000000000	5.54356138036785e-07\\
0.505100000000000	1.06695921514073e-06\\
0.506100000000000	1.82568898547209e-06\\
0.507100000000000	2.87880558495777e-06\\
0.508100000000000	4.27456914952473e-06\\
0.509100000000000	6.06123981509990e-06\\
0.510100000000000	8.28707771761004e-06\\
0.511100000000000	1.10003429929824e-05\\
0.512100000000000	1.42492957771438e-05\\
0.513100000000000	1.80821962060211e-05\\
0.514100000000000	2.25473044155414e-05\\
0.515100000000000	2.76928805416311e-05\\
0.516100000000000	3.35671847202179e-05\\
0.517100000000000	4.02184770872284e-05\\
0.518100000000000	4.76950177785896e-05\\
0.519100000000000	5.60450669302284e-05\\
0.520100000000000	6.53168846780710e-05\\
0.521100000000000	7.55587311580458e-05\\
0.522100000000000	8.68188665060789e-05\\
0.523100000000000	9.91455508580974e-05\\
0.524100000000000	0.000112587044350028\\
0.525100000000000	0.000127191607117797\\
0.526100000000000	0.000143007499297333\\
0.527100000000000	0.000160082981024562\\
0.528100000000000	0.000178466312435412\\
0.529100000000000	0.000198205753665808\\
0.530100000000000	0.000219349564851676\\
0.531100000000000	0.000241946006128947\\
0.532100000000000	0.000266043337633546\\
0.533100000000000	0.000291689819501399\\
0.534100000000000	0.000318933711868434\\
0.535100000000000	0.000347823274870576\\
0.536100000000000	0.000378406768643755\\
0.537100000000000	0.000410732453323897\\
0.538100000000000	0.000444848589046928\\
0.539100000000000	0.000480803435948775\\
0.540100000000000	0.000518645254165363\\
0.541100000000000	0.000558422303832624\\
0.542100000000000	0.000600182845086482\\
0.543100000000000	0.000643975138062864\\
0.544100000000000	0.000689847442897697\\
0.545100000000000	0.000737848019726905\\
0.546100000000000	0.000788025128686421\\
0.547100000000000	0.000840427029912169\\
0.548100000000000	0.000895101983540076\\
0.549100000000000	0.000952098249706068\\
0.550100000000000	0.00101146408854607\\
0.551100000000000	0.00107324776019601\\
0.552100000000000	0.00113749752479183\\
0.553100000000000	0.00120426164246943\\
0.554100000000000	0.00127358837336475\\
0.555100000000000	0.00134552597761372\\
0.556100000000000	0.00142012271535227\\
0.557100000000000	0.00149742684671632\\
0.558100000000000	0.00157748663184179\\
0.559100000000000	0.00166035033086462\\
0.560100000000000	0.00174606620392073\\
0.561100000000000	0.00183468251114605\\
0.562100000000000	0.00192624751267651\\
0.563100000000000	0.00202080946864803\\
0.564100000000000	0.00211841663919654\\
0.565100000000000	0.00221911728445796\\
0.566100000000000	0.00232295966456824\\
0.567100000000000	0.00242999203966328\\
0.568100000000000	0.00254026266987902\\
0.569100000000000	0.00265381981535138\\
0.570100000000000	0.00277071173621629\\
0.571100000000000	0.00289098669260968\\
0.572100000000000	0.00301469294466748\\
0.573100000000000	0.00314187875252561\\
0.574100000000000	0.00327259237632000\\
0.575100000000000	0.00340688207618657\\
0.576100000000000	0.00354479611226126\\
0.577100000000000	0.00368638274467999\\
0.578100000000000	0.00383169023357868\\
0.579100000000000	0.00398076683909327\\
0.580100000000000	0.00413366082135967\\
0.581100000000000	0.00429042044051383\\
0.582100000000000	0.00445109395669167\\
0.583100000000000	0.00461572963002910\\
0.584100000000000	0.00478437572066207\\
0.585100000000000	0.00495708048872647\\
0.586100000000000	0.00513389219435827\\
0.587100000000000	0.00531485909769338\\
0.588100000000000	0.00550002945886773\\
0.589100000000000	0.00568945153801724\\
0.590100000000000	0.00588317359527783\\
0.591100000000000	0.00608124389078545\\
0.592100000000000	0.00628371068467601\\
0.593100000000000	0.00649062223708544\\
0.594100000000000	0.00670202680814967\\
0.595100000000000	0.00691797265800461\\
0.596100000000000	0.00713850804678622\\
0.597100000000000	0.00736368123463041\\
0.598100000000000	0.00759354048167310\\
0.599100000000000	0.00782813404805022\\
0.600100000000000	0.00806751019389769\\
0.601100000000000	0.00831171717935146\\
0.602100000000000	0.00856080326454744\\
0.603100000000000	0.00881481670962157\\
0.604100000000000	0.00907380577470976\\
0.605100000000000	0.00933781871994793\\
0.606100000000000	0.00960690380547203\\
0.607100000000000	0.00988110929141799\\
0.608100000000000	0.0101604834379217\\
0.609100000000000	0.0104450745051192\\
0.610100000000000	0.0107349307531462\\
0.611100000000000	0.0110301004421388\\
0.612100000000000	0.0113306318322329\\
0.613100000000000	0.0116365731835644\\
0.614100000000000	0.0119479727562693\\
0.615100000000000	0.0122648788104834\\
0.616100000000000	0.0125873396063427\\
0.617100000000000	0.0129154034039831\\
0.618100000000000	0.0132491184635405\\
0.619100000000000	0.0135885330451510\\
0.620100000000000	0.0139336954089503\\
0.621100000000000	0.0142846538150744\\
0.622100000000000	0.0146414565236593\\
0.623100000000000	0.0150041517948409\\
0.624100000000000	0.0153727878887551\\
0.625100000000000	0.0157474130655379\\
0.626100000000000	0.0161280755853250\\
0.627100000000000	0.0165148237082526\\
0.628100000000000	0.0169077056944565\\
0.629100000000000	0.0173067698040726\\
0.630100000000000	0.0177120642972369\\
0.631100000000000	0.0181236374340853\\
0.632100000000000	0.0185415374747537\\
0.633100000000000	0.0189658126793780\\
0.634100000000000	0.0193965113080943\\
0.635100000000000	0.0198336816210383\\
0.636100000000000	0.0202773718783461\\
0.637100000000000	0.0207276303401535\\
0.638100000000000	0.0211845052665965\\
0.639100000000000	0.0216480449178110\\
0.640100000000000	0.0221182975539330\\
0.641100000000000	0.0225953114350983\\
0.642100000000000	0.0230791348214429\\
0.643100000000000	0.0235698159731027\\
0.644100000000000	0.0240674031502137\\
0.645100000000000	0.0245719446129118\\
0.646100000000000	0.0250834886213328\\
0.647100000000000	0.0256020834356127\\
0.648100000000000	0.0261277773158875\\
0.649100000000000	0.0266606185222931\\
0.650100000000000	0.0272006553149655\\
0.651100000000000	0.0277479359540404\\
0.652100000000000	0.0283025086996538\\
0.653100000000000	0.0288644218119418\\
0.654100000000000	0.0294337235510402\\
0.655100000000000	0.0300104621770850\\
0.656100000000000	0.0305946859502120\\
0.657100000000000	0.0311864431305571\\
0.658100000000000	0.0317857819782564\\
0.659100000000000	0.0323927507534459\\
0.660100000000000	0.0330073977162613\\
0.661100000000000	0.0336297711268385\\
0.662100000000000	0.0342599192453135\\
0.663100000000000	0.0348978903318223\\
0.664100000000000	0.0355437326465009\\
0.665100000000000	0.0361974944494851\\
0.666100000000000	0.0368592240009107\\
0.667100000000000	0.0375289695609138\\
0.668100000000000	0.0382067793896303\\
0.669100000000000	0.0388927017471962\\
0.670100000000000	0.0395867848937473\\
0.671100000000000	0.0402890770894195\\
0.672100000000000	0.0409996265943488\\
0.673100000000000	0.0417184816686712\\
0.674100000000000	0.0424456905725226\\
0.675100000000000	0.0431813015660389\\
0.676100000000000	0.0439253629093558\\
0.677100000000000	0.0446779228626096\\
0.678100000000000	0.0454390296859359\\
0.679100000000000	0.0462087316394710\\
0.680100000000000	0.0469870769833505\\
0.681100000000000	0.0477741139777105\\
0.682100000000000	0.0485698908826868\\
0.683100000000000	0.0493744559584153\\
0.684100000000000	0.0501878574650323\\
0.685100000000000	0.0510101436626732\\
0.686100000000000	0.0518413628114743\\
0.687100000000000	0.0526815631715714\\
0.688100000000000	0.0535307930031003\\
0.689100000000000	0.0543891005661973\\
0.690100000000000	0.0552565341209979\\
0.691100000000000	0.0561331419276383\\
0.692100000000000	0.0570189722462542\\
0.693100000000000	0.0579140733369817\\
0.694100000000000	0.0588184934599569\\
0.695100000000000	0.0597322808753153\\
0.696100000000000	0.0606554838431932\\
0.697100000000000	0.0615881506237262\\
0.698100000000000	0.0625303294770504\\
0.699100000000000	0.0634820686633019\\
0.700100000000000	0.0644434164426163\\
0.701100000000000	0.0654144210751299\\
0.702100000000000	0.0663951308209781\\
0.703100000000000	0.0673855939402973\\
0.704100000000000	0.0683858586932233\\
0.705100000000000	0.0693959733398919\\
0.706100000000000	0.0704159861404392\\
0.707100000000000	0.0714459453550010\\
0.708100000000000	0.0724858992437131\\
0.709100000000000	0.0735358960667119\\
0.710100000000000	0.0745959840841328\\
0.711100000000000	0.0756662115561121\\
0.712100000000000	0.0767466267427855\\
0.713100000000000	0.0778372779042889\\
0.714100000000000	0.0789382133007585\\
0.715100000000000	0.0800494811923300\\
0.716100000000000	0.0811711298391395\\
0.717100000000000	0.0823032075013226\\
0.718100000000000	0.0834457624390154\\
0.719100000000000	0.0845988429123541\\
0.720100000000000	0.0857624971814743\\
0.721100000000000	0.0869367735065121\\
0.722100000000000	0.0881217201476031\\
0.723100000000000	0.0893173853648836\\
0.724100000000000	0.0905238174184895\\
0.725100000000000	0.0917410645685565\\
0.726100000000000	0.0929691750752208\\
0.727100000000000	0.0942081971986180\\
0.728100000000000	0.0954581791988843\\
0.729100000000000	0.0967191693361556\\
0.730100000000000	0.0979912158705676\\
0.731100000000000	0.0992743670622566\\
0.732100000000000	0.100568671171358\\
0.733100000000000	0.101874176458008\\
0.734100000000000	0.103190931182343\\
0.735100000000000	0.104518983604498\\
0.736100000000000	0.105858381984610\\
0.737100000000000	0.107209174582814\\
0.738100000000000	0.108571409659246\\
0.739100000000000	0.109945135474043\\
0.740100000000000	0.111330400287340\\
0.741100000000000	0.112727252359273\\
0.742100000000000	0.114135739949977\\
0.743100000000000	0.115555911319590\\
0.744100000000000	0.116987814728246\\
0.745100000000000	0.118431498436083\\
0.746100000000000	0.119887010703235\\
0.747100000000000	0.121354399789838\\
0.748100000000000	0.122833713956029\\
0.749100000000000	0.124325001461944\\
0.750100000000000	0.125828310567718\\
0.751100000000000	0.127343689533487\\
0.752100000000000	0.128871186619388\\
0.753100000000000	0.130410850085555\\
0.754100000000000	0.131962728192126\\
0.755100000000000	0.133526869199236\\
0.756100000000000	0.135103321367021\\
0.757100000000000	0.136692132955617\\
0.758100000000000	0.138293352225159\\
0.759100000000000	0.139907027435784\\
0.760100000000000	0.141533206847628\\
0.761100000000000	0.143171938720827\\
0.762100000000000	0.144823271315516\\
0.763100000000000	0.146487252891831\\
0.764100000000000	0.148163931709909\\
0.765100000000000	0.149853356029886\\
0.766100000000000	0.151555574111896\\
0.767100000000000	0.153270634216077\\
0.768100000000000	0.154998584602563\\
0.769100000000000	0.156739473531492\\
0.770100000000000	0.158493349262999\\
0.771100000000000	0.160260260057219\\
0.772100000000000	0.162040254174289\\
0.773100000000000	0.163833379874345\\
0.774100000000000	0.165639685417523\\
0.775100000000000	0.167459219063958\\
0.776100000000000	0.169292029073787\\
0.777100000000000	0.171138163707145\\
0.778100000000000	0.172997671224169\\
0.779100000000000	0.174870599884994\\
0.780100000000000	0.176756997949755\\
0.781100000000000	0.178656913678591\\
0.782100000000000	0.180570395331635\\
0.783100000000000	0.182497491169024\\
0.784100000000000	0.184438249450894\\
0.785100000000000	0.186392718437381\\
0.786100000000000	0.188360946388621\\
0.787100000000000	0.190342981564749\\
0.788100000000000	0.192338872225902\\
0.789100000000000	0.194348666632216\\
0.790100000000000	0.196372413043826\\
0.791100000000000	0.198410159720868\\
0.792100000000000	0.200461954923479\\
0.793100000000000	0.202527846911794\\
0.794100000000000	0.204607883945949\\
0.795100000000000	0.206702114286081\\
0.796100000000000	0.208810586192324\\
0.797100000000000	0.210933347924816\\
0.798100000000000	0.213070447743691\\
0.799100000000000	0.215221933909086\\
0.800100000000000	0.217387854681137\\
0.801100000000000	0.219568258319979\\
0.802100000000000	0.221763193085749\\
0.803100000000000	0.223972707238583\\
0.804100000000000	0.226196849038616\\
0.805100000000000	0.228435666745984\\
0.806100000000000	0.230689208620824\\
0.807100000000000	0.232957522923271\\
0.808100000000000	0.235240657913461\\
0.809100000000000	0.237538661851531\\
0.810100000000000	0.239851582997615\\
0.811100000000000	0.242179469611850\\
0.812100000000000	0.244522369954372\\
0.813100000000000	0.246880332285317\\
0.814100000000000	0.249253404864821\\
0.815100000000000	0.251641635953019\\
0.816100000000000	0.254045073810048\\
0.817100000000000	0.256463766696043\\
0.818100000000000	0.258897762871141\\
0.819100000000000	0.261347110595477\\
0.820100000000000	0.263811858129187\\
0.821100000000000	0.266292053732408\\
0.822100000000000	0.268787745665275\\
0.823100000000000	0.271298982187924\\
0.824100000000000	0.273825811560491\\
0.825100000000000	0.276368282043111\\
0.826100000000000	0.278926441895922\\
0.827100000000000	0.281500339379058\\
0.828100000000000	0.284090022752656\\
0.829100000000000	0.286695540276852\\
0.830100000000000	0.289316940211781\\
0.831100000000000	0.291954270817580\\
0.832100000000000	0.294607580354384\\
0.833100000000000	0.297276917082329\\
0.834100000000000	0.299962329261552\\
0.835100000000000	0.302663865152188\\
0.836100000000000	0.305381573014373\\
0.837100000000000	0.308115501108243\\
0.838100000000000	0.310865697693934\\
0.839100000000000	0.313632211031582\\
0.840100000000000	0.316415089381323\\
0.841100000000000	0.319214381003293\\
0.842100000000000	0.322030134157627\\
0.843100000000000	0.324862397104462\\
0.844100000000000	0.327711218103933\\
0.845100000000000	0.330576645416177\\
0.846100000000000	0.333458727301329\\
0.847100000000000	0.336357512019526\\
0.848100000000000	0.339273047830902\\
0.849100000000000	0.342205382995595\\
0.850100000000000	0.345154565773740\\
0.851100000000000	0.348120644425474\\
0.852100000000000	0.351103667210930\\
0.853100000000000	0.354103682390247\\
0.854100000000000	0.357120738223560\\
0.855100000000000	0.360154882971004\\
0.856100000000000	0.363206164892716\\
0.857100000000000	0.366274632248832\\
0.858100000000000	0.369360333299487\\
0.859100000000000	0.372463316304818\\
0.860100000000000	0.375583629524959\\
0.861100000000000	0.378721321220049\\
0.862100000000000	0.381876439650221\\
0.863100000000000	0.385049033075613\\
0.864100000000000	0.388239149756360\\
0.865100000000000	0.391446837952597\\
0.866100000000000	0.394672145924462\\
0.867100000000000	0.397915121932089\\
0.868100000000000	0.401175814235615\\
0.869100000000000	0.404454271095176\\
0.870100000000000	0.407750540770908\\
0.871100000000000	0.411064671522946\\
0.872100000000000	0.414396711611426\\
0.873100000000000	0.417746709296485\\
0.874100000000000	0.421114712838259\\
0.875100000000000	0.424500770496882\\
0.876100000000000	0.427904930532492\\
0.877100000000000	0.431327241205224\\
0.878100000000000	0.434767750775214\\
0.879100000000000	0.438226507502598\\
0.880100000000000	0.441703559647512\\
0.881100000000000	0.445198955470092\\
0.882100000000000	0.448712743230473\\
0.883100000000000	0.452244971188792\\
0.884100000000000	0.455795687605185\\
0.885100000000000	0.459364940739787\\
0.886100000000000	0.462952778852735\\
0.887100000000000	0.466559250204164\\
0.888100000000000	0.470184403054210\\
0.889100000000000	0.473828285663010\\
0.890100000000000	0.477490946290699\\
0.891100000000000	0.481172433197413\\
0.892100000000000	0.484872794643288\\
0.893100000000000	0.488592078888460\\
0.894100000000000	0.492330334193065\\
0.895100000000000	0.496087608817238\\
0.896100000000000	0.499863951021117\\
0.897100000000000	0.503659409064836\\
0.898100000000000	0.507474031208531\\
0.899100000000000	0.511307865712339\\
0.900100000000000	0.515160960836396\\
0.901100000000000	0.519033364840837\\
0.902100000000000	0.522925125985798\\
0.903100000000000	0.526836292531415\\
0.904100000000000	0.530766912737825\\
0.905100000000000	0.534717034865163\\
0.906100000000000	0.538686707173565\\
0.907100000000000	0.542675977923166\\
0.908100000000000	0.546684895374104\\
0.909100000000000	0.550713507786513\\
0.910100000000000	0.554761863420530\\
0.911100000000000	0.558830010536290\\
0.912100000000000	0.562917997393930\\
0.913100000000000	0.567025872253586\\
0.914100000000000	0.571153683375393\\
0.915100000000000	0.575301479019487\\
0.916100000000000	0.579469307446005\\
0.917100000000000	0.583657216915082\\
0.918100000000000	0.587865255686854\\
0.919100000000000	0.592093472021457\\
0.920100000000000	0.596341914179027\\
0.921100000000000	0.600610630419700\\
0.922100000000000	0.604899669003612\\
0.923100000000000	0.609209078190899\\
0.924100000000000	0.613538906241695\\
0.925100000000000	0.617889201416139\\
0.926100000000000	0.622260011974366\\
0.927100000000000	0.626651386176510\\
0.928100000000000	0.631063372282710\\
0.929100000000000	0.635496018553099\\
0.930100000000000	0.639949373247815\\
0.931100000000000	0.644423484626993\\
0.932100000000000	0.648918400950769\\
0.933100000000000	0.653434170479280\\
0.934100000000000	0.657970841472659\\
0.935100000000000	0.662528462191045\\
0.936100000000000	0.667107080894573\\
0.937100000000000	0.671706745843378\\
0.938100000000000	0.676327505297598\\
0.939100000000000	0.680969407517366\\
0.940100000000000	0.685632500762820\\
0.941100000000000	0.690316833294097\\
0.942100000000000	0.695022453371329\\
0.943100000000000	0.699749409254656\\
0.944100000000000	0.704497749204211\\
0.945100000000000	0.709267521480132\\
0.946100000000000	0.714058774342554\\
0.947100000000000	0.718871556051613\\
0.948100000000000	0.723705914867445\\
0.949100000000000	0.728561899050185\\
0.950100000000000	0.733439556859971\\
0.951100000000000	0.738338936556937\\
0.952100000000000	0.743260086401220\\
0.953100000000000	0.748203054652955\\
0.954100000000000	0.753167889572279\\
0.955100000000000	0.758154639419327\\
0.956100000000000	0.763163352454236\\
0.957100000000000	0.768194076937141\\
0.958100000000000	0.773246861128178\\
0.959100000000000	0.778321753287483\\
0.960100000000000	0.783418801675192\\
0.961100000000000	0.788538054551442\\
0.962100000000000	0.793679560176367\\
0.963100000000000	0.798843366810104\\
0.964100000000000	0.804029522712789\\
0.965100000000000	0.809238076144557\\
0.966100000000000	0.814469075365545\\
0.967100000000000	0.819722568635889\\
0.968100000000000	0.824998604215724\\
0.969100000000000	0.830297230365186\\
0.970100000000000	0.835618495344412\\
0.971100000000000	0.840962447413537\\
0.972100000000000	0.846329134832697\\
0.973100000000000	0.851718605862029\\
0.974100000000000	0.857130908761667\\
0.975100000000000	0.862566091791749\\
0.976100000000000	0.868024203212409\\
0.977100000000000	0.873505291283784\\
0.978100000000000	0.879009404266010\\
0.979100000000000	0.884536590419222\\
0.980100000000000	0.890086898003557\\
0.981100000000000	0.895660375279151\\
0.982100000000000	0.901257070506139\\
0.983100000000000	0.906877031944658\\
0.984100000000000	0.912520307854842\\
0.985100000000000	0.918186946496829\\
0.986100000000000	0.923876996130755\\
0.987100000000000	0.929590505016753\\
0.988100000000000	0.935327521414963\\
0.989100000000000	0.941088093585518\\
0.990100000000000	0.946872269788555\\
0.991100000000000	0.952680098284210\\
0.992100000000000	0.958511627332618\\
0.993100000000000	0.964366905193917\\
0.994100000000000	0.970245980128240\\
0.995100000000000	0.976148900395725\\
0.996100000000000	0.982075714256508\\
0.997100000000000	0.988026469970724\\
0.998100000000000	0.994001215798510\\
0.999100000000000	1\\
};

\coordinate (x0) at (axis cs:-0.01,0);
\node at (axis cs:0.1,-0.1) {\footnotesize\textcolor{white!15!black}{$W_{\max}$}};
\node at (axis cs:0.25,0.35) {\footnotesize\textcolor{white!15!black}{Steady state}};
\node at (axis cs:0.75,-0.35) {\footnotesize\textcolor{white!15!black}{Max probing}};
\node at (axis cs:0.1,-0.1) {\footnotesize\textcolor{white!15!black}{$W_{\max}$}};
\coordinate (x1) at (axis cs:1,0);

\coordinate (y0) at (axis cs:-0.01,0.25);
\coordinate (y1) at (axis cs:0.5,0.25);
\coordinate (y2) at (axis cs:0.5,-0.25);
\coordinate (y3) at (axis cs:1,-0.25);

 \draw[gray,-,line width=1pt,dashed] (x0) -- (x1);
 \draw[gray,-,line width=1pt,dashed] (y1) -- (y2);
  \draw[arrows={->}] (y0) -- (y1);
  \draw[arrows={->}] (y1) -- (y0); 
  \draw[arrows={->}] (y2) -- (y3);
  \draw[arrows={->}] (y3) -- (y2);

\end{axis}
\end{tikzpicture}

%% file: img/multipath_wifilte.tex
\begin{tikzpicture}
\pgfplotsset{every tick label/.append style={font=\scriptsize}}
\tikzstyle{dotted}= [dash pattern=on \pgflinewidth off 0.5mm] 
\tikzstyle{dashed}= [dash pattern=on 7.5*0.8*0.8pt off 7.5*0.4*0.8pt]
\tikzstyle{dashdotted} = [dash pattern=on 7.5*0.8*0.6pt off 7.5*0.8*0.3pt on \the\pgflinewidth off 7.5*0.8*0.3pt]
\tikzstyle{dotted2} = [dash pattern=on 7.5*0.8*0.3pt off 7.5*0.8*0.2pt]

\begin{axis}[%
width=\fwidth,
height=\fheight,
at={(0\fwidth,0\fheight)},
scale only axis,
xmode=log,
log ticks with fixed point,
xmin=0.002,
xmax=1,
xlabel near ticks,
xlabel style={font=\footnotesize\color{white!15!black}},
xlabel={Fraction of late traffic (latency over 100 ms)},
ylabel near ticks,
ymin=0,
ymax=20,
ylabel style={font=\footnotesize\color{white!15!black}},
ylabel={Average throughput (Mb/s)},
axis background/.style={fill=white},
xmajorgrids,
ymajorgrids,
]

\addplot [color=violet, line width=0.8pt,  mark=x, mark size=2, mark repeat = 5, mark phase = 1]
  table[row sep=crcr]{%
0.141 7.163\\
};

\addplot [color=orange_D, line width=0.8pt,  mark=triangle*, mark size=1.5, mark repeat = 5, mark phase = 1]
  table[row sep=crcr]{%
0.129 4.672\\
};

\node at (axis cs:0.129,4.66) [anchor=north] {\footnotesize\textcolor{white!15!black}{Sprout}};

\node at (axis cs:0.129,14.53) [anchor=south] {\footnotesize\textcolor{white!15!black}{Verus}};

\addplot [color=green, line width=0.8pt,  mark=diamond*, mark size=1.5, mark repeat = 1, mark phase = 1]
  table[row sep=crcr]{%
0.007 3.223\\
0.009 3.894\\
0.013 4.389\\
0.018 5.139\\
0.026 6.436\\
0.039 7.425\\
0.058 8.2643\\
0.074 9.008\\
0.105 9.660\\
0.138 10.264\\
0.164 10.828\\
0.213 11.849\\
0.249 13.694\\
0.347 14.659\\
0.424 15.591\\
0.516 16.73\\
};

\addplot [color=blue, line width=0.8pt,  mark=star, mark size=1.8, mark repeat = 5, mark phase = 1]
  table[row sep=crcr]{%
0.988 5.406\\
};

\node at (axis cs:1,14) [anchor=east] {\footnotesize\textcolor{white!15!black}{MPTCP Balia}};

\addplot [color=magenta, line width=0.8pt,  mark=square, mark size=1.8, mark repeat = 5, mark phase = 1]
  table[row sep=crcr]{%
0.999 14.265\\
};

\node at (axis cs:1,5.5) [anchor=east] {\footnotesize\textcolor{white!15!black}{MPTCP Cubic}};

\addplot [color=red, line width=0.8pt,  mark=+, mark size=1.8, mark repeat = 5, mark phase = 1]
  table[row sep=crcr]{%
0.129 14.530\\
};

\node at (axis cs:0.141,7.18) [anchor=south] {\footnotesize\textcolor{white!15!black}{Vegas}};

\node at (axis cs:0.018,5.139) [anchor=south east] {\footnotesize\textcolor{white!15!black}{\gls{leap}}};

\end{axis}
\end{tikzpicture}